\documentclass[12pt,oneside]{article}
\usepackage{amsmath,bbm}
\usepackage{breqn}

\begin{document}

\begin{center}
{\large\bf $d = 11$ supergravity on almost flat
  $\mathbf{{{R}^4}}$ times a\\
  compact hyperbolic 7-manifold, and the dip\\
  and bump seen in ATLAS-CONF-2010-088\\}
\vspace{0.14cm}
\vspace*{.05in}
{Chris Austin\footnote{Email: chris@chrisaustin.info}\\
\small 33 Collins Terrace, Maryport, Cumbria CA15 8DL, England\\
}
\end{center}

\begin{center}
{\bf Abstract}
\end{center}

\noindent Rough estimates are presented to show that the bump at 1.7 to 1.9 TeV
seen in
ATLAS-CONF-2010-088 could arise from about $10^{30}$ approximately degenerate
Kaluza-Klein states of the $d = 11$ supergravity multiplet in the $s$ channel,
that could arise from compactification of $d = 11$ supergravity on a 7-manifold
with a compact hyperbolic Cartesian factor of intrinsic volume around
$10^{34}$ and curvature radius an inverse TeV. \ A first hypothesis that the
modes in the bump arise from a large degeneracy that restores agreement
between the spectral staircase and the Weyl asymptotic formula immediately
above the spectral gap gives a number of modes that is too large by a factor
of around 60000. \ An alternative hypothesis that the modes in the bump arise
from harmonic forms on the compact 7-manifold that are classically massless
and acquire approximately equal masses from the leading quantum corrections to
the CJS action naturally explains the slight reduction on a logarithmic scale
in the number of modes relative to the first hypothesis, and predicts that the
bump is spin 0 if the compact hyperbolic factor of large intrinsic volume is
7-dimensional, and a mixture of spins 0 and 1 if it is 5-dimensional or
3-dimensional. \ Even dimensions probably give too many modes. \ A
provisional solution of the quantum-corrected $d = 11$ Einstein equations on a
compact hyperbolic 7-manifold times 4 almost flat extended dimensions whose de
Sitter radius can easily be as large as the observed value is considered, and
a Ho\v{r}ava-Witten boundary is introduced to accommodate the Standard Model
fields.
\vspace{2.0cm}
\tableofcontents

\section{Introduction}
\label{Introduction}

The ATLAS collaboration has presented in ATLAS-CONF-2010-088 {\cite{AC1088}} a
search for events with at least 3 Standard Model objects in the final state,
invariant mass above 800 GeV, and total transverse momentum above 700 GeV,
produced in 295 per nb of proton-proton collisions at centre-of-mass energy
$\sqrt{s} = 7$ TeV. \ Although not commented on in the report, the graphs of
the numbers of events against invariant mass $M_{\mathrm{{inv}}}$, which
are the right-hand graphs on pages 8 and 9 of the report, show that the
measured QCD background decreases more rapidly than the Monte Carlo over the
range $M_{\mathrm{{inv}}} = 1.1$ to 1.6 TeV, and is just half the
predicted background at 1.5 to 1.6 TeV, and there is then a bump at 1.7 to 1.9
TeV, in which the number of events per energy bin increases by a factor of
two.

The data is 3 sigma below the Monte Carlo in the $M_{\mathrm{{inv}}} =
1.5$ to 1.6 TeV bin, and although the data is only 1 sigma above the Monte
Carlo in the 1.8 to 1.9 TeV bin, the data follow a very smooth curve, with
almost no scatter, which suggests that the overall significance of the
combined dip and bump is well above 3 sigma. \ If we assume that the curve
followed by the data below and above the bump is the correct QCD background,
then the total number of events in the bump above this measured background is
8, so the bump is just under 3 sigma evidence for some sort of object in the
$s$ channel with a mass of 1.8 TeV, that produces the excess of events over
the measured background.

No update to ATLAS-CONF-2010-088 has yet been published, but since ATLAS had
collected 45 per pb of collisions by the end of 2010, the dip and bump will by
now be around 36 sigma, if the trend in the graphs has continued.

A natural interpretation of the more rapid decrease of the QCD background than
predicted by the Monte Carlo is that it represents the onset of
Dienes-\hspace{-1pt}Dudas-Gherghetta or Arkani-Hamed-Cohen-Georgi accelerated
unification of the Standard Model coupling constants {\cite{DDG1, DDG2, ACG}},
which begins at about a third of the energy at which compact extra dimensions
accessible to the Standard Model gauge fields start to become visible.

In this article I shall present rough estimates to show that the bump could
arise from about $10^{30}$ approximately degenerate
Kaluza-Klein states of the $d = 11$ supergravity multiplet
in the $s$ channel, which according to a conjecture by Brooks, reported in
\cite{KMST}, on the
lowest nonzero eigenvalue of the Laplace-Beltrami operator on generic compact
hyperbolic manifolds of large intrinsic volume, might arise from
compactification of $d = 11$ supergravity on a manifold with a compact
hyperbolic Cartesian factor of intrinsic volume around $10^{34}$, and
curvature radius around an inverse TeV.

In section \ref{The bulk}, on page \pageref{The bulk}, I shall look for
solutions of the quantum-corrected Einstein equations of $d = 11$ supergravity
{\cite{CJS}} that have 4 flat extended dimensions and a compact 7-manifold
with one or more compact hyperbolic factors, in the presence of magnetic
4-form fluxes whose bilinears are on average covariantly constant. \ The
necessary flux-dependent terms in the dimension 8 local term in the $d = 11$
quantum effective action are not yet known, and I shall use a simple guess for
them for the case when the flux is covariantly constant, which on reduction to
10 dimensions with NS 3-form flux agrees with the Kehagias-Partouche
conjecture {\cite{KP}}, supported by recent calculations by Richards
{\cite{Richards 2}}, up to correction terms that contain factors that
occur in the classical Einstein equation in 10 dimensions.

Using this action solutions are proved impossible for the $\bar{H}^7$,
$\bar{H}^5 \times \bar{H}^2$, and $\bar{H}^4 \times \bar{H}^3$ cases even when
5-branes are added whose world-volumes lie along the 4 extended dimensions and
wrap 2-cycles of the compact manifold $\mathcal{M}^7$, and various types of
search have found no solutions in the $\bar{H}^5 \times S^2$, $\bar{H}^4
\times S^3$, $\bar{H}^3 \times S^4$, and $\bar{H}^3 \times S^2 \times S^2$
cases. \ Here $\bar{H}^n$ denotes a smooth compact quotient of $n$-dimensional
hyperbolic space $H^n$ by a freely-acting discrete subgroup of
$\mathrm{{SO}} \left( n, 1 \right)$, which is the isometry group of
$H^n$. \ However at this level of approximation, where only the leading
quantum corrections to the classical action of $d = 11$ supergravity are
included, the existence of solutions is affected by field redefinitions, which
should have no effect when all orders of perturbation theory are included. \
An admissible field redefinition results in the existence of a solution in the
$\bar{H}^7$ case, and probably also in the other cases. \ A Ho\v{r}ava-Witten
boundary to accommodate the strong/electroweak Standard Model fields is
introduced in section \ref{The HW boundary}, on page \pageref{The HW
boundary}.

The motivation for considering such solutions of quantum-corrected $d = 11$
supergravity is that the space-time we live in is approximately flat up to
distances larger, by a factor of around $10^{61}$, than the radius of
curvature that would be expected from combining the Standard Model (SM) of the
strong/electroweak interactions with Einstein gravity in $3 + 1$ dimensions. \
Rigid extended objects such as the jib of a crane, a wing or the hull of an
aeroplane, or an oil tanker, which are much longer in one dimension than in
their other two dimensions, have a structure that makes the fullest possible
use of their extension in their two short dimensions, in order to stiffen them
in their long dimension. \ These examples suggest that the large size and
flatness of the space-time we live in are supported by structures that make
use of the extension of the universe in small additional spatial dimensions.

Compact hyperbolic manifolds $\bar{H}^n$ with $n \geq 3$ have an ideal
structure for stiffening the extended dimensions in this way, because when the
metric on them is locally symmetric, and their curvature is fixed, they are
completely rigid {\cite{Mostow 1, Mostow 2, Prasad, Thurston, Gromov
Hyperbolic}}. \ Their shape and size is fixed by their topology, and they can
be arbitrarily large. \ In particular, the volume $\bar{V}$ of a compact
hyperbolic manifold $\bar{H}^n$ when its sectional curvature is equal to $-
1$, which I shall call its \emph{intrinsic volume}, is fixed by its
topology, and can be arbitrarily large.

Supergravity in $10 + 1$ dimensions is the unique quantum field theory that
includes gravity, and whose gauge invariances force the cosmological constant
to vanish in its defining dimension {\cite{CJS, Nicolai Townsend van
Nieuwenhuizen, Sagnotti Tomaras, Bautier Deser Henneaux Seminara}}, and $10 +
1$ is the largest number of dimensions in which supergravity can exist if
there is just one time dimension and flat space-time is required to be a
solution of the classical field equations {\cite{Nahm}}. \ Thus a solution of
the quantum corrected field equations of $d = 11$ supergravity on an almost
flat $\mathbf{R}^4$ times a compact hyperbolic 7-manifold $\bar{H}^7$ of
large intrinsic volume, or failing that a compact 7-manifold $\mathcal{M}^7$
with at least one Cartesian factor $\bar{H}^n$ with $n \geq 3$ and large
intrinsic volume, might provide a reasonable candidate model for the
space-time we live in.

$d = 11$ supergravity can only exist on manifolds whose topology allows them
to have a spin structure. \ Such manifolds are called \emph{spin
manifolds}. \ All orientable manifolds of dimensions $n \leq 3$ are spin
manifolds, and Cartesian products of spin manifolds are spin manifolds
{\cite{Michelson Lawson}}. \ A manifold $\mathcal{M}$ is spin if and only if
its second Stiefel-Whitney class $w_2 \left( \mathcal{M} \right) \in H^2
\left( \mathcal{M}; \mathbf{Z}_2 \right)$ vanishes {\cite{Hirzebruch et
al}}, where $H^2 \left( \mathcal{M}; \mathbf{Z}_2 \right)$ is the second
cohomology group with $\mathbf{Z}_2$ coefficients {\cite{Milnor Stasheff}}.
\ Thus in the absence of additional information, the chance that an
$n$-dimensional orientable manifold $\mathcal{M}^n$ with $n \geq 4$ is spin is
$2^{- B_{2, \mathbf{Z}_2}}$, where $B_{2, \mathbf{Z}_2}$, the second Betti
number with $\mathbf{Z}_2$ coefficients, is the number of $\mathbf{Z}_2$
factors in $H^2 \left( \mathcal{M}; \mathbf{Z}_2 \right)$. \ The sum of the
Betti numbers of an $n$-dimensional compact hyperbolic manifold $\bar{H}^n$
with intrinsic volume $\bar{V}$ is bounded above by $b \bar{V}$ for arbitrary
coefficients, where $b$ is a constant that depends only on $n$ and the
coefficient ring {\cite{Gromov Volume Bounded Cohomology, Gromov Volume and
Bounded Cohomology}}, and the number
of topologically distinct $\bar{H}^n$ with intrinsic volume $\bar{V}$ less
than a fixed value $V$ is finite for fixed $n \geq 4$ {\cite{Wang}}, but grows
with $V$ as $V^{cV}$ for sufficiently large $V$, where $c > 0$ is a constant
that depends only on $n$ {\cite{Burger Gelander Lubotzky Mozes}}. \ The Davis
manifold, which is an $\bar{H}^4$, is spin {\cite{Davis manifold, Ratcliffe
Tschantz Davis manifold}}, so we may reasonably expect that for all $n \geq
2$, there are spin $\bar{H}^n$, and that for $n \geq 4$, their number will
grow with $V$ as $V^{cV}$ for sufficiently large $V$, while for $n = 3$, their
number is already infinite for $V \geq 2.03$ {\cite{Cao Meyerhoff, Thurston}}.

Supergravity does not couple to any matter fields in $10 + 1$ smooth
dimensions, but if the $10 + 1$ dimensional space-time has a $9 + 1$
dimensional smooth ``boundary'', which could perhaps better be described as a
flexible mirror, because all the fields on one side of the mirror are exactly
copied, up to sign, on the other side of the mirror, then supergravity in the
$10 + 1$ dimensional ``bulk'' couples to a supersymmetric Yang-Mills multiplet
{\cite{Wess Zumino, Gliozzi Scherk Olive}}, with gauge group $E_8$, on the $9
+ 1$ dimensional ``boundary''. \ The Yang-Mills multiplet is adjacent to the
flexible mirror, but infinitesimally displaced from it, so that it has its own
reflection infinitesimally on the other side of the mirror {\cite{Lu}}. \ This
is called Ho\v{r}ava-Witten (HW) theory {\cite{Horava Witten 1, Horava Witten 2,
Moss 1, Moss 2, Moss 3, Moss 4}}, and I shall assume that we live on an HW
boundary, while the graviton is part of the supergravity multiplet in the $10
+ 1$ dimensional bulk.

An $n$-dimensional compact hyperbolic space $\bar{H}^n$, $n \geq 2$, is a
quotient of $n$-dimen-sional hyperbolic space, $H^n$, by a discrete subgroup
$\Gamma$ of the symmetry group $\mathrm{{SO}} \left( n, 1 \right)$ of
$H^n$, such that $\Gamma$ acts freely on $H^n$, which means that no element of
$\Gamma$ other than the identity leaves any point of $H^n$ invariant. \ This
is equivalent to the requirement that $\Gamma$ have no nontrivial finite
subgroup, which in the language of group theory is expressed by saying that
$\Gamma$ has no torsion. \ The construction of a large family of $\bar{H}^n$,
called arithmetic manifolds {\cite{Borel Harish Chandra}}, is reviewed in
{\cite{Witte Morris}} and section 3 of {\cite{CCHT}}. \ Non-arithmetic
manifolds were constructed from suitable pairs of arithmetic manifolds for all
$n \geq 2$ in {\cite{Gromov Piatetski Shapiro}}. \ It is the rapid growth with
intrinsic volume $\bar{V}$ of the number of topologically distinct finite
coverings of the non-arithmetic manifolds that results in the $V^{cV}$ growth
of the number of topologically distinct $\bar{H}^n$ with $\bar{V} \leq V$
{\cite{Burger Gelander Lubotzky Mozes}}.

For the special case of $n = 3$, an infinite number of topologically distinct
$\bar{H}^3$ can be constructed from each finite-volume cusped hyperbolic
3-manifold $\hat{H}^3$ by applying an operation called Dehn filling to each
cusp, and each $\bar{H}^3$ constructed in this way has $\bar{V} < \hat{V}$,
where $\hat{V}$ is the intrinsic volume of the initial $\hat{H}^3$
{\cite{Thurston}}. \ A large number of $\bar{H}^3$ of small $\bar{V}$ have
been constructed and catalogued in this way, and their properties can be
studied using computer programs {\cite{SnapPea, SnapPy, Snap}}. \ The
Cartesian product of two $\bar{H}^3$ of small $\bar{V}$ might be a possible
topology for the compact Cartesian factor of the HW boundary that accommodates
the Standard Model fields. \ A \emph{cusp} of a finite-volume hyperbolic
$n$-manifold $\hat{H}^n$ is a region of the manifold whose topology is the
Cartesian product of an infinite half-line and a flat $\left( n - 1
\right)$-manifold, for example an $\left( n - 1 \right)$-torus, such that the
area of the cross section decreases so rapidly along the half-line that the
volume of the cusp is finite. \ The number of cusps is always finite.

The $n$-dimensional hyperbolic space $H^n$ can be realized as the hypersurface
$t^2 - \vec{x}^2 = 1$, $t > 0$, in $n + 1$ dimensional Minkowski space. \
Using spherical polar coordinates for $\vec{x}$, the equation reduces to $t^2
- r^2 = 1$, $t > 0$, so we can choose $t = \mathrm{\cosh} \rho$, $r =
\mathrm{\sinh} \rho$, and $\rho$ is then the geodesic distance of the point
$\left( \rho, \theta_1, \ldots, \theta_{n - 1} \right)$ from the origin
{\cite{Costa}}. \ The Riemann tensor of $H^n$ has the form:
\begin{equation}
  \label{Riemann tensor of Hn} R_{i j k l} = g_{i l} g_{j k} - g_{i k} g_{j
  l},
\end{equation}
where $g_{i j}$ is the metric, which means that $H^n$ has constant sectional
curvature equal to $- 1$. \ The Ricci tensor of $H^n$ is then $R_{i j} = -
\left( n - 1 \right) g_{i j}$. \ The Riemann tensor of a $d$-dimensional space
or space-time, with metric $G_{I J}$, is defined in general by:
\begin{equation}
  \label{Riemann tensor} R_{I J} \, \!^K \, \!_L = \partial_I \Gamma_J \, \!^K
  \, \!_L - \partial_J \Gamma_I \, \!^K \, \!_L + \Gamma_I \, \!^K \, \!_M
  \Gamma_J \, \!^M \, \!_L - \Gamma_J \, \!^K \, \!_M \Gamma_I \, \!^M \,
  \!_L,
\end{equation}
and the Ricci tensor and scalar are defined by $R_{I J} = R_{K I} \, \!^K \,
\!_J$, and $R = G^{I J} R_{I J}$. \ The standard Christoffel connection is
$\Gamma_I \, \!^J \, \!_K = \frac{1}{2} G^{J L} \left( \partial_I G_{L K} +
\partial_K G_{L I} - \partial_L G_{I K} \right)$.

I shall use units such that $\hbar = c = 1$. \ The gravitational coupling
constant in 11 dimensions is $\kappa_{11}$. \ The metric is mostly $+$. \
Coordinate indices $I, J, K, \ldots$ run over all 11 dimensions, and
coordinate indices $\mu, \nu, \sigma, \ldots$ are tangent to the four observed
space-time dimensions. \ The bosonic fields of supergravity in 11 dimensions
are the graviton $G_{I J}$, and a 3-index antisymmetric tensor Abelian gauge
field $C_{I J K}$, called a 3-form gauge field, with 4-form field strength
$H_{I J K L} = \partial_I C_{J K L} - \partial_J C_{K L I} + \partial_K C_{L I
J} - \partial_L C_{I J K}$.

The bosonic part of the classical action of $d = 11$ supergravity is
{\cite{CJS}}:
\begin{equation}
  \label{CJS action} S^{\left( \mathrm{{bos}}
  \right)}_{\mathrm{{CJS}}} = \frac{1}{2 \kappa_{11}^2}
  \int_{\mathcal{B}} d^{11} xe \left( R - \frac{1}{48} H_{I J K L} H^{I J K L}
  - \frac{1}{144^2} \epsilon^{I_1 \ldots I_{11}}_{11} C_{I_1 I_2 I_3} H_{I_4
  \ldots I_7} H_{I_8 \ldots I_{11}} \right),
\end{equation}
where $\mathcal{B}$ means the bulk. \ The sign of the Einstein term follows
from the Riemann and Ricci tensor conventions stated above, and the
normalization of the 4-form kinetic term, and the definition of $H_{I J K L}$
given above, are chosen to agree with {\cite{Hyakutake Ogushi 2}}. \ $e =
\sqrt{- G}$ is the determinant of the vielbein $e_{I \hat{J}}$, where $G$ is
the determinant of the metric $G_{I J}$, and the antisymmetric tensor
$\epsilon^{I_1 \ldots I_{11}}_{11}$ is related to the $\mathrm{{SO}}
\left( 10, 1 \right)$ invariant tensor $\epsilon^{\hat{I}_1 \ldots
\hat{I}_{11}}_{11}$, with components $0, \pm 1$, by $\epsilon^{I_1 \ldots
I_{11}}_{11} = e^{I_1} \, \!_{\hat{J}_1} \ldots e^{I_{11}} \,
\!_{\hat{J}_{11}} \epsilon^{\hat{J}_1 \ldots \hat{J}_{11}}_{11}$. \ Hatted
indices are local Lorentz indices. \ $C_{I J K}$ here is related to $C_{I J
K}$ of {\cite{Horava Witten 2, Moss 1, Moss 2, Moss 3, Moss 4}} by $C_{I J K
\left( \mathrm{{here}} \right)} = 6 \sqrt{2} C_{I J K \left(
\mathrm{{HW}} \right)}$, and $H_{I J K L}$ is related to $G_{I J K L}$ of
{\cite{Horava Witten 2, Moss 1, Moss 2, Moss 3, Moss 4}} by $H_{I J K L} =
\sqrt{2} G_{I J K L}$.

I assume the unperturbed metric in 11 dimensions has the form:
\begin{equation}
  \label{metric ansatz} ds_{11}^2 = G_{IJ} dx^I dx^J = a \left( x^C \right)^2
  \eta_{\mu \nu} dx^{\mu} dx^{\nu} + h_{AB} dx^A dx^B
\end{equation}
where $\eta_{\mu \nu} = \mathrm{{diag}} \left( - 1, 1, 1, 1 \right)$ is
the metric on $\left( 3 + 1 \right)$-dimensional Minkowski space, and $h_{AB}$
is the metric on the compact 7-dimensional manifold-with-boundary
$\mathcal{M}^7$. \ Coordinate indices $A, B, C, \ldots$ are tangential to
$\mathcal{M}^7$. \ The ``warp factor'' $a \left( x^C \right)$ is equal to a
constant value $A$ everywhere on $\mathcal{M}^7$ except in the immediate
vicinity of the HW boundary, and it is equal to 1 on the HW boundary. \ I
shall show in section \ref{The HW boundary}, starting on page \pageref{The HW
boundary}, that a consistent semiclassical treatment of the boundary region is
possible if $A$ is $< 1$ but comparable to 1, for example $A = 0.7$. \ I
assume that $\mathcal{M}^7$ has an $n$-dimensional compact hyperbolic
Cartesian factor $\breve{H}^n$ of large intrinsic volume $\breve{V}_n$, which
has one boundary of small intrinsic area $\bar{A}_{n - 1}$, where $\bar{A}_{n
- 1}$ is defined using the metric of sectional curvature $- 1$ on
$\breve{H}^n$, and that the 10-dimensional HW boundary is the Cartesian
product of that boundary, and the remaining Cartesian factor $\mathcal{M}^{7 -
n}$ of $\mathcal{M}^7$, and the 4 extended dimensions. \ $\breve{H}^n$ is
constructed by cutting a complete $n$-dimensional compact hyperbolic manifold
$\bar{H}^n$ of large intrinsic volume $\bar{V}^n$ along a small $\left( n - 1
\right)$-cycle that separates it into two disconnected parts. \ The smaller of
the two parts is discarded, and the other becomes $\breve{H}^n$.

The Einstein action in the 4 extended dimensions has the form:
\begin{equation}
  \label{Einstein action} S_{\mathrm{{Ein}}} = \frac{1}{16 \pi G_N} \int
  d^4 x \sqrt{- g} g^{\mu \nu} R_{\mu \nu} \left( g \right),
\end{equation}
where $g_{\mu \nu}$ differs from $\eta_{\mu \nu}$ by a small perturbation,
that depends on the coordinates $x^{\sigma}$ on the 4 extended dimensions, but
not on the coordinates $x^A$ on $\mathcal{M}^7$. \ $G_N$ is Newton's constant,
with the value {\cite{PDG}}:
\begin{equation}
  \label{Newtons constant} G_N = 6.7087 \times 10^{- 33}
  \textrm{{  TeV}}^{- 2},
\end{equation}
so that $\sqrt{G_N} = 8.1907 \times 10^{- 17}$ TeV$^{- 1} \hspace{3pt} =
\hspace{3pt} 1.6160 \times 10^{- 35}$ metres. \ Comparing with (\ref{CJS
action}), and noting that $R_{I J} \, \!^K \, \!_L$ and hence $R_{I J}$ are
unaltered by rescaling the metric by a constant factor, so that $\sqrt{- G}
G^{\mu \nu} R_{\mu \nu} \left( G \right) = A^4 \sqrt{- g} \sqrt{h}
\frac{1}{A^2} g^{\mu \nu} R_{\mu \nu} \left( g \right)$ everywhere on
$\mathcal{M}^7$ except in the immediate vicinity of the HW boundary, where
$G_{I J}$ here represents the metric obtained from (\ref{metric ansatz}) by
replacing $\eta_{\mu \nu}$ by $g_{\mu \nu}$, we find, in the approximation of
neglecting the volume of the region where $a \left( x^C \right)$ differs
appreciably from $A$, that the volume $V_7 \equiv \int_{\mathcal{M}^7} d^7 x^C
\sqrt{h}$ of $\mathcal{M}^7$ is given by {\cite{ADD1, AADD, ADD2, RS1}}:
\begin{equation}
  \label{V7 in terms of GN} \frac{A^2 V_7}{2 \kappa^2_{11}} = \frac{1}{16 \pi
  G_N}
\end{equation}
The experimental limits on the gravitational coupling constant in $D$
dimensions are expressed in terms of a mass $M_D$, such that for $D = 11$,
$M_{11} = \left( 2 \pi \right)^{7 / 9} \kappa^{- 2 / 9}_{11} = 4.1764
\kappa^{- 2 / 9}_{11}$. \ The latest limit on $M_{11}$ for 7 flat extra
dimensions, from {\cite{Franceschini et al}}, including bounds from
non-observation of effects of virtual graviton exchange and graviton emission
at the LHC, is that $M_{11}$ is larger than about 1.5 TeV, so that $\kappa^{-
2 / 9}_{11}$ is larger than about 0.36 TeV. \ The limit might be less
stringent for hyperbolic large extra dimensions, since the light Kaluza-Klein
modes will be far fewer or absent. \ The dimension $n$ of the large intrinsic
volume hyperbolic Cartesian factor $\breve{H}^n$ of $\mathcal{M}^7$ might be
less than 7, but the lower limit on $M_D$ found in {\cite{Franceschini et al}}
is about 1.5 TeV for all $D \geq 6$. \ If $\kappa^{- 2 / 9}_{11}$ was about
0.36 TeV and $A$ was 1, then from (\ref{Newtons constant}) and (\ref{V7 in
terms of GN}), $V_7$ would be about $5.8 \times 10^{34}$ TeV$^{- 7}$.

The classical field equations of $d = 11$ supergravity do not have any
solutions with 4 flat extended dimensions if $\mathcal{M}^7$ has a compact
hyperbolic factor, for any arrangement of 4-form fluxes such that the flux
bilinears are on average covariantly constant, but in section \ref{The bulk} I
shall show that when the leading quantum corrections to the classical action
are taken into account, together with the freedom to make redefinitions of the
fields of the form $G_{I J}, C_{I J K} \rightarrow G_{I J} + X_{I J}, C_{I J
K} + Y_{I J K}$, where $X_{I J}$ and $Y_{I J K}$ are polynomials in the fields
and their derivatives, with sufficiently small coefficients, then there is a
solution with an almost flat $\mathbf{R}^4$ times a compact hyperbolic
7-manifold $\bar{H}^7$ of large intrinsic volume, with magnetic 4-form fluxes
wrapping 4-cycles of the $\bar{H}^7$. \ Field redefinitions of this type are
like a change of coordinates in ``field space'', so they do not change the
physical content of the theory. \ They do not alter the $S$-matrix
{\cite{Diagrammar, Lee Les Houches, Tseytlin Field Redefinitions}}, and when
the leading higher-order corrections to the classical CJS action are
calculated by requiring anomaly cancellation on 5-branes {\cite{Duff Liu
Minasian, Witten 5 branes, Freed Harvey Minasian Moore, Bilal Metzger,
Harvey}} and in Ho\v{r}ava-Witten theory {\cite{Horava Witten 2, de Alwis, Conrad,
Faux Lust Ovrut, Lu, Bilal Derendinger Sauser, Harmark, Bilal Metzger,
Meissner Olechowski, Moss 3}}, and supersymmetry {\cite{Hyakutake Ogushi 1,
Hyakutake Ogushi 2, Hyakutake, Hyakutake 2,
Metsaev}}, the result is only determined up to the freedom to add
arbitrary linear combinations of terms that vanish when the classical field
equations are satisfied, because the coefficients of such linear combinations
of terms can be adjusted arbitrarily by small amounts, by field redefinitions
of this type.

The leading quantum corrections are local terms that depend on the curvature
of $\mathcal{M}^7$ but not directly on its topology, so the curvature radius
$B$ of the $\bar{H}^7$ is around $\kappa^{2 / 9}_{11}$. \ The value of $B$
depends on the field redefinition parameter, but using the ``principle of
minimal sensitivity'' advocated for dealing with the renormalization scheme
dependence of low order calculations in perturbative QCD {\cite{Stephenson}},
the best value of $B$ is around $0.43 \kappa^{2 / 9}_{11}$. \ Thus if
$\kappa^{- 2 / 9}_{11}$ was about 0.36 TeV, $B$ would be around 1.2 TeV$^{-
1}$, so if $A$ was 1, so that $V_7$ was about $5.8 \times 10^{34}$ TeV$^{-
7}$, the intrinsic volume $\bar{V}$ of the $\bar{H}^7$ would be around $1.6
\times 10^{34}$. \ The $\mathbf{R}^4$ could be exactly flat but for the
quantization of the 4-form fluxes {\cite{Rohm Witten, Witten flux
quantization}}, and for an $\bar{H}^7$ of intrinsic volume $\bar{V}^7 \sim
10^{34}$, the de Sitter radius of the $\mathbf{R}^4$ can easily be as large
as the observed de Sitter radius, which is $16.0$ Gyr $= \hspace{1pt}
\hspace{1pt} 1.51 \times 10^{26}$ metres $= \hspace{1pt} \hspace{1pt}
\hspace{2pt} 0.94 \times 10^{61} \sqrt{G_N}$ {\cite{HST, 0302209 Spergel et
al, de Sitter radius}}.

The diameter $L$ of a compact manifold is by definition the maximum over all
pairs of points of the manifold of the shortest geodesic distance between
them, and I shall call $\bar{L}$, the diameter of a compact hyperbolic
manifold $\bar{H}^n$ when its sectional curvature is equal to $- 1$, its
\emph{intrinsic diameter}. \ If $\bar{H}^n$ is reasonably isotropic, in
the sense that it has a fundamental domain in $H^n$ that is approximately
spherical, then using spherical polar coordinates on $H^n$ as above, the
intrinsic volume $\bar{V}_n$ of $\bar{H}^n$ is approximately related to its
intrinsic diameter $\bar{L}$ by
\begin{equation}
  \label{V bar n} \bar{V}_n \simeq S_{n - 1} \int^{\bar{L} / 2}_0
  \mathrm{\sinh}^{n - 1} \rho d \rho,
\end{equation}
where $S_{n - 1}$ is the area of the $\left( n - 1 \right)$-sphere of unit
radius. \ Thus if $\bar{H}^n$ is reasonably isotropic and has large intrinsic
volume $\bar{V}_n$, the relation between $\bar{V}_n$ and $\bar{L}$ is
approximately
\begin{equation}
  \label{V bar n for large volume} \bar{V}_n \simeq \frac{S_{n - 1}}{2^{n - 1}
  \left( n - 1 \right)} e^{\left( n - 1 \right) \frac{\bar{L}}{2}} .
\end{equation}
Thus for reasonably isotropic $\bar{H}^7$ with volume $V_7 \simeq 5.8 \times
10^{34}$ TeV$^{- 7}$ and curvature radius $B$ around 1.2 TeV$^{- 1}$, so that
$\bar{V}_7 \simeq 1.6 \times 10^{34}$, we find from $S_6 = \frac{16}{15}
\pi^3$ that the intrinsic diameter of $\bar{H}^7$ is $\bar{L} \simeq 27$, so
its actual diameter is $L \simeq 32 \textrm{{  TeV}}^{- 1}$.

On a flat Riemannian manifold of diameter $L$, the lowest non-zero eigenvalue
of the negative of the Laplace-Beltrami operator $\Delta \equiv
\frac{1}{\sqrt{g}} \partial_i \left( \sqrt{g} g^{i j} \partial_j \right)$,
which is the generally covariant form of the Laplacian, is typically $\left(
\frac{2 \pi}{L} \right)^2$. \ However there is some evidence that the lowest
non-zero eigenvalue $\lambda_1$ of $- \Delta$ on a reasonably isotropic
compact hyperbolic manifold $\bar{H}^n$ with sectional curvature $- 1$ and
large diameter $\bar{L}$ will be $\sim 1$ rather than $\sim
\frac{1}{\bar{L}^2}$ {\cite{KMST}}. \ Firstly, the spectrum of $- \Delta$ on
$H^n$, which is completely continuous, starts at $\lambda = \frac{\left( n - 1
\right)^2}{4}$, rather than 0 {\cite{Yau Isoperimetric, Agmon}}. \ Secondly,
for $n = 2$, Brooks and Makover have constructed very large families of
compact Riemann surfaces with sectional curvature $- 1$ and large genus, all
of which have $\lambda_1 > \frac{3}{16}$ {\cite{Brooks Makover 1, Brooks
Makover 2, Brooks Makover 3, Brooks Makover 4}}.

And thirdly, we can use Cheeger's inequality {\cite{Cheeger, Cheeger
Wikipedia}}, which states that for a compact Riemannian manifold
$\mathcal{M}$, the lowest non-zero eigenvalue $\lambda_1$ of $- \Delta$ on
$\mathcal{M}$ is bounded below by:
\begin{equation}
  \label{Cheegers inequality} \lambda_1 \left( \mathcal{M} \right) \geq
  \frac{h^2 \left( \mathcal{M} \right)}{4},
\end{equation}
where $h \left( \mathcal{M} \right)$, called the Cheeger isoperimetric
constant of $\mathcal{M}$, is defined by:
\begin{equation}
  \label{Cheeger constant} h \left( \mathcal{M} \right) \equiv
  \textrm{min}_{\mathcal{N}} \left( \frac{A \left( \mathcal{N}
  \right)}{\mathrm{min} \left( V \left( \mathcal{M}_1 \right), V \left(
  \mathcal{M}_2 \right) \right)} \right),
\end{equation}
where $\mathcal{N}$ denotes a hypersurface in $\mathcal{M}$ that divides
$\mathcal{M}$ into two disconnected parts $\mathcal{M}_1$ and $\mathcal{M}_2$,
$V \left( \mathcal{M}_i \right)$ is the volume of $\mathcal{M}_i$, and $A
\left( \mathcal{N} \right)$ is the area of $\mathcal{N}$, or in other words,
if $\mathcal{M}$ is $n$-dimensional, the $\left( n - 1 \right)$-volume of
$\mathcal{N}$, calculated in the given metric on $\mathcal{M}$.

To apply Cheeger's inequality to a large volume compact hyperbolic manifold
$\bar{H}^n$ with sectional curvature $- 1$ that is reasonably isotropic, in
the sense that it has an approximately spherical fundamental domain, we note
that a totally geodesic hypersurface through the centre of the approximately
spherical fundamental domain will not, in general, divide the manifold into
two disconnected parts, because the surface of the fundamental domain consists
of a large number of $\left( n - 1 \right)$-dimensional polyhedral faces, each
of which is identified with another face typically roughly on the opposite
side of the fundamental domain. \ To minimize the value of $\frac{A \left(
\mathcal{N} \right)}{\mathrm{\min} \left( V \left( \mathcal{M}_1 \right), V
\left( \mathcal{M}_2 \right) \right)}$, we first look for minimal area
topologically non-trivial hypersurfaces $\mathcal{N}$ that divide $\bar{H}^n$
into two disconnected parts of approximately equal volume, and we see that a
typical such $\mathcal{N}$ will consist of one totally geodesic hypersurface
through the centre of the fundamental domain, plus polyhedral pieces that
cover exactly half the surface of the fundamental domain. \ Using spherical
polar coordinates as before, we see that for $\bar{H}^n$ of large intrinsic
volume and diameter $\bar{L}$, the area of the part of $\mathcal{N}$ that
covers half the surface of the fundamental domain is
\begin{equation}
  \label{A of mathcal N} A \left( \mathcal{N}_{\mathrm{{half}}
  \,\,\mathrm{{sphere}}} \right) \simeq \frac{S_{n - 1}}{2^n}
  e^{\left( n - 1 \right) \frac{\bar{L}}{2}} \simeq \frac{\left( n - 1
  \right)}{2} \bar{V}_n,
\end{equation}
and the area of the part of $\mathcal{N}$ that is a totally geodesic
hypersurface through the origin is negligible in comparison to this,
suggesting that the Cheeger constant of $\bar{H}^n$ is $h \left( \bar{H}^n
\right) \simeq n - 1$. \ As a check on this, we consider spherical
$\mathcal{N}$ of radius $\rho$ centred at the centre of the fundamental
domain. \ For all $\rho \gg 1$ the area of such an $\mathcal{N}$ is $n - 1$
times the volume it encloses, so for all $\rho \gg 1$, $\rho \leq \bar{L} -
\frac{2 \mathrm{\ln} 2}{n - 1}$, the value of $\frac{A \left( \mathcal{N}
\right)}{\mathrm{\min} \left( V \left( \mathcal{M}_1 \right), V \left(
\mathcal{M}_2 \right) \right)}$ is again $n - 1$. \ Thus it looks likely that
for all reasonably isotropic compact hyperbolic manifolds $\bar{H}^n$ with
sectional curvature $- 1$ and large intrinsic volume, the lowest nonzero
eigenvalue $\lambda_1$ of $- \Delta$ on $\bar{H}^n$ will by Cheeger's
inequality be bounded below by:
\begin{equation}
  \label{lower bound on lambda 1} \lambda_1 \geq \frac{\left( n - 1
  \right)^2}{4}
\end{equation}
This is the value of $\lambda$ at which the spectrum starts on $H^n$
\cite{Donnelly}, and the
lower bound on $\lambda$ for $H^n$ could be obtained from Cheeger's inequality
using spheres of radius $\rho \gg 1$ as above, if Cheeger's inequality could
be applied to $H^n$.

Furthermore, considering now $\bar{H}^n$ with sectional curvature $- 1$ and
large intrinsic volume that depart somewhat from being reasonably isotropic,
there is for $n = 3$ a family of manifolds called Bianchi manifolds of
arbitrarily large but finite intrinsic volume, that depart to the greatest
possible extent from being reasonably isotropic, namely they have a finite
number of cusps as described just before (\ref{Riemann tensor of Hn}) above,
for which $\lambda_1$ has been bounded below by $21 / 25 = 0.84$ and is
conjectured to be equal to 1 {\cite{Sarnak, Luo, Rudnick Sarnak, Orlando
Park}}. \ The diameter of a cusped hyperbolic $n$-manifold $\hat{H}^n$ of
finite volume and sectional curvature $- 1$ is infinite, and each cusp
contributes a continuous part to the spectrum of $- \Delta$ on $\hat{H}^n$. \
However just as for $H^n$, the continuous part of the spectrum starts at
$\lambda = \frac{\left( n - 1 \right)^2}{4}$, rather than 0 {\cite{Mazzeo
Phillips, Donnelly Xavier}}. \ For $n = 3$ the Dehn filling construction
produces from any cusped $\hat{H}^n$ of finite volume an infinite sequence of
compact $\bar{H}_i^n$ that converge to $\hat{H}^n$ as $i \rightarrow \infty$
in the sense that the intrinsic volumes converge, and the $\bar{H}_i^n$ look
almost identical to $\hat{H}^n$ except for regions that move further and
further out along the cusps of $\hat{H}^n$ as $i \rightarrow \infty$
{\cite{Thurston}}, and the eigenvalues of $- \Delta$ below the bottom of the
continuous spectrum of $\hat{H}^n$ are limits of the eigenvalues of the
$\bar{H}_i^n$ {\cite{Colbois Courtois 1, Colbois Courtois 2}}. \ Thus for $n =
3$ there are infinite sequences of compact $\bar{H}_i^n$ with sectional
curvature $- 1$ that converge to fixed cusped Bianchi manifolds, which have
arbitrarily large but finite intrinsic volumes, and these $\bar{H}_i^n$ have
$\lambda_1$ converging to a number that is $\geq 0.84$ and conjectured to be
1, while their diameters tend to $\infty$ and their intrinsic volumes tend to
fixed finite values, that can, however, be arbitrarily large.

So it seems possible, at least for $n = 2$ and $n = 3$, that a compact
hyperbolic manifold $\bar{H}^n$ with sectional curvature $- 1$ and arbitrarily
large intrinsic volume will have $\lambda_1 \simeq \frac{\left( n - 1
\right)^2}{4}$ even if it departs substantially from being isotropic. \ This
is in agreement with a conjecture by Brooks reported in {\cite{KMST}}. \ I
shall now assume that this is the case, which means that classically, the
lightest massive spin 2 Kaluza-Klein graviton would have mass $\sim \kappa^{-
2 / 9}_{11}$ and thus around a TeV in the geometry considered above.

In contrast with the spectrum of $- \Delta$, the spectrum of the Dirac operator
on the $n$-dimensional hyperbolic space $H^n$ is the whole real line
{\cite{Bunke, Bar}}. \ Thus in view of the correspondence between the spectrum
of $- \Delta$ on $H^n$ and on a compact hyperbolic manifold $\bar{H}^n$ of
large intrinsic volume $\bar{V}$ and intrinsic diameter $\bar{L}$ as
considered above, it seems likely that classically, the lightest gravitino
modes would have masses $\sim \frac{2 \pi}{\bar{L} B}$, and thus around 0.2
TeV in the geometry considered above, but just as in the case of large flat
extra dimensions, their effects would not show up at the LHC below energies
around $M_{11}$, which is about 1.5 TeV in this example.

The $\mathbf{R}_{\mathrm{{flat}}}^4 \times \bar{H}^7$ solution of the
quantum-corrected field equations of $d = 11$ supergravity found in section
\ref{The bulk} depends essentially on the leading quantum corrections to the
classical action, so there will be corresponding corrections $\sim \kappa^{- 2
/ 9}_{11}$ to the masses of all the Kaluza-Klein modes. \ I shall assume
provisionally that the effect of these corrections is to increase the masses
of the Kaluza-Klein modes. \ This is particularly significant for the 3-form
gauge field $C_{I J K}$, since a compact hyperbolic manifold $\bar{H}^n$ of
large intrinsic volume $\bar{V}$ has Betti numbers $B_1, \ldots, B_{n - 1}$
that in a particular family of examples are bounded below by constants times
powers $>0$ and $<1$ of $\bar{V}$ {\cite{Xue}}, and for $n = 7$ these would
lead classically to $B_1$ massless 2-form Abelian gauge fields, each equivalent
in $3 + 1$ dimensions to a massless scalar \cite{Cremmer Scherk, Kalb Ramond},
$B_2$ massless Abelian vector gauge fields, and $B_3$ massless scalars. \ There
would also be $B_1$ additional massless Abelian vector gauge fields arising
from
$G_{I J}$. \ I shall assume provisionally that all these classically massless
modes aquire masses $\sim \kappa^{- 2 / 9}_{11}$ from the leading quantum
corrections.

If we write the eigenvalues of $- \Delta$ on an $n$-dimensional compact
manifold $\mathcal{M}^n$, with volume $V_n$, as $k^2$, then for large
$\bar{k}$, the number of modes with $k \leq \bar{k}$ is approximately given by
the Weyl asymptotic formula as
\begin{equation}
  \label{number of modes up to k bar} \frac{V_n}{\left( 2 \pi \right)^n} S_{n
  - 1} \frac{\bar{k}^n}{n},
\end{equation}
which corresponds to a density of states $\frac{V_n}{\left( 2 \pi \right)^n}$
in momentum space. \ In numerical studies of the spectrum of $- \Delta$ on
compact hyperbolic 3-manifolds of small intrinsic volume, Inoue found that the
Weyl asymptotic formula is actually approximately valid down to $\bar{k}^2
\simeq \lambda_1$, the lowest non-zero eigenvalue, so that if $\lambda_1$
occurs at a larger value of $\bar{k}$ than would be expected from the Weyl
formula, there is a degeneracy or approximate degeneracy of eigenvalues near
$\lambda_1$, that restores agreement with the Weyl formula for $\bar{k}^2$
above $\lambda_1$ {\cite{Inoue}}.

It seems reasonable to expect that a similar approximate degeneracy of
eigenvalues near $\lambda_1$, that restores agreement with the Weyl asymptotic
formula (\ref{number of modes up to k bar}) for $\bar{k}^2 > \lambda_1$, will
also occur for compact hyperbolic manifolds of large intrinsic volume, since
the number of modes up to $\bar{k} = \sqrt{\lambda_1}$ given by (\ref{number
of modes up to k bar}) is already large, comparable to the intrinsic volume
$\bar{V}_n$. \ I shall provisionally assume that this happens, which means that
for the
geometry described above, we expect in the region of $10^{34}$ approximately
degenerate Kaluza-Klein modes of the gravitational multiplet, close to a mass
$\simeq \frac{n - 1}{2 B} = \frac{3}{B} \simeq 2.5$ TeV.
In partial support of this hypothesis,
every $\bar{H}^n$ has pairs of finite covers of arbitrarily large volume
ratio, whose sets of eigenvalues of $-\Delta$, ignoring multiplicities, are
identical \cite{Leininger et al}, so since the Weyl asymptotic formula
(\ref{number of modes up to k bar}) is certainly valid for sufficiently large
$\bar{k} = \sqrt{\lambda}$,
every $\bar{H}^n$ has finite covers whose eigenvalues have arbitrarily large
multiplicities, for sufficiently large $\bar{k}$.

I shall now provisionally assume that the bump at 1.7 to 1.9 TeV seen in the
right-hand
graphs on pages 8 and 9 of ATLAS-CONF-2010-088 arises from the production and
decay of these approximately $10^{34}$ approximately degenerate gravitational
Kaluza-Klein modes in the $s$ channel, and give rough estimates to test
this interpretation. \ I shall show in section
\ref{The HW boundary}, starting on page \pageref{The HW boundary}, that a mode
on $\bar{H}^7$ of intrinsic mass $k$ has mass $\frac{Ak}{B}$ as measured on
the HW boundary, where $B \simeq 0.43 \kappa^{2 / 9}_{11}$ is the curvature
radius of the $\bar{H}^7$, and $A$ is the value of the ``warp factor'' $a
\left( x^C \right)$ everywhere on the $\bar{H}^7$ except in the immediate
vicinity of the HW boundary, as described after (\ref{metric ansatz}). \ Thus
if the bump centred at 1.8 TeV corresponds to approximately degenerate modes
of intrinsic mass $\frac{n - 1}{2} = 3$ on the $\bar{H}^7$, and
$\kappa^{- 2 / 9}_{11}$ is at its current lower bound for 2 or more flat extra
dimensions, $\kappa^{- 2 / 9}_{11} \simeq 0.36$ TeV \cite{Franceschini et al},
as in the example above, then $A =
\frac{1.8 \hspace{0.3em} \mathrm{{TeV}}}{2.5 \hspace{0.3em}
\mathrm{{TeV}}} = 0.72$. \ This then revises $V_7$ from the value $5.8
\times 10^{34}$ TeV$^{- 7}$ calculated from (\ref{V7 in terms of GN}) assuming
$A = 1$, to $1.1 \times 10^{35}$ TeV$^{- 7}$, and the intrinsic volume
$\bar{V}_7$ is revised to $3.1 \times 10^{34}$.

For an order of magnitude estimate, I shall treat each Kaluza-Klein mode of
the graviton multiplet, with mass around 1.8 TeV, as an independent scalar
particle $\varphi$. \ Then the total cross section for the process $gg
\rightarrow \varphi \rightarrow u \bar{u}$, for example, where $g$ represents
a gluon, is given near 1.8 TeV by the Breit-Wigner cross-section
{\cite{Perkins}}, and in order of magnitude is:
\begin{equation}
  \label{Breit Wigner cross section} \sigma \left( g g \rightarrow \varphi
  \rightarrow u \bar{u} \right) \sim \frac{\Gamma_{g g} \Gamma_{u
  \bar{u}}}{E^2} \cdot \frac{1}{\left( \left( E - M_{\varphi} \right)^2 +
  \Gamma^2 / 4 \right)},
\end{equation}
where $\Gamma_{gg}$ and $\Gamma_{u \bar{u}}$ are the partial widths for
$\varphi$ to decay to $g g$ and $u \bar{u}$ respectively, $E$ is the invariant
mass of the $g g$ system, $M_{\varphi} \simeq 1.8$ TeV is the mass of
$\varphi$, and $\Gamma$ is the total width of $\varphi$. \ The partial widths
are estimated in order of magnitude by {\cite{Giudice Rattazzi Wells}}:
\begin{equation}
  \label{partial widths} \Gamma_{g g} \sim \Gamma_{g g g} \sim \ldots \sim
  \Gamma_{u \bar{u}} \sim \Gamma_{g u \bar{u}} \sim \ldots \sim
  \Gamma_{\nu_{\tau} \bar{\nu}_{\tau}} \sim M_{\varphi}^3 G_N \simeq 3.9
  \times 10^{- 32} \textrm{{TeV}}
\end{equation}
The only factor in the cross-section (\ref{Breit Wigner cross section}) that
varies significantly with energy over the energy range of interest is the
final factor, and this arises from the squared magnitude of the
energy-dependent factor:
\begin{equation}
  \label{energy dependent factor} \frac{1}{E - M_{\varphi} + i \Gamma / 2}
\end{equation}
in the corresponding amplitude. \ The Kaluza-Klein graviton is not present in
the final state, so we have to sum the amplitude over the contributions of all
the different Kaluza-Klein graviton modes, all of which give the same
contribution to the amplitude, apart from slight differences in their masses.

The number of bosonic modes in the $d = 11$ supergravity multiplet is $128$,
so the number of approximately degenerate bosonic Kaluza-Klein states of mass
$\simeq 1.8$ TeV, which is the number of states up to 1.8 TeV as given by the
Weyl asymptotic formula (\ref{number of modes up to k bar}), is around:
\begin{equation}
  \label{number of degenerate KK modes} 128 \frac{\bar{V}_7}{7 \left( 2 \pi
  \right)^7} S_6 \,\, 3^7 \simeq 1.1 \times 10^{35}
\end{equation}
where I used $\bar{V}_7 \simeq 3.1 \times 10^{34}$ from just before
(\ref{Breit Wigner cross section}), and $S_6 = \frac{16}{15} \pi^3$, and
(\ref{lower bound on lambda 1}) with $ n=7 $. \ I shall
assume that these states are distributed over a smooth peak in the density of
states from about 1.7 TeV to 1.9 TeV.

The total width $\Gamma$ of $\varphi$ is the product of the partial width
(\ref{partial widths}) and the number of Standard Model states, which is 36,
so $\Gamma$ is extremely small. \ We can replace the sum over the individual
Kaluza-Klein modes by an integral over their masses {\cite{Franceschini et
al}}, and to a first approximation, when convoluted with a smooth density of
states, the energy-dependent factor (\ref{energy dependent factor}) is
effectively $\simeq - i \pi \delta \left( E - M \right)$, because the $i
\Gamma$ means that the integration path has to go around the singularity in
the lower half of the complex $M$ plane. \ Choosing the integration path to be
along the real axis except for a small semicircle centred at $M = E$, the
contributions from the real axis approximately cancel, and the semicircle
gives $- i \pi$ times the density of states evaluated at $M = E$. \ Thus after
summing the amplitude over the Kaluza-Klein modes, the energy-dependent factor
in the amplitude is approximately replaced by:
\begin{equation}
  \label{energy dependent factor summed over modes} - 1.1 \times 10^{35} i \pi
  f \left( E \right),
\end{equation}
where $f \left( E \right)$ has a smooth peak from about 1.7 TeV to 1.9 TeV,
and $\int f \left( E \right) dE = 1$. \ Thus the cross-section is:
\begin{equation}
  \label{cross section summed over KK modes} \sigma \left( g g \rightarrow
  \varphi \rightarrow u \bar{u} \right) \sim \frac{\Gamma_{g g} \Gamma_{u
  \bar{u}}}{E^2} \times 1.2 \times 10^{70} \pi^2 f \left( E \right)^2 \sim 5.6
  \times 10^7 f \left( E \right)^2 .
\end{equation}

A monoenergetic high energy beam of protons with $N$ protons of energy $E$ per
unit area per unit time is equivalent to a beam of partons, such that the
number of $u$ quarks per unit area per unit time with energy between $xE$ and
$\left( x + dx \right) E$ is $f_u \left( x \right) dx$, and similarly for the
other types of parton, where $f_p \left( x \right)$, $p = u, d, g, \bar{u},
\bar{d}, \ldots$ are the parton distribution functions (PDFs). \ The PDFs
evolve logarithmically with $Q^2$, the square of the momentum transferred in a
scattering process, and for a rough estimate I shall use the gluon
distribution $f_g \left( x \right)$ from the plot {\cite{MSTW}} with $Q^2 =
10^4$ GeV$^2$. \ To produce $\varphi$ with mass 1.8 TeV at rest with 3.5 TeV
per proton beam, each gluon needs $x = 0.26$. \ If the 4-momenta of the beams
are $\left( P, 0, 0, P \right)$ and $\left( P, 0, 0, - P \right)$ and the
momentum fractions of the gluons are $x_1$ and $x_2$, then their Mandelstam
$s$ is $P^2 \left( \left( x_1 + x_2 \right)^2 - \left( x_1 - x_2 \right)^2
\right) = 4 P^2 x_1 x_2$. \ From the plot we find that $f_g \left( x \right)
\simeq 0.060 x^{- 2.17}$ for $0.05 \leq x \leq 0.2$, but substantially smaller
than this for $x \geq 0.3$. \ Then the total cross section to produce a
$\varphi$ within the bump is roughly:
\[ \int^{0.3}_0 dx_1 \int^{0.3}_0 dx_2  0.060^2 \left( x_1 x_2
   \right)^{- 2.17} \times 5.6 \times 10^7 f \left( 2 P \sqrt{x_1 x_2}
   \right)^2 \]
\begin{equation}
  \label{cross section for bump} \simeq \frac{2.0 \times 10^5}{\left( 0.2
  \textrm{{  TeV}} \right)^2} \times \int^{0.3}_{0.23} \frac{dx_1}{x_1}
  \int^{0.27}_{0.24} \frac{2 xdx}{x^{4.34}} \simeq 3.0 \times 10^6
  \textrm{{  nb}},
\end{equation}
where I approximated $f \left( E \right)$ as a constant from 1.7 TeV to 1.9
TeV and 0 outside this interval. \ Multiplying by $4 \pi$, and by 75 for the
number of Standard Model helicity states, and by $\frac{1}{8}$ for the
probability that the two initial gluons can form a colour singlet, and by
$\frac{1}{4}$ for the average over the helicity states of the initial gluons,
gives a final estimate of $8.8 \times 10^7$ nb for the total cross section to
produce a $\varphi$ within the bump that decays to two Standard Model objects,
if $\kappa^{- 2 / 9}_{11} \simeq 0.36$ TeV, and all the bosonic modes of the
$d = 11$ supergravity multiplet that according to the Weyl asymptotic formula
(\ref{number of modes up to k bar}) would have masses up to 1.8 TeV, are in
fact approximately degenerate, with masses in the range 1.7 TeV to 1.9 TeV.

To reduce the background, the ATLAS measurement selected events with at least
3 Standard Model objects in the final state, and total transverse momentum
above 700 GeV. \ However the partial widths for $\varphi$ to decay to 2, 3, or
4 Standard Model states are all of the same order of magnitude (\ref{partial
widths}). \ And for an order of magnitude estimate, the decay of $\varphi$ is
effectively isotropic, and the 700 GeV lower limit on the total tranverse
momentum will not result in missing a high percentage of the decays of the 1.8
TeV $\varphi$ into 3 or 4 Standard Model states. \ The number of different
types of 3 object final states is roughly 10 times the number of different
types of 2 object final states, and omitting this factor could very roughly
compensate for neglecting the effect of the 700 GeV lower limit on the total
tranverse momentum. \ Thus we find an estimate of $10^8$ nb for the total
cross section to produce a $\varphi$ within the bump that is counted as an
event in the ATLAS measurement, if $\kappa^{- 2 / 9}_{11} \simeq 0.36$ TeV,
and all the bosonic modes of the $d = 11$ supergravity multiplet that
according to (\ref{number of modes up to k bar}) would have masses up to 1.8
TeV, are in fact approximately degenerate, with masses in the range 1.7 TeV to
1.9 TeV. \ So in 295 per nb of proton-proton collisions there should be about
$3 \times 10^{10}$ events above background in the bump. \ If we assume that the
curve followed by the data below and above the bump is the correct QCD
background, then the total number of events in the bump above this measured
background is 8.

Thus the estimate of the number of modes in the bump, (\ref{number of
degenerate KK modes}), is too large by a factor of around 60000,
so the correct number is around $1.8 \times 10^{30} $. \ From
(\ref{V7 in terms of GN}) and (\ref{number of degenerate KK modes})
and the relation $V_7 = B^7 \bar{V}_7$, where
$B \simeq 0.43 \kappa^{2 / 9}_{11}$
is the curvature radius of the $\bar{H}^7$, and the
relation $\frac{3 A}{B} = 1.8$ TeV, where $A$ is the value of the ``warp
factor'' $a \left( x^C \right)$ everywhere on the $\bar{H}^7$ except in the
immediate vicinity of the HW boundary, as described after (\ref{metric
ansatz}), we see that the intrinsic volume
$\bar{V}_7$ of $\bar{H}_7$, which determines the number of modes in the bump
if all the modes expected from (\ref{number of modes up to k bar}) up to the
expected intrinsic $\lambda_1 = 3^2$ are approximately degenerate at 1.8 TeV,
is independent of $\kappa^{2 / 9}_{11}$, so the discrepancy cannot be
corrected by reducing $\kappa^{2 / 9}_{11}$.

However the expected number of
modes in the bump would be reduced by a factor of $3^{- 7} = \frac{1}{2187}$
to around $ 5.0 \times 10^{31} $ if for some reason the intrinsic $\lambda_1$
was 1 rather than $3^2$ as expected from Cheeger's inequality. \ This might
happen if the modes in the bump originate from modes of the 3-form $C_{I J K}$
that are 2-forms on $\bar{H}^7$ and vectors along the extended dimensions,
because for $p < \frac{n - 1}{2}$, $\lambda_1$ for a $p$-form on $H^n$ is
$\frac{\left( n - 1 - 2 p \right)^2}{4}$ {\cite{Donnelly Xavier}}.
In this case the relation $\frac{3 A}{B} = 1.8$ TeV would be modified to
$\frac{A}{B} = 1.8$ TeV, so since $A \leq 1$ by section \ref{The HW boundary},
starting on page \pageref{The HW boundary}, we would find from $B \simeq 0.43
\kappa^{2 / 9}_{11}$ that $\kappa^{- 2 / 9}_{11} \geq 0.78
\textrm{{  TeV}}$, hence $M_{11} \geq 3.3$ TeV.

If this is the correct interpretation, then the number of modes would actually
be reduced still further, possibly to around the correct value, because the
factor 128 in (\ref{number of degenerate KK modes}) would be replaced by a
smaller value, representing the fact that the modes in the bump represent only
modes of $C_{I J K}$ that are 2-forms on $\bar{H}^7$ and vectors along the
extended dimensions. \ In this case, there might be a second bump at around $2
\times 1.8 \textrm{{  TeV}} = 3.6 \textrm{{  TeV}}$ from modes of
$C_{I J K}$ that are 1-forms on $\bar{H}^7$ and 2-form Abelian gauge fields,
equivalent in 4 dimensions to scalars \cite{Cremmer Scherk, Kalb Ramond}, along
the extended dimensions, and from
modes of the metric $G_{I J}$ that are vectors on $\bar{H}^7$ and vectors along
the extended dimensions, and a
third bump at around $3 \times 1.8 \textrm{{  TeV}} = 5.4
\textrm{{  TeV}}$ from the KK modes of the graviton that are scalars on
$\bar{H}^7$ and traceless symmetric tensors along the extended dimensions.
However the position is made less clear by the fact that, as discussed just
before (\ref{number of modes up to k bar}), on page \pageref{number of modes
up to k bar}, I had to assume that the classically massless modes of
$ C_{I J K} $ and $G_{I J}$
resulting from the Hodge - de Rham harmonic forms on $\bar{H}^7$ all acquire
masses $ \sim \kappa_{11}^{-2/9} $ from the leading quantum corrections to
the classical CJS action (\ref{CJS action}).

If the intrinsic $\lambda_1$ is $3^2$, as assumed on the basis of Cheeger's
bound for scalars on $\bar{H}^7$ in the initial estimate above of the expected
number of events above background in the bump, then we would have to conclude
that just a fraction around $\frac{1}{60000}$ of
the bosonic gravitational KK modes expected from (\ref{number of modes up to k
bar}) are approximately degenerate at 1.8 TeV.

An alternative hypothesis is that notwithstanding the rapid restoration of
agreement between the spectral staircase
and the Weyl asymptotic formula (\ref{number of modes up to k bar}) found for
small intrinsic volume $\bar{H}^3$'s in \cite{Inoue}, and the possibility of
arbitrarily large degeneracies of eigenvalues of $-\Delta$ for large intrinsic
volume $\bar{H}^n$'s indicated by \cite{Leininger et al}, the restoration of
agreement with (\ref{number of modes up to k bar}) for $\bar{H}^7$ occurs over
a wider range of energies than 0.2 TeV, and the modes responsible for the bump
from 1.7 to 1.9 TeV are instead the Hodge - de Rham harmonic 1-forms, 2-forms,
and 3-forms on $\bar{H}^7$, which have all acquired approximately equal masses
$\sim \kappa_{11}^{-2/9}$ from the $(DH)^2 R^2$ and $H^2 R^3$ terms in the
leading quantum corrections to the CJS action (\ref{CJS action}).

The number of linearly independent Hodge - de Rham harmonic $p$-forms, $0 \leq
p \leq n$, on a compact orientable $n$-manifold, is equal to the $p$th Betti
number with integer coefficients, $B_p$, which is called the $p$th Betti
number.  The sum of all the Betti numbers of an $\bar{H}^n$ of intrinsic
volume $\bar{V}_n$ is bounded above by $c \bar{V}_n$, where $c$ is a constant
that depends only on $n$ \cite{Gromov Volume Bounded Cohomology, Gromov Volume
and Bounded Cohomology}.  The $p$th Betti number $B_p$ of finite coverings of
an $\bar{H}^n$ grows linearly with the inrinsic volume of the covering if and
only if there are square-integrable harmonic $p$-forms on the covering space
$H^n$ \cite{Lueck, Clair
Whyte}, which by \cite{Donnelly Xavier} means if and only if $n$ is even and
$p=\frac{n}{2}$.  In the other cases, the rate of growth of $B_p$ with the
intrinsic volume of the covering is bounded above by a rate of growth that is
strictly less than linear, by an amount that depends on the rate of growth of
the
\emph{injectivity radius} of the covering.   The injectivity radius
$\mathrm{Inj}\left(\mathcal{M},x\right)$ at a point
$x$ of a Riemannian manifold $\mathcal{M}$ is the largest radius for which the
map from the tangent space at $x$, to $\mathcal{M}$, defined by mapping each
tangent vector $v$ at $x$ to the point reached by starting at $x$ and going out
a distance $|v|$ along the geodesic starting in the direction $v$ from $x$, is
a diffeomorphism, and $\mathrm{Inj}\left(\mathcal{M}\right)$ is the minimum of
$\mathrm{Inj}\left(\mathcal{M},x\right)$ over all the points $x$ of
$\mathcal{M}$ \cite{Wikipedia injectivity radius}.

If $\bar{H}^n$ is a compact hyperbolic $n$-manifold, and $X$ is a finite regular covering manifold of $\bar{H}^n$ with intrinsic volume
$\mathrm{Vol}\left(X \right)$ and intrinsic injectivity radius
$\mathrm{Inj}\left(X \right)$, then by Theorem 0.3 of
\cite{Clair Whyte}, there are constants $C > 0$ and $\beta_p > 0$ such that:
\begin{itemize}
 \item For $n$ odd:
\begin{itemize}
 \item If $p \neq \frac{n \pm 1}{2}$, then $B_p\left(X \right) \leq C
\displaystyle{
\frac{\mathrm{Vol}\left(X \right)}{e^{\beta_p \mathrm{Inj}\left(X \right)}}}$.
 \item For $p = \frac{n \pm 1}{2}$, $B_p\left(X \right) \leq C \displaystyle{
\frac{\mathrm{Vol}\left(X \right) \cdot \mathrm{log}\,\,\mathrm{Inj}\left(X
\right)}{\mathrm{Inj}\left(X \right)}}$.
\end{itemize}
\item For $n$ even:
\begin{itemize}
 \item If $p \neq \frac{n}{2}$, then $B_p\left(X \right) \leq C
\displaystyle{
\frac{\mathrm{Vol}\left(X \right)}{e^{\beta_p \mathrm{Inj}\left(X \right)}}}$.
\end{itemize}
\end{itemize}

By the generalized Gauss-Bonnet theorem {\cite{Allendoerfer Weil, Chern}}, the
Euler
characteristic, or Euler number, $\chi \left( \mathcal{M}^{2 q} \right) \equiv
\sum^{2 q}_{p = 0} \left( - \right)^p B_p$, of an arbitrary smooth $2
q$-manifold $\mathcal{M}^{2 q}$, is given by:
\begin{equation}
  \label{generalized Gauss Bonnet theorem} \chi \left( \mathcal{M}^{2 q}
  \right) = \frac{1}{\left( 8 \pi \right)^q q!} \int_{\mathcal{M}^{2 q}} d^{2
  q} x \sqrt{g} \epsilon_{j_1 \ldots j_{2 q}} \epsilon^{i_1 \ldots i_{2 q}}
  R_{i_1 i_2} \, \!^{j_1 j_2} \ldots R_{i_{2 q - 1} i_{2 q}} \, \!^{j_{2 q -
  1} j_{2 q}} .
\end{equation}
Thus for a $2 q$-dimensional compact hyperbolic manifold $\bar{H}^{2 q}$ with
sectional curvature $- 1$:
\begin{equation}
  \label{Euler number of H bar 2q} \chi \left( \bar{H}^{2 q} \right) = \left(
  - \right)^q \frac{\left( 2 q \right)\! !}{\left( 2 \pi \right)^q 2^q q!}
\bar{V}_{2 q} = \left( - \right)^q \frac{\left(2 q - 1\right)\! !\! !}{\left( 2
  \pi \right)^q} \bar{V}_{2 q} = \left( - \right)^q \frac{2}{S_{2 q}}
  \bar{V}_{2 q},
\end{equation}
where $S_{2 q} = \frac{2 \left( 2 \pi \right)^q}{\left( 2 q - 1 \right) !!}$
is the $2 q$-volume of the $2 q$-sphere of unit radius. \ In particular, $\chi
\left( \bar{H}^6 \right) = - \frac{15}{8 \pi^3} \bar{V}_6 \simeq - 0.060
\bar{V}_6$, $\chi \left( \bar{H}^4 \right) = \frac{3}{4 \pi^2} \bar{V}_4
\simeq 0.076 \bar{V}_4$, and $\chi \left( \bar{H}^2 \right) = - \frac{1}{2
\pi} \bar{V}_2 \simeq - 0.159 \bar{V}_2$. \ Thus since by {\cite{Lueck, Clair
Whyte}} and {\cite{Donnelly Xavier}}, $B_p \left( X \right)$ for a finite
regular covering $X$ of $\bar{H}^{2 q}$ can grow linearly with the intrinsic
volume $\mathrm{{Vol}} \left( X \right)$ if and only if $p = q$, $\left(
- \right)^q \frac{B_q \left( X \right)}{\chi \left( X \right)} \rightarrow 1$
as $\mathrm{{Vol}} \left( X \right) \rightarrow \infty$, hence $\frac{B_q
\left( X \right)}{\mathrm{{Vol}} \left( X \right)} \rightarrow
\frac{2}{S_{2 q}}$ as $\mathrm{{Vol}} \left( X \right) \rightarrow
\infty$.

The relatively strong suppression of $B_p\left(X \right)$ relative to
$\mathrm{Vol}\left(X \right)$ for $p\neq\frac{n\pm 1}{2}$ when $n$ is odd and
for $p\neq\frac{n}{2}$ when $n$ is even arises from the fact that in these
cases, the spectrum of the Hodge - de Rham $-\Delta$ for $p$-forms on $H^n$
has a gap \cite{Donnelly Xavier}.  For $p = \frac{n\pm 1}{2}$ when $n$ is odd
the spectrum of $-\Delta$ on $H^n$ has no gap, but there are no
square-integrable harmonic $p$-forms on $H^n$, while for $p = \frac{n}{2}$
when $n$ is even, there are an infinite number of linearly independent
square-integrable harmonic $p$-forms on $H^n$ \cite{Donnelly Xavier}.

For finite coverings $X$ of $\bar{H}^n$ that are reasonably isotropic, in the
sense that they have a fundamental domain that is approximately spherical, we
might expect that the intrinsic injectivity radius $\mathrm{Inj}\left(X
\right)$, which is half the intrinsic length of the shortest closed geodesic on
$X$, is
roughly a fixed fraction of the intrinsic diameter of $X$, in which case for
for $p\neq\frac{n\pm 1}{2}$ when $n$ is odd and for $p\neq\frac{n}{2}$ when $n$
is even, the above bounds on $B_p\left(X \right)$ become upper bounds by
constant multiples of powers strictly less than 1 of $\mathrm{Vol}\left(X
\right)$.  In the special case of $\bar{H}^n$ whose fundamental groups are
arithmetic discrete subgroups $\Gamma$ of $\mathrm{SO}\left(n,1\right)$
\cite{Borel Harish Chandra}, and finite coverings $X$ of $\bar{H}^n$ whose
fundamental groups are congruence subgroups of $\Gamma$, which are roughly the
subgroups obtained by Selberg's Lemma \cite{Selberg}, reviewed in subsection
3.1.2 of \cite{CCHT}, there are lower bounds on $B_p\left(X \right)$ of this
form \cite{Xue}.

If we assume that the above results for finite coverings $X$ of a fixed
$\bar{H}^n$ roughly describe the dependence of the Betti numbers $B_p\left(
\bar{H}^n\right)$ on the intrinsic volume $\bar{V}$ of $\bar{H}^n$ for typical
$\bar{H}^n$ of large $\bar{V}$, at least for reasonably isotropic $\bar{H}^n$,
then for $p\neq\frac{n\pm 1}{2}$ when $n$ is odd and for $p\neq\frac{n}{2}$
when $n$ is even, $B_p\left(\bar{H}^n\right)$ is roughly a constant times a
power $<1$ of $\bar{V}$, while for $p = \frac{n\pm 1}{2}$ when $n$ is odd,
$B_p\left(\bar{H}^n\right)$ is roughly a constant times $\frac{\bar{V}}{
\mathrm{ln}\,\bar{V}}$, and for $p = \frac{n}{2}$ when $n$ is even, $B_p\left(
\bar{H}^n \right)$ is roughly $\frac{2}{S_n}\bar{V}$.

Thus if the $\bar{H}^n$ Cartesian factor of
$\mathcal{M}^7$ of large $\bar{V}$ had $n$ even, the number of modes in the bump
according to the alternative hypothesis would probably be too large by a factor
of at least 1000.  For the $\bar{H}^7$ case, the number of modes would
be about right if $B_3$ and $B_4$ were around $\frac{1}{200}\,\frac{\bar{V}}{
\mathrm{ln}\,\bar{V}}$.

If this is the correct hypothesis, then the bump will be spin 0 if the
compact hyperbolic factor of large intrinsic volume is 7-dimensional, and a
mixture of spins 0 and 1 if it is 5-dimensional or
3-dimensional.  This could be tested by plotting the energies of the decay
products in 3-body decays of candidate bump particles, in the reconstructed
rest frame of the candidate bump particle, on a Dalitz plot \cite{Dalitz,
Fabri, Perkins}.

Confirmation that the bump particles decay democratically to all Standard
Model particles, and are thus gravity-like, but have only spin 0, would
provide evidence for the existence of the 3-form gauge field $C_{I J K}$ of
$d = 11$ supergravity \cite{CJS}, because in the $\bar{H}^7$ case, the bump
particles arise only from $C_{I J K}$.
\vspace{0.3cm}

\section{Ho\v{r}ava-Witten theory}
\label{Horava Witten theory}

I shall use Moss's improved version of Ho\v{r}ava-Witten theory {\cite{Moss 1,
Moss 2, Moss 3, Moss 4}}. \ In the region of the HW boundary, the coordinates
$x^I$ have the form $\left( \tilde{x}^U, y \right)$, where indices $U, V, W,
\ldots$ are tangential to a family of hypersurfaces foliating the $\left( 10 +
1 \right)$-dimensional manifold-with-boundary, one of these hypersurfaces
coinciding with the boundary, and $y$ takes a constant value on each of these
hypersurfaces, with the value of $y$ distinguishing the hypersurfaces. \ $y$
takes the value $y_1$ on the boundary, and $y > y_1$ in the bulk. \ The symbol
$y$ is also used as the
coordinate index for the $y$ coordinate. \ The bosonic part of the
semiclassical action is:
\begin{equation}
  \label{Horava Witten action} S^{\left( \mathrm{{bos}}
  \right)}_{\mathrm{{HW}}} = S^{\left( \mathrm{{bos}}
  \right)}_{\mathrm{{CJS}}} + S^{\left( \mathrm{{bos}}
  \right)}_{\mathrm{{GH}}} + S^{\left( \mathrm{{bos}}
  \right)}_{\mathrm{{YM}}}
\end{equation}
The supergravity term is (\ref{CJS action}), on page \pageref{CJS action}.

The Gibbons-Hawking term is:
\begin{equation}
  \label{Gibbons Hawking term} S^{\left( \mathrm{{bos}}
  \right)}_{\mathrm{{GH}}} = \frac{1}{\kappa^2_{11}} \int_{\beta} d^{10}
  \tilde{x}  \tilde{e} K
\end{equation}
Here $\beta$ denotes the $9 + 1$ dimensional boundary at $y = y_1$, and on the
boundary, $\tilde{e} = \sqrt{- \tilde{G}}$ denotes the square root of minus
the determinant of the induced metric $\tilde{G}_{U V}$, which is obtained
from $G_{I J}$ by dropping the row and column with an index $y$. \ $K = G^{I
J} K_{I J}$ is the trace of the extrinsic curvature $K_{I J} \equiv \left(
\delta_I \, \!^K - n_I n^K \right)\! \left( \delta_J \, \!^L - n_J n^L \right)
D_K n_L$, where $n_I$ is the outward unit normal, and $D_K$ is the covariant
derivative. \ Thus since $n_I$ is a scalar factor times $\partial_I y$, the
only nonvanishing component of $n_I$ is $n_y = - \frac{1}{\sqrt{G^{y y}}}$, so
only the $D_U n_V = - \Gamma_U \, \!^y \, \!_V n_y$ components of $D_K n_L$
contribute to $K_{I J}$, so $K_{I J}$ is symmetric, and $K =
\frac{1}{\sqrt{G^{y y}}} \left( G^{U V} - \frac{G^{U y} G^{V y}}{G^{y y}}
\right) \Gamma_U \, \!^y \, \!_V$.

The Yang-Mills term is:
\begin{equation}
  \label{Yang Mills term} S^{\left( \mathrm{{bos}}
  \right)}_{\mathrm{{YM}}} = - \frac{1}{16 \pi \kappa^2_{11}} \left(
  \frac{\kappa_{11}}{4 \pi} \right)^{2 / 3} \int_{\beta} d^{10} \tilde{x} 
  \tilde{e} \left( \frac{1}{30} \mathrm{{tr}} F_{U V} F^{U V} -
  \frac{1}{2} \bar{R}_{U V \hat{W}  \hat{X}} \bar{R}^{U V \hat{W}  \hat{X}}
  \right)
\end{equation}
Here $F_{U V} = \partial_U A_V - \partial_V A_U + i \left[ A_U, A_V \right]$
is the field strength of an $E_8$ Yang-Mills gauge field $A_U =
T^{\mathcal{A}} A_U^{\mathcal{A}}$ localized on the boundary. \
$\mathrm{{tr}}$ means the ordinary trace, not the modified trace used by
HW. \ Indices $\mathcal{A}, \mathcal{B}, \ldots$ run over the 248 generators
of $E_8$, and the hermitian generators $T^{\mathcal{A}}$ in the
fundamental/adjoint of $E_8$ satisfy $\mathrm{{tr}} T^{\mathcal{A}}
T^{\mathcal{B}} = 30 \delta^{\mathcal{A}\mathcal{B}}$. \ In the
$\mathrm{{SO}} \left( 16 \right)$ basis for $E_8$, the $T^{\mathcal{A}}$
are $- \frac{1}{2} i$ times the generators in Appendix 6.A of {\cite{Green
Schwarz Witten}} or subsection 2.1 of {\cite{CCHT}}, and in the
$\mathrm{{SU}} \left( 9 \right)$ basis for $E_8$, the $T^{\mathcal{A}}$
are the generators in subsection 5.2 of {\cite{CCHT}}. \ The coefficient of the
first term in
(\ref{Yang Mills term}) is fixed by anomaly cancellation {\cite{Horava Witten
2, de Alwis, Conrad, Faux Lust Ovrut, Lu, Bilal Derendinger Sauser, Harmark,
Bilal Metzger, Meissner Olechowski, Moss 3}} and has the value found by
Conrad {\cite{Conrad}}, which is slightly different from the original value
found by HW. \ The $\bar{R}_{U V \hat{W}  \hat{X}} \bar{R}^{U V \hat{W} 
\hat{X}}$ term was derived by Moss {\cite{Moss 4}}, with:
\begin{equation}
  \label{R bar} \bar{R}_{U V} \, \!^{\hat{W}} \, \!_{\hat{X}} = \partial_U
  \bar{\omega}_V \, \!^{\hat{W}} \, \!_{\hat{X}} - \partial_V \bar{\omega}_U
  \, \!^{\hat{W}} \, \!_{\hat{X}} + \bar{\omega}_U \, \!^{\hat{W}} \,
  \!_{\hat{Y}} \bar{\omega}_V \, \!^{\hat{Y}} \, \!_{\hat{X}} - \bar{\omega}_V
  \, \!^{\hat{W}} \, \!_{\hat{Y}} \bar{\omega}_U \, \!^{\hat{Y}} \,
  \!_{\hat{X}},
\end{equation}
where
\begin{equation}
  \label{omega bar} \bar{\omega}_{U \hat{V}  \hat{W}} = \tilde{\omega}_{U
  \hat{V}  \hat{W}} \pm \frac{1}{2} H_{\hat{y} U \hat{V}  \hat{W}}
\end{equation}
and $\tilde{\omega}_{U \hat{V}  \hat{W}} = e^X \, \!_{\hat{W}} \left(
\tilde{\Gamma}_U \, \!^Y \, \!_X e_{Y \hat{V}} - \partial_U e_{X \hat{V}}
\right)$ is the Levi-Civita connection for the vielbein $\tilde{e}_{U
\hat{V}}$, that satisfies $\tilde{e}_{U \hat{W}} \tilde{e}_{V \hat{X}}
\eta^{\hat{W}  \hat{X}} = \tilde{G}_{U V}$, where $\eta^{\hat{U}  \hat{V}}$ is
the Minkowski metric on the $9 + 1$ dimensional boundary. \ The sign choice in
(\ref{omega bar}) is correlated with the chirality conditions on the
gravitino, gaugino, and supersymmetry variation parameter on the boundary.

Variation of the metric in $S^{\left( \mathrm{{bos}}
\right)}_{\mathrm{{HW}}}$ leads to the Einstein equations:
\begin{equation}
  \label{Einstein equations} R_{I J} - \frac{1}{2} RG_{I J} - \frac{1}{12} H_I
  \, \!^{K L M} H_{J K L M} + \frac{1}{96} H^{K L M N} H_{K L M N} G_{I J} = 0
\end{equation}
and on the boundary to the Israel boundary conditions {\cite{Israel, Chamblin
Reall, Dyer Hinterbichler, Moss 2}}:
\begin{equation}
  \label{Israel boundary conditions} K^{U V} - K \tilde{G}^{U V} -
  \kappa_{11}^2 \tilde{T}^{\left( \mathrm{{bos}} \right) U V} = 0,
\end{equation}
where
\begin{equation}
  \label{T tilde i U V} \tilde{T}^{\left( \mathrm{{bos}} \right) U V} =
  \frac{2}{\tilde{e}}  \frac{\delta S^{\left( \mathrm{{bos}}
  \right)}_{\mathrm{{YM}}}}{\delta \tilde{G}_{U V}} .
\end{equation}

The boundary is equivalent to a double-sided mirror at $y = y_1$, because all
the fields on one side of the mirror are exactly copied, up to sign, on the
other side of the mirror. \ The Yang-Mills multiplet is adjacent to the
mirror, but infinitesimally displaced from it, so that it has its own
reflection infinitesimally on the other side of the mirror {\cite{Lu}}. \ The
bulk and its reflection on the other side of the mirror form two parts of a
single closed manifold, and the formula (\ref{CJS action}) for the classical
action in the bulk applies on both sides of the mirror, with the \emph{same}
$\mathrm{SO}\left(10,1\right) $-invariant tensor $\epsilon^{\hat{I}_1 \ldots
\hat{I}_{11}}_{11}$ occurring in the definition of $\epsilon^{I_1 \ldots
I_{11}}_{11}$ on both sides of the mirror.  Thus $\epsilon^{\hat{I}_1 \ldots
\hat{I}_{11}}_{11}$ transforms as a \emph{pseudo-tensor} under reflection in
the mirror, which means that its components are multiplied by an extra factor
of $-1$ under the reflection, relative to what they would have become if it
had transformed as a tensor, because it would have been multiplied by $-1$
under the reflection, if it had transformed as a tensor.

The metric $G_{I J}$ transforms as a tensor under the reflection, because if
it transformed as a pseudo-tensor, the measure factor $\sqrt{-G}$ would become
imaginary under the reflection, and the Ricci scalar would change sign under
the reflection.  The Yang-Mills gauge field $A_U^{\mathcal{A}}$ also
transforms as a tensor under the reflection, because it only has components
tangential to the boundary, and if it transformed as a pseudo-tensor, its
components would have to vanish identically, if they were continuous across
the mirror.
However to preserve the sign of the last term, called the Chern-Simons term,
of the classical action in the bulk (\ref{CJS action}) under reflection in the
mirror, the 3-form $C_{I J K}$ has to transform, like $\epsilon^{\hat{I}_1
\ldots \hat{I}_{11}}_{11}$, as a pseudo-tensor under the reflection.

Thus $C_{U V W}$ and $H_{U V W X}$
change sign on reflection in the mirror, so if $H_{U V W X}$ is continuous
across the mirror, $H_{U V W X} = 0$ at the mirror. \ Thus the boundary
conditions for $C_{I J K}$ are controlled not by the action, but by the
pseudo-tensor transformation rule of $C_{I J K}$ under reflection in the
mirror, and continuity. \ Therefore variations of $C_{I
J K}$ in $S_{\mathrm{{HW}}}$ have $\delta C_{U V W} = 0$ at the boundary,
and lead only to the 3-form field equations, whose bosonic terms are:
\begin{equation}
  \label{3 form field equations} D_L H^{L I J K} - \frac{1}{3456} \epsilon^{I
  J K L M N O P Q R S}_{11} H_{L M N O} H_{P Q R S} = 0
\end{equation}

At zeroth order in $\kappa_{11}$, $S_{\mathrm{{YM}}}$ is absent and $H_{U
V W X}$ is continuous across the mirror, so the boundary condition is $H_{U V
W X} = 0$. \ At order $\kappa^{2 / 3}_{11}$, the $E_8$ Yang-Mills multiplet is
required to cancel the gravitational anomalies resulting from the chirality of
the gravitino boundary condition, and $H_{U V W X}$ is no longer continuous
across the mirror. \ The boundary condition for $H_{I J K L}$ at the surface
of the mirror is determined by anomaly cancellation and supersymmetry, and the
bosonic terms are:
\begin{equation}
  \label{bc for H} H_{U V W X} = \pm \frac{3}{2 \pi} \left(
  \frac{\kappa_{11}}{4 \pi} \right)^{2 / 3} \left( \frac{1}{30}
  \mathrm{{tr}} F_{\left[ U V \right.} F_{\left. W X \right]} -
  \frac{1}{2} \bar{R}_{\left[ U V \right.} \, \!^{\hat{Y} \hat{Z}}
  \bar{R}_{\left. W X \right] \hat{Y} \hat{Z}} - \frac{1}{3} \partial_{\left[
  U \right.} \Omega_{\left. V W X \right]}^{\left( \mathrm{H} \right)} \right)
\end{equation}
where
\[ \Omega_{U V W}^{\left( \mathrm{H} \right)} = 3 H_{y \left[ U \right.} \,
   \!^{X_1 X_2} D_V H_{\left. W \right] X_1 X_2 y} + 3 H_{y X_1 X_2 \left[ U
   \right.} D^{X_1} H^{X_2} \, \!_{\left. V W \right] y} \]
\begin{equation}
  \label{Omega H} \hspace{1cm} - \frac{1}{36} \varepsilon_{U V W} \, \!^{X_1
  X_2 X_3 X_4 X_5 X_6 X_7} H_{y X_1 X_2 X_3} D_{X_4} H_{X_5 X_6 X_7 y}
\end{equation}
The sign choice in (\ref{bc for H}) is correlated with the chirality
conditions on the gravitino, gaugino, and supersymmetry variation parameter on
the boundary. \ Square brackets around two or more indices denote
antisymmetrization with unit total weight.

The boundary condition (\ref{bc for H}) can be integrated to:
\begin{equation}
  \label{bc for C} C_{U V W} = \pm \frac{1}{4 \pi} \left( \frac{\kappa_{11}}{4
  \pi} \right)^{2 / 3} \left( \frac{1}{30} \Omega_{U V W}^{\left( \mathrm{Y}
  \right)} - \frac{1}{2} \Omega_{U V W}^{\left( \mathrm{L} \right)} -
  \frac{1}{2} \Omega_{U V W}^{\left( \mathrm{H} \right)} \right) + \lambda_{U
  V W},
\end{equation}
where
\begin{equation}
  \label{Omega Y} \Omega_{U V W}^{\left( \mathrm{Y} \right)} = 6
  \mathrm{{tr}} \left( A_{\left[ U \right.} \partial_V A_{\left. W
  \right]} + \frac{2}{3} iA_{\left[ U \right.} A_V A_{\left. W \right]}
  \right)
\end{equation}
and
\begin{equation}
  \label{Omega L} \Omega_{U V W}^{\left( \mathrm{L} \right)} = - 6 \left(
  \bar{\omega}_{\left[ U \right.} \, \!^{\hat{X}_1 \hat{X}_2} \partial_V
  \bar{\omega}_{\left. W \right] \hat{X}_2 \hat{X}_1} + \frac{2}{3}
  \mathcal{A}_{\left\{ U, V, W \right\}} \bar{\omega}_U \, \!^{\hat{X}_1} \,
  \!_{\hat{X}_2} \bar{\omega}_V \, \!^{\hat{X}_2} \, \!_{\hat{X}_3}
  \bar{\omega}_W \, \!^{\hat{X}_3} \, \!_{\hat{X}_1} \right)
\end{equation}
are Yang-Mills and Lorentz Chern-Simons 3-forms respectively, $\lambda_{U V
W}$ is an arbitrary closed 3-form on the boundary, and the notation
$\mathcal{A}_{\left\{ U, V, W \right\}}$ indicates that what follows it is to
be antisymmetrized in the indices $U, V, W$, with unit total weight.

The classical field equations of $d = 11$ supergravity do not have any
solutions with 4 flat extended dimensions if $\mathcal{M}^7$ has a compact
hyperbolic factor, for any arrangement of 4-form fluxes such that the flux
bilinears are on average covariantly constant, so it is necessary to include
the leading quantum corrections to the classical action in the bulk. \ The
quantum-corrected field equations have the form $\frac{\delta
\Gamma_{\mathrm{{HW}}}}{\delta \Phi} = 0$, where $\Phi$ represents all
the fields, and the bosonic part of the quantum effective action
$\Gamma_{\mathrm{{HW}}}$ {\cite{Schwinger, DeWitt, Nambu Jona Lasinio 1,
Nambu Jona Lasinio 2}} has the form:
\begin{equation}
  \label{Gamma HW} \Gamma^{\left( \mathrm{{bos}}
  \right)}_{\mathrm{{HW}}} = S^{\left( \mathrm{{bos}}
  \right)}_{\mathrm{{HW}}} + \Gamma^{\left( 8, \mathrm{{bos}}
  \right)}_{\mathrm{{SG}}} + \Gamma^{\left( \mathrm{h.o., {bos}}
  \right)}_{\mathrm{{HW}}} .
\end{equation}
Here $\Gamma^{\left( 8, \mathrm{{bos}} \right)}_{\mathrm{{SG}}}$ is
a dimension 8 local polynomial in the fields and derivatives in the bulk, of
order $\kappa^{4 / 3}_{11}$ relative to $S^{\left( \mathrm{{bos}}
\right)}_{\mathrm{{CJS}}}$, and $\Gamma^{\left( \mathrm{h.o., {bos}}
\right)}_{\mathrm{{HW}}}$, representing the higher order corrections,
contains both local and non-local terms, and is of order $\kappa^2_{11}$ or
higher relative to $S^{\left( \mathrm{{bos}}
\right)}_{\mathrm{{CJS}}}$ in the bulk, and of order $\kappa^{4 /
3}_{11}$ or higher relative to $S^{\left( \mathrm{{bos}}
\right)}_{\mathrm{{GH}}}$ on the boundary. \ The dimension of a local
monomial in the fields and derivatives is defined by counting 0 for each
metric or 3-form, $\frac{1}{2}$ for each gravitino, and 1 for each derivative,
and on the boundary also counting 1 for each Yang-Mills gauge field and
$\frac{3}{2}$ for each gaugino.

The leading quantum corrections to $d = 11$ supergravity in the bulk have the
form {\cite{Tseytlin}}:
\begin{equation}
  \label{Gamma 8 SG} \Gamma^{\left( 8, \mathrm{{bos}}
  \right)}_{\mathrm{{SG}}} = \frac{1}{147456 \pi^2 \kappa_{11}^2} \left(
  \frac{\kappa_{11}}{4 \pi} \right)^{4 / 3} \int_{\mathcal{B}} d^{11} xe
  \left( t_8 t_8 R^4 - \frac{1}{4!} \epsilon_{11} \epsilon_{11} R^4 -
  \frac{1}{6} \epsilon_{11} t_8 CR^4 + \Xi^{\left( \mathrm{{flux}}
  \right)} \right)
\end{equation}
Here for an arbitrary tensor $X_{I J K L}$, that is antisymmetric in its first
two indices and antisymmetric in its last two indices:
\begin{equation}
  \label{t8 t8 R4} t_8 t_8 X^4 \equiv t^{I_1 I_2 J_1 J_2 K_1 K_2 L_1 L_2}_8
  t^{M_1 M_2 N_1 N_2 O_1 O_1 P_1 P_2}_8 X_{I_1 I_2 M_1 M_2} X_{J_1 J_2 N_1
  N_2} X_{K_1 K_2 O_1 O_2} X_{L_1 L_2 P_1 P_2},
\end{equation}
where $t^{I J K L M N O P}_8$ is a tensor built from $G^{I J}$ and
antisymmetric in each successive pair of indices, such that for antisymmetric
tensors $A_{I J}$, $B_{I J}$, $C_{I J}$, and $D_{I J}$:
\[ t^{I J K L M N O P}_8 A_{I J} B_{K L} C_{M N} D_{O P} = \hspace{6cm}  \]
\[ = 8 \left( \mathrm{{tr}} \left( ABCD \right) + \mathrm{{tr}}
   \left( ACBD \right) + \mathrm{{tr}} \left( ACDB \right) \right) \]
\[ - 2 \left( \mathrm{{tr}} \left( AB \right) \mathrm{{tr}} \left(
   CD \right) + \mathrm{{tr}} \left( AC \right) \mathrm{{tr}} \left(
   BD \right) + \mathrm{{tr}} \left( AD \right) \mathrm{{tr}} \left(
   BC \right) \right) = \]
\[ = 8 \left( A_{I J} B_{J K} C_{K L} D_{L I} + A_{I J} C_{J K} B_{K L} D_{L
   I} + A_{I J} C_{J K} D_{K L} B_{L I} \right) \]
\begin{equation}
  \label{t8 A B C D} - 2 \left( A_{I J} B_{J I} C_{K L} D_{L K} + A_{I J} C_{J
  I} B_{K L} D_{L K} + A_{I J} D_{J I} B_{K L} C_{L K} \right)
\end{equation}
Here and in the following, repeated lower coordinate indices are understood to
be contracted with an inverse metric, for example $A_{I J} B_{J K} \equiv A_I
\, \!^J B_{J K} = G^{J L} A_{I L} B_{J K}$. \ Thus for an arbitrary such
tensor $X_{I J K L}$:
\begin{dmath}[compact, spread=3pt]
  \label{t8 t8 X4} t_8 t_8 X^4 = 12 \hspace{0.25em} X_{I J K L} X_{J I L K}
  X_{M N O P} X_{N M P O} + 24 \hspace{0.25em} X_{I J K L} X_{J I M N} X_{O P
  L K} X_{P O N M} - 96 \hspace{0.25em} X_{I J K L} X_{J I L M} X_{N O M P}
  X_{O N P K} - 48 \hspace{0.25em} X_{I J K L} X_{J I M N} X_{O P L M} X_{P O
  N K} - 96 \hspace{0.25em} X_{I J K L} X_{J M L K} X_{M N O P} X_{N I P O} -
  48 \hspace{0.25em} X_{I J K L} X_{J M N O} X_{M P L K} X_{P I O N} + 192
  \hspace{0.25em} X_{I J K L} X_{J M L N} X_{M O N P} X_{O I P K} + 384
  \hspace{0.25em} X_{I J K L} X_{J M L N} X_{M O P K} X_{O I N P}
\end{dmath}
And similarly, for an arbitrary such tensor $X_{I J K L}$:
\begin{dmath}[compact, spread=3pt]
  \label{eps11 eps11 X4} \frac{1}{4!} \epsilon_{11} \epsilon_{11} X^4 \equiv
  \frac{1}{4!} \epsilon_{11}^{I J K L_1 L_2 M_1 M_2 N_1 N_2 O_1 O_2} \left(
  \epsilon_{11} \right)_{I J K P_1 P_2 Q_1 Q_2 R_1 R_2 S_1 S_2} \times
  X_{L_1 L_2} \, \!^{P_1 P_2} X_{M_1 M_2} \, \!^{Q_1 Q_2} X_{N_1 N_2} \,
  \!^{R_1 R_2} X_{O_1 O_2} \, \!^{S_1 S_2} = - \frac{8!}{4} X_{L_1 L_2} \,
  \!^{\left[ L_1 L_2 \right.} X_{M_1 M_2} \, \!^{M_1 M_2} X_{N_1 N_2} \,
  \!^{N_1 N_2} X_{O_1 O_2} \, \!^{\left. O_1 O_2 \right]}
\end{dmath}
And similarly:
\begin{dmath}
  \label{eps11 t8 C R4} \frac{1}{6} \epsilon_{11} t_8 CR^4 \equiv 4
  \epsilon^{I_1 \ldots I_{11}} C_{I_1 I_2 I_3} R_{I_4 I_5 J K} R_{I_6 I_7 K L}
  R_{I_8 I_9 L M} R_{I_{10} I_{11} M J} - \epsilon^{I_1 \ldots I_{11}} C_{I_1
  I_2 I_3} R_{I_4 I_5 J K} R_{I_6 I_7 K J} R_{I_8 I_9 L M} R_{I_{10} I_{11} M
  L}
\end{dmath}
The relative coefficients of all terms in (\ref{Gamma 8 SG}) are fixed by
supersymmetry, up to the fact that arbitrary multiples of linear combinations
of terms that vanish when the classical field equations (\ref{Einstein
equations}) and (\ref{3 form field equations}) are satisfied can be added,
because the overall coefficients of such linear combinations of terms can be
adjusted arbitrarily by small amounts, by making redefinitions of the fields
of the form $G_{I J}, C_{I J K} \rightarrow G_{I J} + X_{I J}, C_{I J K} +
Y_{I J K}$, where $X_{I J}$ and $Y_{I J K}$ are polynomials in the fields and
their derivatives, with small coefficients {\cite{Hyakutake Ogushi 2, Hyakutake,
Metsaev}}. \ The coefficient of the $\epsilon_{11} t_8 CR^4$ term, which is
known as the Green-Schwarz term because of its role in anomaly cancellation
{\cite{Green Schwarz}}, is fixed absolutely by anomaly cancellation on
five-branes {\cite{Duff Liu Minasian, Witten 5 branes, Freed Harvey Minasian
Moore, Bilal Metzger, Harvey}}, and confirmed by anomaly cancellation in
Ho\v{r}ava-Witten theory {\cite{de Alwis, Conrad, Faux Lust Ovrut, Lu, Bilal
Derendinger Sauser, Harmark, Bilal Metzger, Meissner Olechowski, Moss 3}}.

The flux term $\Xi^{\mathrm{{flux}}}$ in (\ref{Gamma 8 SG}) is a linear
combination of dimension 8 coordinate scalar monomials built from $R_{I J K
L}$, $G^{I J}$, $\epsilon^{I_1 \ldots I_{11}}_{11}$, $H_{I J K L}$, and the
covariant derivative $D_I$, such that $H_{I J K L}$ occurs at least once in
each monomial, and the number of $R_{I J K L}$ plus the number of $H_{I J K
L}$ is at least 4. \ $\epsilon^{I_1 \ldots I_{11}}_{11}$ occurs once in each
monomial with an odd number of $D_I$, and is absent from monomials with an
even number of $D_I$. \ The only building blocks of odd dimension are $H_{I J
K L}$ and $D_I$, so each occurs an odd number of times if the other does. \ If
$H_{I J K L}$ occurs exactly once then there are three $R_{I J K L}$'s and one
$D_I$, hence also an $\epsilon^{I_1 \ldots I_{11}}_{11}$, and it is impossible
to build a coordinate scalar that does not vanish identically by a Bianchi
identity or the Ricci cyclic identity {\cite{Deser Seminara 1, Deser Seminara
2}}, so in fact there are at
least two $H_{I J K L}$'s in each monomial.

The terms in $\Xi^{\mathrm{{flux}}}$ can be classified by the numbers of
$R_{I J K L}$ and $H_{I J K L}$ in them. \ For general fluxes $H_{I J K L}$
only the $R^2 H^2$, $RH^3$, and $H^4$ terms are known at present {\cite{Deser
Seminara 1, Deser Seminara 2, Peeters Plefka Stern, Metsaev}}. \ These terms
simplify substantially in the special case where there is a particular
coordinate, which for convenience of notation I shall call $y$, although in
general there is no HW boundary nearby, such that all nonvanishing components
of $H_{I J K L}$ have an index $y$, all the fields are independent of $y$,
and, borrowing the notation used above for HW theory, the metric components
$G_{U y}$ are 0. \ Then the $RH^3$ terms vanish, because they have two $H_{I J
K L}$'s contracted with an $\epsilon^{I_1 \ldots I_{11}}_{11}$ {\cite{Deser
Seminara 1, Deser Seminara 2}}, and by the correspondence with type IIA
superstring theory {\cite{Green Schwarz IIA, Witten Various Dimensions}}, in
the special case where only the fields of the type I supergravity multiplet in
10 dimensions are nonzero, the 4-field terms in $\Gamma^{\left( 8,
\mathrm{{bos}} \right)}_{\mathrm{{SG}}}$ must reduce to the 4-field
terms in:
\begin{dmath}
  \label{Gamma I y} \left. \Gamma^{\left( 8, \mathrm{{bos}}
  \right)}_{\mathrm{{SG}}} \right|_{\mathrm{I}, y} = \frac{1}{147456
  \pi^2 \kappa_{11}^2} \left( \frac{\kappa_{11}}{4 \pi} \right)^{4 / 3}
  \int_{\mathcal{B}} d^{11} xe \left( t_8 t_8 \bar{R}^4 - \frac{1}{4!}
  \epsilon_{11} \epsilon_{11} \bar{R}^4 \right),
\end{dmath}
where $\bar{R}_{U V W X} = e_W \, \!^{\hat{Y}} e_X \, \!^{\hat{Z}} \bar{R}_{U
V \hat{Y}  \hat{Z}}$, with $\bar{R}_{U V \hat{W}  \hat{X}}$ defined by (\ref{R
bar}) and $\bar{\omega}_{U \hat{V}  \hat{W}}$ defined by (\ref{omega bar}),
$\bar{R}_{y U V W} = \bar{R}_{V W y U} = R_{y U V W} = 0$, $\bar{R}_{y U y V}
= R_{y U y V}$, and either sign in (\ref{omega bar}) may be used, since with
$\tilde{H}_{U V W} \equiv H_{\hat{y} U V W}$, the $\tilde{R}_{U V W X}$ and\\
$-
\frac{1}{4} \tilde{G}^{Y Z} \left( \tilde{H}_{U W Y} \tilde{H}_{V X Z} -
\tilde{H}_{V W Y} \tilde{H}_{U X Z} \right)$ terms in $\bar{R}_{U V W X}$ are
unaltered by swapping $U V$ with $W X$, while the $\pm \frac{1}{2} \left(
\tilde{D}_U \tilde{H}_{V W X} - \tilde{D}_V \tilde{H}_{U W X} \right)$ terms
change sign due to the Bianchi identity for $H_{I J K L}$, so reversing the
sign of $\tilde{H}_{U V W}$ is equivalent to swapping the two $t_8$'s or the
two $\epsilon_{11}$'s in (\ref{Gamma I y}), so (\ref{Gamma I y}) only contains
even powers of $\tilde{H}_{U V W}$. \ The coefficient of $\tilde{H}_{U V W}$
in $\bar{\omega}_{U \hat{V}  \hat{W}}$ is fixed to $\pm \frac{1}{2}$, in
agreement with (\ref{omega bar}), by comparison with equations (2.8), (2.12),
and (2.13) of {\cite{Gross Sloan}}, noting the unusual normalization of
antisymmetrization brackets around indices used in that article, as shown by a
statement after their equation (2.9), or equivalently by comparison with
equations (5.10), (5.18), and (5.36) of {\cite{Policastro Tsimpis}}, who use
the standard normalization of antisymmetrization brackets, as defined after
equation (\ref{Omega H}) above.

Richards {\cite{Richards 2}} has found evidence that for the field
configurations specified above, the formula (\ref{Gamma I y}) for $\left.
\Gamma^{\left( 8, \mathrm{{bos}} \right)}_{\mathrm{{SG}}}
\right|_{\mathrm{I}, y}$ might also be valid beyond the 4 field level, in
agreement with an earlier suggestion by Kehagias and Partouche {\cite{KP}}.
However the following example shows that (\ref{Gamma I y}) as it stands cannot
be oxidized to a generally covariant formula in 11 dimensions.

We consider the Cartesian product of 7-dimensional Minkowski space and a
2-sphere of radius $a$ and a 2-sphere of radius $b$. \ We choose spherical
polar coordinates $\left( \theta, \varphi \right)$ on the first 2-sphere, and
spherical polar coordinates $\left( \eta, \xi \right)$ on the second 2-sphere,
so the metric is:
\begin{equation}
  \label{metric for KP example} ds_{11}^2 = G_{IJ} dx^I dx^J = \eta_{\alpha
  \beta} dx^{\alpha} dx^{\beta} + a^2 d \theta^2 + a^2 \mathrm{\sin}^2 \theta
  d \varphi^2 + b^2 d \eta^2 + b^2 \mathrm{\sin}^2 \eta d \xi^2,
\end{equation}
where $\eta_{\alpha \beta} = \mathrm{{diag}} \left( - 1, 1, \ldots, 1
\right)$ is the metric on $\left( 6 + 1 \right)$-dimensional Minkowski space.
\ The non-vanishing components of $H_{I J K L}$, up to index permutations,
are:
\begin{equation}
  \label{flux for KP example} H_{\theta \varphi \eta \xi} = h\,
\varepsilon_{\theta \varphi}\, \varepsilon_{\eta \xi} =h\, \mathrm{\sin} \theta
  \, \mathrm{\sin} \eta,
\end{equation}
where if indices $i, j, \ldots$ are tangential to the first 2-sphere and
indices $p, q, \ldots$ are tangential to the second 2-sphere, $\varepsilon_{i
j}$ is a covariantly constant harmonic 2-form on the first 2-sphere, with
non-vanishing components $\varepsilon_{\theta \varphi} = -
\varepsilon_{\varphi \theta} = \mathrm{\sin} \theta$, and $\varepsilon_{p q}$
is a covariantly constant harmonic 2-form on the second 2-sphere, with
non-vanishing components $\varepsilon_{\eta \xi} = - \varepsilon_{\xi \eta} =
\mathrm{\sin} \eta$. \ The flux (\ref{flux for KP example}) is a covariantly
constant harmonic 4-form on the Cartesian product of the two 2-spheres.

For the metric (\ref{metric for KP example}) and the flux (\ref{flux for KP
example}), either $\varphi$ or $\xi$ could be the special coordinate $y$ in
(\ref{Gamma I y}). The $\epsilon_{11} \epsilon_{11} \bar{R}^4$ term gives 0
because $\bar{R}_{I J K L}$ is 0 unless all its indices are tangential to the
Cartesian product of the two 2-spheres, and if we choose $\varphi$ as $y$ then
using Cadabra {\cite{Cadabra 1, Cadabra 2, Cadabra 3, Cadabra 4, Cadabra 5,
Cadabra 6}} we find:
\begin{equation}
  \label{t8 t8 for KP example} t_8 t_8 \bar{R}^4 = \frac{576}{a^8} +
  \frac{384}{a^4 b^4} + \frac{576}{b^8} - \frac{192 h^2}{a^8 b^6} - \frac{576
  h^2}{a^4 b^{10}} + \frac{72 h^4}{a^{12} b^8} + \frac{264 h^4}{a^8 b^{12}} -
  \frac{60 h^6}{a^{12} b^{14}} + \frac{45 h^8}{4 a^{16} b^{16}}
\end{equation}
Choosing $\xi$ as $y$ gives (\ref{t8 t8 for KP example}) with $a$ and $b$
swapped, so since (\ref{t8 t8 for KP example}) is not symmetrical in $a$ and
$b$, (\ref{Gamma I y}) cannot be oxidized to a generally covariant expression
in 11 dimensions, because any such expression would give a result symmetrical
in $a$ and $b$.

Thus (\ref{Gamma I y}) must be subject to a correction if used beyond the
4-field level. \ I shall consider only the case where $H_{I J K L}$ is
covariantly constant. \ For a metric $G_{I J}$ as specified before (\ref{Gamma
I y}), the non-vanishing Christoffel symbols are:
\[ \Gamma_U \, \!^V \, \!_W = \frac{1}{2} G^{V X} \left( \partial_U G_{X W} +
   \partial_W G_{X U} - \partial_X G_{U W} \right) = \tilde{\Gamma}_U \, \!^V
   \, \!_W \]
\begin{equation}
  \label{Christoffel symbols for KP case} \Gamma_y \, \!^y \, \!_W =
  \frac{1}{2} G^{y y} \partial_W G_{y y}, \hspace{1cm} \Gamma_y \, \!^V \,
  \!_y = - \frac{1}{2} G^{V X} \partial_X G_{y y}, \hspace{1cm} \Gamma_U \,
  \!^y \, \!_y = \frac{1}{2} G^{y y} \partial_U G_{y y}
\end{equation}
Thus with $H_{\hat{y} V W X} = \frac{1}{\sqrt{G_{y y}}} H_{y V W X}$:
\begin{equation}
  \label{covariant constancy in 10 dims and 11 dims} \tilde{D}_U H_{\hat{y} V
  W X} = \frac{1}{\sqrt{G_{y y}}} D_U H_{y V W X}
\end{equation}
So $H_{\hat{y} V W X}$ is covariantly constant as a 3-form in the
10-dimensional sense, with the $\hat{y}$ index ignored, if and only if $H_{y V
W X}$ is covariantly constant as a 4-form in the 11-dimensional sense.

If the 4-form is covariantly constant, then from (\ref{R bar}) and (\ref{omega
bar}):
\begin{equation}
  \label{R bar for covariantly constant H} \bar{R}_{U V W X} = R_{U V W X} -
  \frac{1}{4} G^{y y} G^{Y Z} H_{y Y U W} H_{y Z V X} + \frac{1}{4} G^{y y}
  G^{Y Z} H_{y Y V W} H_{y Z U X}
\end{equation}
Thus a simple guess for the oxidation of (\ref{Gamma I y}) to 11 dimensions
when the 4-form is covariantly constant would be to replace $R_{I J K L}$ in
(\ref{Gamma 8 SG}) by:
\begin{equation}
  \label{X I J K L} X_{I J K L} = R_{I J K L} - \frac{1}{8} H_{I K M N} H_{J
  L} \, \!^{M N} + \frac{1}{8} H_{J K M N} H_{I L} \, \!^{M N},
\end{equation}
and discard $\Xi^{\left( \mathrm{{flux}} \right)}$, because terms with a
covariant derivative acting on the 4-form now vanish, and terms without
covariant derivatives are absorbed in the $X^4$ terms. \ This gives a
correction to (\ref{Gamma I y}) when only the fields of the type I
supergravity multiplet in 10 dimensions are nonzero, as required by the above
example, because we then have:
\begin{equation}
  \label{X y U y V} X_{y U y V} = R_{y U y V} + \frac{1}{8} H_{U y M N} H_{y
  V} \, \!^{M N}
\end{equation}
The correction term $ \frac{1}{8} H_{U y M N} H_{y V} \, \!^{M N} $ occurs
multiplied by $2 G^{y y}$ as a source term in the
classical Einstein equations (\ref{Einstein equations}) for $G_{U V}$, so
terms containing it might arise from a modified treatment of the
classical Einstein equation in extracting the $d = 10$ effective action from
the superstring scattering amplitude.

\section{The bulk}
\label{The bulk}

I shall now look for solutions of the quantum-corrected $d = 11$ Einstein
equations that have 4 flat extended dimensions and a compact 7-manifold
$\mathcal{M}^7$ with one or more compact hyperbolic factors, in the presence
of magnetic 4-form fluxes whose bilinears are on average covariantly constant.
\ The averages of terms containing covariant derivatives of the 4-form could
in principle be non-zero, but I shall assume that these terms can be neglected
for a first approximation. \ As a guess for the approximate oxidation of the
Kehagias-Partouche form (\ref{Gamma I y}) when covariant derivatives of the
4-form can be neglected, I shall use (\ref{Gamma 8 SG}) with $R_{I J K L}$
replaced by the $X_{I J K L}$ defined in (\ref{X I J K L}), and $\Xi^{\left(
\mathrm{{flux}} \right)}$ discarded. \ I shall consider $\mathcal{M}^7$
that are Cartesian products of compact hyperbolic and spherical factors, and
work in the leading order of the Lukas-Ovrut-Waldram harmonic expansion of the
energy-momentum tensor on $\mathcal{M}^7$ {\cite{Lukas Ovrut Waldram}}.

There is then one independent Einstein equation for each Cartesian factor of
$\mathcal{M}^7$, plus one Einstein equation for the 4 flat extended
dimensions, because no source terms for the off-diagonal blocks of the Ricci
tensor can be built from the covariantly constant average flux bilinears, and
the source term for each diagonal block of the Ricci tensor is proportional to
the corresponding diagonal block of the metric. \ The Einstein equations are
consistent with the symmetry of the metric ansatz, so by Palais's Principle of
Symmetric Criticality {\cite{Palais, Deser Franklin, Torre}}, the Einstein
equations can be derived by varying the action with respect to an independent
scale factor introduced for each Cartesian factor of $\mathcal{M}^7$, and a
scale factor for the 4 flat extended dimensions.

Within these approximations we can also introduce 5-branes {\cite{Guven}},
which are magnetic sources such that the 5-form dual to the 6-dimensional
world-volume of the 5-brane provides a source for the Bianchi identity of the
4-form, localized on the 5-brane world-volume. \ To preserve Lorentz
invariance 4 dimensions of the 5-brane world-volume must lie along the 4
extended dimensions, and the remaining two dimensions of the 5-brane
world-volume wrap a 2-cycle of $\mathcal{M}^7$. \ I do not consider 2-branes
{\cite{Duff Stelle}} since they are electric sources and their presence would
violate Lorentz invariance, and I do not consider a flux $H_{\mu \nu \sigma
\tau}$ proportional to $\epsilon_{\mu \nu \sigma \tau}$ with indices along the
4 extended dimensions, because by the HW boundary condition (\ref{bc for H})
it would require the presence either of Lorentz-violating Yang-Mills fields,
or Lorentz-violating fluxes, or fluxes that violate translation-invariance
along the extended dimensions, and it is not in general simultaneously
measurable with the magnetic fluxes \cite{Moore}.

The simplest case is when $H^7$ is a compact hyperbolic 7-manifold
$\bar{H}^7$. \ The metric ansatz (\ref{metric ansatz}) now takes the form:
\begin{equation}
  \label{metric ansatz for H7} ds_{11}^2 = G_{IJ} dx^I dx^J = A^2 \eta_{\mu
  \nu} dx^{\mu} dx^{\nu} + B^2 g_{i j} dx^i dx^j,
\end{equation}
where indices $i, j, \ldots$ are tangential to the $\bar{H}^7$, and $g_{i j}$
is a metric of sectional curvature $- 1$ on the $\bar{H}^7$. \ We introduce
fluxes $H_{i j k l}$ wrapping 4-cycles of $\bar{H}^7$ such that on average:
\begin{equation}
  \label{flux bilinears} H_{i j m n} H_{k l o p} G^{m o} G^{n p} =
  \frac{h^2}{B^8} \left( G_{i k} G_{j l} - G_{j k} G_{i l} \right),
\end{equation}
where $h$ is a constant. \ The dependence on $B$ is fixed by the fact that
$H_{i j m n}$ is independent of $B$, because the integral of $H_{i j k l} dx^i
dx^j dx^k dx^l$ over a 4-cycle of $\bar{H}^7$ is quantized, and independent of
$B$. \ From (\ref{flux bilinears}) and (\ref{X I J K L}) we find:
\begin{equation}
  \label{X i j k l} X_{i j k l} = \left( G_{i l} G_{j k} - G_{j l} G_{i k}
  \right) \left( \frac{1}{B^2} + \frac{h^2}{8 B^8} \right)
\end{equation}
The $\epsilon_{11} \epsilon_{11} X^4$ term (\ref{eps11 eps11 X4}) in the
action vanishes because the indices $i, j, \ldots$ run over only 7 dimensions,
and using Cadabra {\cite{Cadabra 1, Cadabra 2, Cadabra 3, Cadabra 4, Cadabra 5,
Cadabra 6}} to evaluate the
$t_8 t_8 X^4$ term (\ref{t8 t8 X4}), the action density is:
\begin{equation}
  \label{action density for H7} \frac{A^4 B^7}{2 \kappa^2_{11}} \left( - 42
  \frac{1}{B^2} - \frac{42}{48}  \frac{h^2}{B^8} - \frac{k^2}{B^5} +
  \frac{254016}{73728 \pi^2} \left( \frac{\kappa_{11}}{4 \pi} \right)^{4 / 3}
  \left( \frac{1}{B^2} + \frac{h^2}{8 B^8} \right)^4 \right)
\end{equation}
The $- \frac{k^2}{B^5}$ term is the classical action of a distribution of
5-branes. \ There is evidence that in HW theory, a 5-brane can only have the
fundamental tension $T_5 = \frac{1}{8 \pi} \left( \frac{4 \pi}{\kappa_{11}}
\right)^{4 / 3}$ {\cite{Lu, Harmark}}. \ The 5-brane extends with any cycle
that it wraps, and the denominator factor is to cancel the measure factor for
each cycle it does not wrap.

To analyse the field equations it is convenient to introduce rescaled
parameters
\begin{equation}
  \label{rescaled parameters} \tilde{B} = \frac{2^{\frac{29}{18}}
  \pi^{\frac{5}{9}}}{21^{\frac{1}{3}}}  \, \frac{B}{\kappa_{11}^{2 / 9}},
  \hspace{1.5cm} \tilde{h} = \frac{2^{\frac{29}{6}} \pi^{\frac{5}{3}}}{21}  \,
  \frac{h}{\kappa_{11}^{2 / 3}}, \hspace{1.5cm} \tilde{k} =
  \frac{2^{\frac{29}{12}} \pi^{\frac{5}{6}}}{21^{\frac{1}{2}}}  \,
  \frac{k}{\kappa_{11}^{1 / 3}},
\end{equation}
in terms of
which the action density becomes a constant multiple of:
\begin{equation}
  \label{with rescaled parameters} A^4 \tilde{B}^7 \left( - 42
  \frac{1}{\tilde{B}^2} - \frac{42}{48}  \frac{\tilde{h}^2}{\tilde{B}^8} -
  \frac{\tilde{k}^2}{\tilde{B}^5} + \left( \frac{1}{\tilde{B}^2} +
  \frac{\tilde{h}^2}{8 \tilde{B}^8} \right)^4 \right)
\end{equation}
The field equations are then:
\begin{equation}
   4096 \tilde{B}^{27}  \tilde{k}^2 - \tilde{h}^8 - 32 \tilde{B}^6 
   \tilde{h}^6 - 384 \tilde{B}^{12}  \tilde{h}^4 + 3584 \tilde{B}^{24} 
   \tilde{h}^2 - 2048 \tilde{B}^{18}  \tilde{h}^2 + 172032 \tilde{B}^{30} -
   4096 \tilde{B}^{24} = 0
\end{equation}
\begin{dmath}[compact, spread=3pt]
   8192 \tilde{B}^{27}  \tilde{k}^2 + 25 \tilde{h}^8 + 608 \tilde{B}^6 
   \tilde{h}^6 + 4992 \tilde{B}^{12}  \tilde{h}^4 - 3584 \tilde{B}^{24} 
   \tilde{h}^2 + 14336 \tilde{B}^{18}  \tilde{h}^2 + 860160 \tilde{B}^{30} +
   4096 \tilde{B}^{24} = 0
\end{dmath}
Adding these two equations, we find:
\begin{equation}
   24 \left( 512 \tilde{B}^{27}  \tilde{k}^2 + \tilde{h}^8 + 24 \tilde{B}^6 
   \tilde{h}^6 + 192 \tilde{B}^{12}  \tilde{h}^4 + 512 \tilde{B}^{18} 
   \tilde{h}^2 + 43008 \tilde{B}^{30} \right) = 0
\end{equation}
Thus there is no solution. \ Consideration of the $\tilde{B}^{24} \tilde{h}^2$
and $\tilde{B}^{24}$ terms shows that this is the only linear combination, up
to an overall factor, that gives a non-negative coefficient to every term.

I then studied the $\bar{H}^5 \times \bar{H}^2$ and $\bar{H}^4 \times
\bar{H}^3$ cases, including all possible types of flux. \ In both cases, there
was a unique linear combination of the 3 field equations, up to an overall
factor, that gave a non-negative coefficient to every term, and thus proved
there was no solution. \ I then studied the $\bar{H}^5 \times S^2$, $\bar{H}^4
\times S^3$, and $\bar{H}^3 \times S^4$ cases, and also the $\bar{H}^3 \times
S^2 \times S^2$ case, with the radii of the two $S^2$'s set equal after
deriving the field equations. \ There was no longer any simple proof that no
solution existed, but searches by various methods found no solution. \ TeXmacs
files containing some of these calculations, together with Cadabra scripts and
source code for a C++ program used in one of the searches, are available on
the web page {\cite{h7bulk}}.

I therefore returned to the $\bar{H}^7$ case, and studied whether a field
redefinition could produce a solution, even though, in principle, field
redefinitions should have no physical effects when all orders of perturbation
theory are included. \ Trying a redefinition that affects only the coefficient
of the highest power of $\frac{1}{\tilde{B}}$ in the action, one finds that as
soon as this coefficient becomes negative, a solution appears. \ One might
suspect that a solution that disappears at the leading order it exists on
performing a field redefinition is unphysical, and that this will show up
through the presence of tachyonic modes or through large corrections at higher
orders in $\kappa^{2 / 3}_{11}$.

However there is no absolutely preferred set of coordinates in ``field
space'', and I will now show that a solution also appears when we make a field
redefinition that has the same form as the redefinition that transforms the
action with the unexpanded $\epsilon_{11} \epsilon_{11} X^4$ term to the form
that occurs naturally when supersymmetry is systematically implemented by the
Noether method {\cite{Hyakutake Ogushi 1, Hyakutake Ogushi 2, Hyakutake,
Hyakutake 2}}, except that instead of
making the redefinition with coefficient 1, we make it with a coefficient $< -
0.34$. \ Thus a solution appears after a redefinition that is much closer to
the identity than a redefinition that is certainly admissible.

To transform the action to the form that would naturally be obtained by the
Noether method, we have to remove all the Ricci terms from the expansion of
$\epsilon_{11} \epsilon_{11} X^4$ by using the classical Einstein equations
(\ref{Einstein equations}). \ Let:
\begin{equation}
  \label{V I K} V_{I K} \equiv X_{I J K} \, \!^J = R_{I K} - \frac{1}{8} H_I
  \, \!^{J M N} H_{K J M N}
\end{equation}
\begin{equation}
  \label{U} U \equiv V_I \, \!^I = R - \frac{1}{8} H^{I J M N} H_{I J M N}
\end{equation}
where I used (\ref{X I J K L}). \ Let $\tilde{V}_{I K}$ and $\tilde{U}$ be the
result of using the classical Einstein equations (\ref{Einstein equations}) to
replace the Ricci terms in $V_{I K}$ and $U$ by $H^2$ terms. \ Then:
\begin{equation}
  \label{V tilde I K} \tilde{V}_{I K} = - \frac{1}{24} H_I \, \!^{J M N} H_{K
  J M N} - \frac{1}{144} H^{O L M N} H_{O L M N} G_{I K}
\end{equation}
\begin{equation}
  \label{U tilde} \tilde{U} = - \frac{17}{144} H^{I J M N} H_{I J M N}
\end{equation}
So for the $\bar{H}^7$ case we find from (\ref{flux bilinears}):
\begin{equation}
  \label{V tilde for H bar 7} \tilde{V}_{\mu \nu} = - \frac{7}{24} 
  \frac{h^2}{B^8} G_{\mu \nu}, \hspace{2cm} \tilde{V}_{i j} = - \frac{13}{24} 
  \frac{h^2}{B^8} G_{i j}
\end{equation}
\begin{equation}
  \label{U tilde for H bar 7} \tilde{U} = - \frac{119}{24}  \frac{h^2}{B^8}
\end{equation}
Now using Cadabra we find that:
\begin{dmath}[compact, spread=3pt]
  \label{minus eps eps X4} - \frac{1}{4!} \epsilon_{11} \epsilon_{11} X^4 = 4
  U^4 - 96 V_{I J} V_{I J} U^2 + 24 X_{I J K L} X_{I J K L} U^2 + 384 X_{I J K
  L} V_{I K} V_{J L} U + 256 V_{I J} V_{I K} V_{J K} U - 384 X_{I J K L} X_{I
  J K M} V_{L M} U + 32 X_{I J K L} X_{I J M N} X_{K L M N} U - 128 X_{I J K
  L} X_{I M K N} X_{J N L M} U + 192 V_{I J} V_{I J} V_{K L} V_{K L} - 96 X_{I
  J K L} X_{I J K L} V_{M N} V_{M N} - 768 X_{I J K L} X_{I M K N} V_{J L}
  V_{M N} - 1536 X_{I J K L} V_{I K} V_{J M} V_{L M} + 768 X_{I J K L} X_{I J
  K M} X_{L N M O} V_{N O} - 384 V_{I J} V_{I K} V_{J L} V_{K L} + 768 X_{I J
  K L} X_{I J K M} V_{L N} V_{M N} + 384 X_{I J K L} X_{I J M N} V_{K M} V_{L
  N} - 384 X_{I J K L} X_{I J M N} X_{K L M O} V_{N O} + 768 X_{I J K L} X_{I
  M K N} V_{J M} V_{L N} + 1536 X_{I J K L} X_{I M K N} X_{J N L O} V_{M O} +
  12 X_{I J K L} X_{I J K L} X_{M N O P} X_{M N O P} - 192 X_{I J K L} X_{I J
  K M} X_{L N O P} X_{M N O P} + 24 X_{I J K L} X_{I J M N} X_{K L O P} X_{M N
  O P} - 384 X_{I J K L} X_{I J M N} X_{K O M P} X_{L P N O} + 192 X_{I J K L}
  X_{I M K N} X_{J O L P} X_{M O N P} - 384 X_{I J K L} X_{I M K N} X_{J O M
  P} X_{L O N P}
\end{dmath}
This used the fact that $X_{I J K L}$ in (\ref{X I J K L}) has the same
monoterm symmetries as the Riemann tensor, namely antisymmetry in the first
two indices, antisymmetry in the last two indices, and symmetry under exchange
of the first two indices with the last two indices, although it does not
satisfy the Ricci cyclic identity, because it has a piece with [1111] Young
tableau symmetry.

Let $Z \equiv - \frac{1}{4!} \epsilon_{11} \epsilon_{11} X^4$, and let
$\tilde{Z}$ be the result of replacing $V_{I J}$ by $\tilde{V}_{I J}$ and $U$
by $\tilde{U}$ in the right-hand side of (\ref{minus eps eps X4}), so that $Z
= \tilde{Z}$ if the classical Einstein equations (\ref{Einstein equations})
are satisfied. \ Then $\left( \tilde{Z} - Z \right)$, with an arbitrary
coefficient that is not too large, is an expression that can be added to the
action at this order by means of a field redefinition. \ To reach the form of
the action that would naturally be obtained by imposing supersymmetry by the
Noether method, one has to add $\left( \tilde{Z} - Z \right)$ with coefficient
1 to the integrand of the dimension 8 local term (\ref{Gamma 8 SG}), on page
\pageref{Gamma 8 SG},
which is here being taken with $\Xi^{\left( \mathrm{{flux}} \right)}$
discarded, and $R_{IJKL}$ replaced by the $X_{I J K L}$ defined in (\ref{X I J
K L}), on page \pageref{X I J K L}.

From (\ref{V tilde for H bar 7}), (\ref{U tilde for H bar 7}), and (\ref{minus
eps eps X4}), we find using Cadabra and Maxima {\cite{Maxima}} that for the
$\bar{H}^7$ case, for which $Z = 0$:
\begin{equation}
  \label{Z tilde minus Z} \tilde{Z} - Z = 7 \left( - \frac{32832}{B^8} +
  \frac{17088 h^2}{B^{14}} + \frac{161 h^4}{2 B^{20}} + \frac{20395 h^6}{24
  B^{26}} + \frac{2225255 h^8}{82944 B^{32}} \right)
\end{equation}
Thus adding $\left( \tilde{Z} - Z \right)$ with a coefficient $c$ to the
integrand of (\ref{Gamma 8 SG}) with the above replacements, we find that
the action density in terms of the rescaled parameters (\ref{rescaled
parameters}), with the 5-brane term omitted, becomes a constant
positive multiple of:
\begin{dmath}[compact, spread=3pt]
  \label{action with field redefinition term} A^4  \tilde{B}^7  \left(
  - \frac{42}{\tilde{B}^2} - \frac{7 \tilde{h}^2}{8 \tilde{B}^8} + \left(
  \frac{1}{\tilde{B}^2} + \frac{\tilde{h}^2}{8 \tilde{B}^8} \right)^4 + c
  \left( - \frac{19}{21 \tilde{B}^8} + \frac{89 \tilde{h}^2}{189
  \tilde{B}^{14}} + \frac{23 \tilde{h}^4}{10368 \tilde{B}^{20}} + \frac{20395
  \tilde{h}^6}{870912 \tilde{B}^{26}} + \frac{2225255 \tilde{h}^8}{3009871872
  \tilde{B}^{32}} \right) \right)
\end{dmath}

Deriving the
field equations, one finds that a flat $\mathbf{R}^4$ times $\bar{H}^7$
solution exists for $c < c_{\mathrm{\max}} \equiv - \frac{734832}{2225255}
\simeq - 0.3302$. \ Thus a solution exists for precisely the range of $c$
for which the coefficient of the highest power of $\frac{1}{\tilde{B}}$ in
the action is negative. \ To study the field
equations, it is convenient to define
$x \equiv \frac{\tilde{h}}{\tilde{B}^3} = \frac{h}{B^3}$. \ The equations
then become two
simultaneous linear equations for $c$ and $\tilde{B}^6$. \ Solving the
equations one finds a hyperbola-like relation between $c$ and $x$, with $x
\rightarrow + \infty$ as $c \rightarrow c_{\mathrm{\max}}$ from below, and $c
\rightarrow - \infty$ as $x \rightarrow x_{\mathrm{\min}} \simeq 0.9364$ from
above. \ The range of $x$ perceived as the ``corner'' of the hyperbola
depends on the range of $x$ for which the curve is plotted, but when the
lower limit of the plot is about $x = 3$, so that $c$ does not drop below
about $2 c_{\mathrm{\max}}$, the ``corner'' of the hyperbola is from about
$x = 3$ to about $x = 8$.

One possibility is that this solution is unphysical, and either has tachyonic
modes or is subject to very large corrections at higher orders in $\kappa^{2 /
3}_{11}$. \ Another possibility is that there might be a ``field redefinition
group'' similar to the renormalization group, and that at low orders of
perturbation theory, we should use ``coordinates in field space'' best suited
to the geometry being studied. \ Doing that would minimize the size of the
higher-order corrections. \ Sensitivity to field redefinitions should decrease
as higher-order corrections are included, so we should use the \emph{principle
of minimal sensitivity} {\cite{Stephenson}} to choose the best field
redefinition at low orders of perturbation theory, as in perturbative
QCD. \ $c$ is the coefficient of the difference from the identity of a
perturbative field redefinition, that we are neglecting terms of quadratic and
higher order in the Taylor expansion of the action with respect to, so the
magnitude of $c$ should also be as small as possible.  Thus we should
choose $c$ at the ``corner'' of the hyperbola in the $\left( x, c \right)$
plane. \ Choosing $x = 5.5$, we find $c \simeq - 0.38$, and $\tilde{B} \simeq
0.91$, which corresponds to
\begin{equation}
\label{best value of B} B \simeq 0.43 \kappa^{2 / 9}_{11}
\end{equation}
as the best
value of the radius of curvature of the $\bar{H}^7$. \ The best value of $h =
B^3 x$ is then $h \simeq 0.45 \kappa^{2 / 3}_{11}$.

The first Pontryagin class of an $\bar{H}^7$ is 0, so
the fluxes $H_{i j k l}$ wrapping individual 4-cycles of the
$\bar{H}^7$\hspace{-2pt} are constrained\hspace{-1pt} by\hspace{-1pt} flux
quantization such that
the integral of $\frac{1}{4!} H_{i j k l} dx^i \wedge dx^j \wedge dx^k \wedge
dx^l$ over a 4-cycle is equal to $\frac{2 \pi}{T_2}$ times an integer, where
$T_2 = \frac{1}{2} \left( \frac{4 \pi}{\kappa_{11}} \right)^{2 / 3}$ is the
fundamental membrane tension
{\cite{Rohm Witten, Witten flux quantization, Bousso Polchinski}}. \ If the
average flux $h$ is not exactly equal to the value required by the
solution of the field equations with flat extended dimensions, then the
extended dimensions will have to be slightly curved to compensate. \ With
$\kappa^{2 / 9}_{11}$ around an inverse TeV, thus around $10^{- 19}$ metres,
the observed de Sitter radius of the extended dimensions,
which is $16.0$ Gyr $= \hspace{1pt}
\hspace{1pt} 1.51 \times 10^{26}$ metres $= \hspace{1pt} \hspace{1pt}
\hspace{2pt} 0.94 \times 10^{61} \sqrt{G_N}$ {\cite{de Sitter radius}},
is around $10^{45} \kappa^{2 /
9}_{11}$, so the curvature of the extended dimensions is around $10^{- 90}
\kappa^{- 4 / 9}_{11}$. \ For an $\bar{H}^7$ with intrinsic volume
$\bar{V}_7$ around $10^{34}$, typical 4-cycles of the $\bar{H}^7$ also have
intrinsic 4-area
around $10^{34}$, so individual flux intensities can be adjusted to a relative
precision of around $10^{- 34}$. \ By choosing incommensurate flux numbers,
the average intensity of the fluxes wrapping any three 4-cycles can be
adjusted to a relative precision of around $10^{- 102}$. \ The number of
independent fluxes is the number of linearly independent harmonic 4-forms,
which is the 4th Betti number, which from the discussion on the pages before
and after (\ref{generalized Gauss Bonnet theorem}), near the end of
section~\ref{Introduction}, is likely to be roughly in proportion to the
intrinsic volume, up to a logarithmic correction factor {\cite{Gromov Volume
Bounded Cohomology, Xue, Clair Whyte}}. \ Thus flux
quantization does not prevent the observed de Sitter radius from being
attained.

The second derivative of (\ref{action with field redefinition term}) with
respect to $\tilde{B}$, at the above solution, is $\simeq - 19000
\tilde{A}^4$, so a deviation of $\tilde{B}$ from its value $\tilde{B} \simeq
0.91$ at the solution has positive energy, and this would normally correspond
to positive mass$^2$ for the mode where the deviation of $\tilde{B}$ from
$0.91$ depends on position in the 4 extended dimensions. \ This mode is a
scalar on the extended dimensions, and its only position dependence is its
position dependence along the extended dimensions. \ For a first approximation
I shall take the derivative part of its kinetic term as given by the Einstein
action plus gauge-fixing terms. \ The Einstein action density, up to a
positive overall factor and a total
derivative term which I shall temporarily drop, is:
\begin{dmath}[compact, spread=3pt]
  \label{spin 2 action density} \sqrt{- G} R = - \frac{1}{4} \sqrt{- G} G^{I
  J} G^{K L} G^{M N} \left( \partial_I G_{K M} \partial_J G_{L N} - 2
  \partial_I G_{N K} \partial_M G_{J L} + 2 \partial_I G_{K L} \partial_M G_{N
  J} - \partial_I G_{K L} \partial_J G_{M N} \right)
\end{dmath}
Extracting the part of this which has $\partial_{\mu}$ derivatives on $B$ in
the metric ansatz (\ref{metric ansatz for H7}), on page \pageref{metric ansatz
for H7}, we see that only the first and last terms in the right-hand side of
(\ref{spin 2 action density}) contribute, giving
\begin{equation}
\label{derivative part of kinetic term}- \frac{1}{4} \sqrt{- G}
\frac{1}{B^2} g^{kl} g^{mn} \left( g_{k m} g_{l n} - g_{kl} g_{mn} \right)
\partial_{\mu} B \partial^{\mu} B = \frac{21}{2} \sqrt{- G} \frac{1}{B^2}
\partial_{\mu} B \partial^{\mu} B.
\end{equation}
Thus this is a mode for which the
kinetic term coming from the Einstein action has the wrong sign, and its
positive mass$^2$ could potentially lead to exponential instead of oscillatory
time dependence. \ However a gauge-fixing term such as
\begin{equation}
\label{gauge fixing term}\frac{1}{2 \alpha}
\sqrt{- G} G^{I J} G^{K L} \left( \partial_K \varphi_{L I} - \beta \partial_I
\varphi_{K L} \right) G^{M N} \left( \partial_M \varphi_{N J} - \beta
\partial_J \varphi_{M N} \right),
\end{equation}
where $\varphi_{I J}$ is the perturbation
of $G_{I J}$ from the solution found above, and any value of $\beta$ except
$\beta = 1$ is allowed, will generically alter the derivative part of the
kinetic term of this mode, and can change its sign. \ ($\beta = 1$ is not
allowed because $G^{K L} \left( \partial_K \varphi_{L I} - \partial_I
\varphi_{K L} \right)$ has a residual invariance under $\varphi_{I J}
\rightarrow \varphi_{I J} + \partial_I \partial_J \lambda$, and can thus not
be set equal to an arbitrary 11-vector by a gauge transformation $\varphi_{I
J} \rightarrow \varphi_{I J} + D_I \varepsilon_J + D_J \varepsilon_I$.) \ Thus
no exponential time dependence of this mode is observable.

\section{The HW boundary}
\label{The HW boundary}

Assuming provisionally that the solution found in the previous subsection is
valid, I shall now show that it might be possible to introduce an HW boundary,
with the region around the boundary treated in a consistent semiclassical
approximation. \ This can at best be only semi-quantitative, because no higher
derivative corrections to the boundary conditions, corresponding to the
leading quantum corrections used in the bulk, are introduced. \ However
higher derivative boundary conditions have multiple solutions, most of which
will be unstable, with tachyonic excitations. \ We have to select the solution
free of tachyons, which will be the smoothest one, for which the higher
derivative corrections have the least possible significance. \ Thus it is
useful to find a solution that satisfies the semiclassical boundary conditions
(\ref{Israel boundary conditions}) and (\ref{bc for H}), since it should
provide a good starting point for finding the correct solution to the boundary
conditions with the higher derivative corrections included. \ I shall assume
that the
boundary condition (\ref{bc for H}) for the fluxes can be satisfied in an
average sense, and consider the boundary condition (\ref{Israel boundary
conditions}) for the metric.

For simplicity I shall consider a boundary whose topology is a smooth compact
quotient $\bar{H}^6$ of $H^6$, although it is not known whether $\bar{H}^6$ of
suffiently small intrinsic volume $\bar{V}_6$ actually exist. \ Better choices
for the boundary topology might be the Cartesian product of two small
intrinsic volume $\bar{H}^3$'s, or an Anderson 6-manifold {\cite{Anderson}},
but it does not appear to be known whether an $\bar{H}^7$ can have
minimal-area 6-cycles of these topologies. \ I shall assume that the boundary
is somewhere near a minimal-area 6-cycle of the closed $\bar{H}^7$ that is cut
into two disconnected parts along the boundary, and use the notation of
section \ref{Horava Witten theory}, starting on page \pageref{Horava Witten
theory}. \ The metric in this region has the form:
\begin{equation}
  \label{metric near boundary} ds_{11}^2 = G_{IJ} dx^I dx^J = a \left( y
  \right)^2 \eta_{\mu \nu} dx^{\mu} dx^{\nu} + b \left( y \right)^2 h_{a b}
  dx^a dx^a + dy^2
\end{equation}
Here $a \left( y \right)$ is as in (\ref{metric ansatz}), on page
\pageref{metric ansatz}, so $a \left( y
\right) \rightarrow A$ away from the boundary, and $a \left( y \right)
\rightarrow 1$ on the boundary. \ Indices $a, b, c, \ldots$ are tangential to
the boundary, so that $x^A$ in (\ref{metric ansatz}) is now $\left( x^a, y
\right)$, and $h_{a b}$ is a metric of sectional curvature $- 1$.
\vspace{0.2cm}

The nonvanishing Riemann tensor components for the metric ansatz (\ref{metric
near boundary}) are:
\[ R_{\mu \nu \sigma \tau} = \frac{\dot{a}^2}{a^2} \left( G_{\mu \tau} G_{\nu
   \sigma} - G_{\mu \sigma} G_{\nu \tau} \right), \]
\[ R_{a b c d} = b^2 R_{a b c d} \left( h \right) + \frac{\dot{b}^2}{b^2}
   \left( G_{a d} G_{b c} - G_{a c} G_{b d} \right), \]
\begin{equation}
  \label{Riemann tensor components} R_{\mu a \nu b} = - \frac{\dot{a}
  \dot{b}}{ab} G_{\mu \nu} G_{a b}, \hspace{1.2cm} R_{\mu y \nu y} = -
  \frac{\ddot{a}}{a} G_{\mu \nu}, \hspace{1.2cm} R_{ayby} = -
  \frac{\ddot{b}}{b} G_{a b},
\end{equation}
together with the components related to these by the antisymmetries of the
Riemann tensor under the interchange of its first two indices, and under the
interchange of its last two indices, where a dot denotes
differentiation with respect to $y$, and $R_{a b c d} \left( h \right) =
h_{a d} h_{b c} - h_{a c} h_{b d}$
denotes the Riemann tensor calculated from the six-dimensional metric $h_{a
b}$.
\vspace{0.2cm}

Thus the nonvanishing Ricci tensor components for the metric ansatz
(\ref{metric near boundary}) are:
\[ R_{\nu \tau} = - 3 \frac{\dot{a}^2}{a^2} G_{\nu \tau} - 6 \frac{\dot{a}
   \dot{b}}{ab} G_{\nu \tau} - \frac{\ddot{a}}{a} G_{\nu \tau} = - \left( 3
   \frac{\dot{a}^2}{a^2} + 6 \frac{\dot{a} \dot{b}}{ab} + \frac{\ddot{a}}{a}
   \right) G_{\nu \tau} \]
\[ R_{b d} = R_{b d} \left( h \right) - 5 \frac{\dot{b}^2}{b^2} G_{b d} - 4
   \frac{\dot{a} \dot{b}}{ab} G_{b d} - \frac{\ddot{b}}{b} G_{b d} = - \left(
   \frac{5}{b^2} + 5 \frac{\dot{b}^2}{b^2} + 4 \frac{\dot{a} \dot{b}}{ab} +
   \frac{\ddot{b}}{b} \right) G_{b d} \]
\begin{equation}
  \label{Ricci tensor components} R_{y y} = - 4 \frac{\ddot{a}}{a} - 6
  \frac{\ddot{b}}{b}
\end{equation}
where I used that $R_{b d} \left( h \right) = - 5 h_{b d}$.
\vspace{1.0cm}

I shall look for a solution where the metric in the region of the boundary is
a small perturbation of what it would have been if the boundary was not there.
\ In the absence of the boundary $a$ is equal to the constant $A$, and $R_{A B
C D}$ is $\frac{1}{B^2} \left( G_{A D} G_{B C} - G_{A C} G_{B D} \right)$, the
Riemann tensor on the closed $\bar{H}^7$, where
$B$ has the value $B \simeq 0.43
\kappa^{2 / 9}_{11}$ found in the previous section. \ Thus from the second
line of (\ref{Riemann tensor components}), we require $\left( \frac{1}{b^2} +
\frac{\dot{b}^2}{b^2} \right) = \frac{1}{B^2}$, and assuming $b \left( y
\right)$ has its minimum value at $y = 0$, the solution of this is $b = B
\mathrm{\cosh} \left( \frac{y}{B} \right)$. \ This then also gives the correct
value $R_{ayby} = - \frac{1}{B^2} G_{a b}$, from the third line of
(\ref{Riemann tensor components}). \ I shall now work out what the effective
energy-momentum tensor components must be, in order for this form of $b \left(
y \right)$, with constant $a\left( y\right) = A$, to solve the classical
Einstein equations.

I assume the effective energy-momentum tensor $T_{I J}$ in the absence of the
boundary has the form:
\begin{equation}
  \label{effective energy momentum tensor} T_{\mu \nu} = t^{\left( 1 \right)}
  \left( y \right) G_{\mu \nu}, \hspace{1cm} T_{a b} = t^{\left( 2 \right)}
  \left( y \right) G_{a b}, \hspace{1cm} T_{y y} = t^{\left( 3 \right)} \left(
  y \right)
\end{equation}
The Einstein equations can be written:
\begin{equation}
  \label{effective Einstein equations} R_{I J} + \kappa^2_{11} \left( T_{I J}
  - \frac{1}{9} G_{I J} G^{K L} T_{K L} \right) = 0,
\end{equation}
and using the Ricci tensor components from (\ref{Ricci tensor components}),
these are:
\begin{equation}
  \label{first Einstein eqn} \frac{\ddot{a}}{a} + 3 \frac{\dot{a}^2}{a^2} + 6
  \frac{\dot{a} \dot{b}}{ab} + \frac{\kappa^2_{11}}{9} \left( 5 t^{\left( 1
  \right)} \left( y \right) - 6 t^{\left( 2 \right)} \left( y \right) -
  t^{\left( 3 \right)} \left( y \right) \right) = 0
\end{equation}
\begin{equation}
  \label{second Einstein eqn} \frac{\ddot{b}}{b} + 5 \frac{\dot{b}^2}{b^2} + 4
  \frac{\dot{a} \dot{b}}{ab} + \frac{5 \lambda}{b^2}
  + \frac{\kappa^2_{11}}{9} \left(
  - 4 t^{\left( 1 \right)} \left( y \right) + 3 t^{\left( 2 \right)} \left( y
  \right) - t^{\left( 3 \right)} \left( y \right) \right) = 0
\end{equation}
\begin{equation}
  \label{third Einstein eqn} 4 \frac{\ddot{a}}{a} + 6 \frac{\ddot{b}}{b} +
  \frac{\kappa^2_{11}}{9} \left( - 4 t^{\left( 1 \right)} \left( y \right) - 6
  t^{\left( 2 \right)} \left( y \right) + 8 t^{\left( 3 \right)} \left( y
  \right) \right) = 0
\end{equation}
In (\ref{second Einstein eqn}) I have introduced a factor $\lambda$
multiplying the $\frac{5}{b^2}$ term as a rough way of allowing for the
possibility that the boundary might be, for example, the Cartesian product of
two $\bar{H}^3$'s, or an Anderson 6-manifold {\cite{Anderson}}. \ Setting
$\lambda = 1$ gives the equation that follows from the Ricci tensor
components (\ref{Ricci tensor components}).

Eliminating the double derivatives between the three Einstein equations, we
find:
\begin{equation}
  \label{no double derivatives} \frac{\dot{a}^2}{a^2} + 4 \frac{\dot{a}
  \dot{b}}{ab} + \frac{5 \dot{b}^2}{2 b^2}
  + \frac{5 \lambda}{2 b^2} - \frac{1}{6}
  \kappa^2_{11} t^{\left( 3 \right)} = 0,
\end{equation}
hence:
\begin{equation}
  \label{eqn for a dot} \frac{\dot{a}}{a} = - 2 \frac{\dot{b}}{b} +
  \frac{1}{2} \sqrt{6 \frac{\dot{b}^2}{b^2} - \frac{10 \lambda}{b^2} +
  \frac{2}{3} \kappa_{11}^2 t^{\left( 3 \right)}},
\end{equation}
where I assumed that $\dot{b}$ does not change sign between the boundary and
the main part of the bulk, so the sign of the square root follows from
$\dot{a} = 0$ and the convention stated at the start of section \ref{Horava
Witten theory}, on page \pageref{Horava Witten theory}, that $y > y_1$ in the
bulk, where $y_1$ is the value of $y$ at the boundary.

The second Einstein equation, (\ref{second Einstein eqn}), now becomes:
\begin{equation}
  \label{eqn for b double dot} \frac{\ddot{b}}{b} - 3 \frac{\dot{b}^2}{b^2} +
  2 \frac{\dot{b}}{b} \sqrt{6 \frac{\dot{b}^2}{b^2} - \frac{10 \lambda}{b^2} +
  \frac{2}{3} \kappa_{11}^2 t^{\left( 3 \right)}} + \frac{5 \lambda}{b^2} +
  \frac{\kappa_{11}^2}{9} \left( - 4 t^{\left( 1 \right)}
  + 3 t^{\left( 2 \right)}
  - t^{\left( 3 \right)} \right) = 0
\end{equation}

From subsection 2.4 of {\cite{CCHT}}, a solution of (\ref{eqn for a dot}) and
(\ref{eqn for b double dot}) solves all three Einstein equations, provided the
energy-momentum tensor is conserved, and the square root does not vanish
identically for the solution. \ The square root does not vanish for the
solution I will consider, so it is sufficient to consider (\ref{eqn for a
dot}) and (\ref{eqn for b double dot}).

Requiring that (\ref{eqn for a dot}) and (\ref{eqn for b double dot}) are
solved by $a \left( y \right) = A$, $b \left( y \right) = B \mathrm{\cosh}
\frac{y}{B}$, we find the effective energy-momentum tensor components:
\begin{equation}
\label{t upper 3} \frac{2}{3} \kappa_{11}^2 t^{\left( 3 \right)} =
  \frac{10\, \mathrm{\sinh}^2 \frac{y}{B}}{B^2 
  \mathrm{\cosh}^2 \frac{y}{B}} + \frac{10 \lambda}{B^2 \mathrm{\cosh}^2
  \frac{y}{B}}
\end{equation}
\begin{equation}
  \label{t1 t2 t3} \frac{\kappa_{11}^2}{9} \left( - 4 t^{\left( 1 \right)} + 3
  t^{\left( 2 \right)} - t^{\left( 3 \right)} \right) = - \left( \frac{1}{B^2}
  + 5 \frac{\mathrm{\sinh}^2 \frac{y}{B}}{B^2  \mathrm{\cosh}^2 \frac{y}{B}} +
  \frac{5 \lambda}{B^2  \mathrm{\cosh}^2 \frac{y}{B}} \right)
\end{equation}

Substituting in these values of the $t^{\left( i \right)}$, the Einstein
equations (\ref{eqn for a dot}) and (\ref{eqn for b double dot}) become:
\begin{equation}
  \label{first eqn in background} \frac{\dot{a}}{a} = - 2 \frac{\dot{b}}{b} +
  \frac{1}{2} \sqrt{6 \frac{\dot{b}^2}{b^2} - \frac{10 \lambda}{b^2} +
  \frac{10 \mathrm{\sinh}^2 \frac{y}{B}}{B^2  \mathrm{\cosh}^2 \frac{y}{B}} +
  \frac{10 \lambda}{B^2  \mathrm{\cosh}^2 \frac{y}{B}}}
\end{equation}
\begin{dmath}[compact, spread=3pt]
  \label{2nd eqn in background} \frac{\ddot{b}}{b} - 3 \frac{\dot{b}^2}{b^2} +
  2 \frac{\dot{b}}{b} \sqrt{6 \frac{\dot{b}^2}{b^2} - \frac{10 \lambda}{b^2} +
  \frac{10 \mathrm{\sinh}^2 \frac{y}{B}}{B^2  \mathrm{\cosh}^2 \frac{y}{B}} +
  \frac{10 \lambda}{B^2  \mathrm{\cosh}^2 \frac{y}{B}}} + \frac{5
  \lambda}{b^2} - \left( \frac{1}{B^2} + 5 \frac{\mathrm{\sinh}^2
  \frac{y}{B}}{B^2  \mathrm{\cosh}^2 \frac{y}{B}} + \frac{5 \lambda}{B^2 
  \mathrm{\cosh}^2 \frac{y}{B}} \right) = 0
\end{dmath}

We now let $a\left(y\right) = \left( 1 + p\left(y\right) \right) A$,
$b\left(y\right) = \left( 1 + q\left(y\right) \right) B
\mathrm{\cosh} \frac{y}{B}$, and expand to first order in $p$ and $q$. \ Then
(\ref{first eqn in background}) and (\ref{2nd eqn in background}) become:
\begin{equation}
  \label{p dot eqn} \dot{p} = - \frac{5}{4} \dot{q} + \frac{5 \lambda q}{4 B
  \mathrm{\sinh} \frac{y}{B} \mathrm{\cosh} \frac{y}{B}}
\end{equation}
\begin{equation}
  \label{q double dot eqn} \ddot{q} + 7 \dot{q}  \frac{\mathrm{\sinh}
  \frac{y}{B}}{B \mathrm{\cosh} \frac{y}{B}} - \frac{5 \lambda q}{B^2
  \mathrm{\cosh}^2 \frac{y}{B}} = 0
\end{equation}

For a first estimate of the boundary conditions (\ref{Israel boundary
conditions}), on page \pageref{Israel boundary conditions}, I shall neglect
the flux terms in $\bar{R}_{U V W X}$, defined
in (\ref{R bar}), by assuming, if necessary, that $H_{\hat{y} U V W}$ is
smaller than its average value, near the boundary. \ The energy-momentum
tensor on the boundary is then:
\begin{dmath}[compact, spread=3pt]
  \label{energy momentum tensor on boundary} \tilde{T}_{U V} = \frac{1}{16 \pi
  \kappa^2_{11}} \left( \frac{\kappa_{11}}{4 \pi} \right)^{2 / 3} \left(
  \frac{4}{30} \mathrm{{tr}} F_{U W} F_V \, \!^W - \frac{1}{30}
  \tilde{G}_{U V} \mathrm{{tr}} F_{W X} F^{W X} - 2 \bar{R}_{U W X Y}
  \bar{R}_V \, \!^{W X Y} + \frac{1}{2} \tilde{G}_{U V} \bar{R}_{W X Y Z}
  \bar{R}^{W X Y Z} \right)
\end{dmath}

I shall work in the leading order of the Lukas-Ovrut-Waldram {\cite{Lukas
Ovrut Waldram}} harmonic expansion of the energy-momentum tensor (\ref{energy
momentum tensor on boundary}) on the boundary, and assume that
$\mathrm{{tr}} F_{a c} F_b \, \!^c$ is a multiple of $\tilde{G}_{a b}$,
and also that $\bar{R}_{a c d e} \bar{R}_b \, \!^{c d e}$ is a multiple of
$\tilde{G}_{a b}$, which it is if the boundary is an $\bar{H}^6$, but is not
if the boundary is an $\bar{H}^3 \times \bar{H}^3$, for example. \ We then
find:
\begin{equation}
  \label{T tilde mu nu} \tilde{T}_{\mu \nu} = \frac{1}{16 \pi \kappa^2_{11}}
  \left( \frac{\kappa_{11}}{4 \pi} \right)^{2 / 3} \left( - \frac{1}{30}
  \mathrm{{tr}} F_{c d} F^{c d} + \frac{1}{2} \bar{R}_{c d e f}
  \bar{R}^{c d e f} \right) \tilde{G}_{\mu \nu}
\end{equation}
\begin{equation}
  \label{T tilde a b} \tilde{T}_{a b} = \frac{1}{16 \pi \kappa^2_{11}} \left(
  \frac{\kappa_{11}}{4 \pi} \right)^{2 / 3} \left( - \frac{1}{90}
  \mathrm{{tr}} F_{c d} F^{c d} + \frac{1}{6} \bar{R}_{c d e f}
  \bar{R}^{c d e f} \right) \tilde{G}_{a b}
\end{equation}

We define $\tilde{t}^{\left( 1 \right)}$ and $\tilde{t}^{\left( 2 \right)}$
by:
\begin{equation}
  \label{definitions of T tildes} \tilde{T}_{\mu \nu} = \tilde{t}^{\left( 1
  \right)} \tilde{G}_{\mu \nu}, \hspace{1.8cm} \tilde{T}_{ab}
  = \tilde{t}^{\left( 2 \right)} \tilde{G}_{ab}
\end{equation}
Then by subsection 2.3.9 of {\cite{CCHT}}, the boundary conditions for
$a\left(y\right)$ and $b\left(y\right)$ are:
\begin{equation}
  \left. \left. \label{bcs in terms of t tildes} \frac{\dot{a}}{a} \right|_{y
  = y_{1 +}} = \frac{\kappa_{11}^2}{18} \left( - 5 \tilde{t}^{\left( 1
  \right)} + 6 \tilde{t}^{\left( 2 \right)} \right), \hspace{6.0ex}
  \frac{\dot{b}}{b} \right|_{y = y_{1 +}} = \frac{\kappa_{11}^2}{18} \left( 4
  \tilde{t}^{\left( 1 \right)} - 3 \tilde{t}^{\left( 2 \right)} \right)
\end{equation}
Thus:
\begin{equation}
  \left. \label{final bc for a} \frac{\dot{a}}{a} \right|_{y = y_{1 +}} = -
  \frac{1}{96 \pi} \left( \frac{\kappa_{11}}{4 \pi} \right)^{2 / 3} \left( -
  \frac{1}{30} \mathrm{{tr}} F_{c d} F^{c d} + \frac{1}{2} \bar{R}_{c d e
  f} \bar{R}^{c d e f} \right)
\end{equation}
\begin{equation}
  \left. \label{final bc for b} \frac{\dot{b}}{b} \right|_{y = y_{1 +}} =
  \frac{1}{96 \pi} \left( \frac{\kappa_{11}}{4 \pi} \right)^{2 / 3} \left( -
  \frac{1}{30} \mathrm{{tr}} F_{c d} F^{c d} + \frac{1}{2} \bar{R}_{c d e
  f} \bar{R}^{c d e f} \right)
\end{equation}

The vacuum Yang-Mills fields on the boundary are quantized, for example by a
Dirac
quantization condition as in subsection 5.3 of {\cite{CCHT}} if they are in
the Cartan subalgebra of $E_8$, so the right-hand sides of (\ref{final bc for
a}) and (\ref{final bc for b}) are proportional to $b^{- 4}_1$, where $b_1
\equiv b \left( y_1 \right)$ is the value of $b$ on the boundary. \ Moss's
derivation of the $R_{U V W X} R^{U V W X}$ term in (\ref{Yang Mills term})
used an expansion scheme in which Ricci tensor and scalar terms, if present,
would only show up at higher orders. \ If the $R_{U V W X} R^{U V W X}$ term
was in fact the first term in a Lovelock-Gauss-Bonnet term of the form $R_{U V
W X} R^{U V W X} - 4 R_{U V} R^{U V} + R^2$, the size of the curvature terms
in (\ref{final bc for a}) and (\ref{final bc for b}) would be increased by a
factor of 6 if the boundary is an $\bar{H}^6$, with the main contribution
coming from the square of the Ricci scalar. \ This would make it easier to
keep the right-hand side of (\ref{final bc for a}) negative and the right-hand
side of (\ref{final bc for b}) positive, while introducing enough vacuum
Yang-Mills fields to break $E_8$ to $\mathrm{{SU}} \left( 3 \right)
\times \mathrm{{SU}} \left( 2 \right) \times \mathrm{U} \left( 1
\right)_Y$ and produce the chiral fermions.

To find the solution of (\ref{p dot eqn}) and (\ref{q double dot eqn}) such
that $p$ and $q$ tend to 0 as $y \rightarrow \infty$, we define
$\mathrm{\tanh} \frac{y}{B} = x$, so that $x \rightarrow 1$ as
$y \rightarrow \infty$. \
The equations then become:
\begin{equation}
  \label{p eqn with x} \frac{dp}{dx} = - \frac{5}{4}  \frac{dq}{dx} + \frac{5
  \lambda q}{4 x}
\end{equation}
\begin{equation}
  \label{q eqn with x} \left( 1 - x^2 \right) \frac{d^2 q}{dx^2} + 5 x
  \frac{dq}{dx} - 5 \lambda q = 0
\end{equation}

For (\ref{q eqn with x}), we try $q \rightarrow \left( 1 - x \right)^{\alpha}$
near $x = 1$, which gives $\alpha = \frac{7}{2}$. \ We then find the
expansion:
\begin{dmath}[compact, spread=3pt]
  \label{expansion of q} q = \left( 1 - x \right)^{\frac{7}{2}} + \frac{\left(
  20 \lambda - 35 \right)  \left( 1 - x \right)^{\frac{9}{2}}}{36} +
  \frac{\left( 400 \lambda^2 - 1240 \lambda + 945 \right)  \left( 1 - x
  \right)^{\frac{11}{2}}}{3168} + \frac{\left( 8000 \lambda^3 - 29200
  \lambda^2 + 32540 \lambda - 10395 \right)  \left( 1 - x
  \right)^{\frac{13}{2}}}{494208} + \cdots
\end{dmath}
For $\lambda = 1$, (\ref{expansion of q}) looks qualitatively like the base of
a parabola centred at $x = 1$, and is $\simeq 0.6184$ for $x = 0$.

The solution of (\ref{p eqn with x}) is $p = - \frac{5}{4} q - \frac{5
\lambda}{4} \int^1_x \frac{q \left( x' \right)}{x'} dx'$. \ Thus $p$ is $\leq
0$ for $0 < x \leq 1$, and has a logarithmic singularity as $x \rightarrow 0$.

We now try to satisfy the boundary conditions (\ref{final bc for a}) and
(\ref{final bc for b}) with $a = \left( 1 + kp \right) A$, $b = \left( 1 + kq
\right) B \mathrm{\cosh} \frac{y}{B}$, where $p$ and $q$ are the solutions
just found, and $k$ is a constant.
We write the boundary conditions (\ref{final bc for a}) and (\ref{final bc for
b}) as:
\begin{equation}
  \label{boundary conditions in terms of rho} \left. \frac{\dot{a}}{a}
  \right|_{y = y_{1 +}} = - \rho \frac{\kappa^{2 / 3}_{11}}{b^4_1},
  \hspace{3.2cm} \left. \frac{\dot{b}}{b} \right|_{y = y_{1 +}} = \rho
  \frac{\kappa^{2 / 3}_{11}}{b^4_1},
\end{equation}
where the number $\rho$ is:
\begin{equation}
  \label{definition of rho} \rho \equiv \frac{1}{\bar{V}_6} \int_{\beta} d^6
  \tilde{x} \sqrt{h} \frac{1}{96 \pi \left( 4 \pi \right)^{2 / 3}} h^{a c}
  h^{b d} \left( - \frac{1}{30} \mathrm{{tr}} F_{a b} F_{c d} +
  \frac{1}{2} \bar{R}_{a b} \, \!^e \, \!_f \bar{R}_{c d e} \, \!^f \right),
\end{equation}
where $\bar{V}_6 \equiv \int_{\beta} d^6 \tilde{x} \sqrt{h}$ is the intrinsic
volume of the boundary $\beta$ in the metric $h_{a b}$, which if the boundary
is an $\bar{H}^6$ is defined to have sectional curvature $- 1$. \ If the
boundary is an $\bar{H}^6$, with $\bar{R}_{a b} \, \!^e \, \!_f = h_{a f}
\delta_b \, \!^e - h_{b f} \delta_a \, \!^e$, then the Riemann$^2$ term in
$\rho$ is $\frac{60}{192 \pi \left( 4 \pi \right)^{2 / 3}} \simeq 0.01840$. \
If the spin connection is naturally embedded in the $E_8$ Yang-Mills gauge
group {\cite{Wilczek embedding spin connection, Charap Duff, Witten Shelter
Island, Candelas Horowitz Strominger Witten}}, which by the Bianchi identity
for the Riemann tensor gives a solution of the classical Yang-Mills equations
if $h_{a b}$ is an Einstein metric, then the Yang-Mills term in $\rho$ is $-
2$ times the Riemann$^2$ term, so $\rho$ is negative.

I shall assume that for moderate values of $\bar{V}_6$ it is possible to
introduce vacuum gauge fields in the $E_8$ Cartan subalgebra, topologically
stabilized by a Dirac quantization condition as in subsection 5.3 of
{\cite{CCHT}}, with field strengths proportional to 4, 5, or 6 of a larger
number of linearly independent harmonic 2-forms on the boundary, such that
$E_8$ is broken to $\mathrm{{SU}} \left( 3 \right) \times
\mathrm{{SU}} \left( 2 \right) \times \mathrm{U} \left( 1 \right)_Y$ and
the Standard Model chiral fermion zero modes are produced, and the Yang-Mills
term in $\rho$ is sufficiently ``diluted'' that $\rho$ is \emph{positive}.
\ If the Riemann$^2$ term in (\ref{Yang Mills term}) is the first term of the
Lovelock-Gauss-Bonnet combination, then the Riemann and Ricci contribution to
$\rho$ for $\bar{H}^6$ would be 6 times larger, so $\rho$ would be positive
even if the spin connection was embedded in the gauge group.

With $\mathrm{\tanh} \frac{y}{B} = x$ as before, and $x_1$ denoting the value
of $x$ at the boundary, the sum of the boundary conditions (\ref{boundary
conditions in terms of rho}) gives:
\begin{equation}
  \label{equation determining k} k \left( 1 - x_1^2 \right) \left( -
  \frac{1}{4}  \left. \frac{dq}{dx} \right|_{x = x_1} + \frac{5 q \left( x_1
  \right)}{4 x_1} \right) + x_1 = 0,
\end{equation}
where (\ref{p eqn with x}) has been used with $\lambda = 1$. \ The logarithmic
singularity of $p$ as $x \rightarrow 0_+$ means that we require $x_1 > 0$ for
the assumption that $\left| kp \right| \ll 1$ to be valid, so since
$\frac{dq}{dx} \leq 0$ and $q \geq 0$ for $0 \leq x \leq 1$, (\ref{equation
determining k}) implies that $k < 0$. \ Using $B \simeq 0.43 \kappa^{2 /
9}_{11}$, from (\ref{best value of B}), on page \pageref{best value of B}, the
second equation of (\ref{boundary conditions in terms of rho}) becomes:
\begin{equation}
  \label{second bc in terms of x} k \left( 1 - x_1^2 \right)  \left.
  \frac{dq}{dx} \right|_{x = x_1} + x_1 \simeq \frac{201 \rho \left( 1 - 4 kq
  \left( x_1 \right) \right)}{\left( \sqrt{\frac{1 + x_1}{1 - x_1}} +
  \sqrt{\frac{1 - x_1}{1 + x_1}} \right)^4}
\end{equation}

Using (\ref{equation determining k}) to express $k$ in terms of $x_1$, we find
that $kq_1 \equiv kq \left( x_1 \right)$, as a function of $x_1$, looks
qualitatively like an upside-down parabola with a peak value of 0 at $x_1 =
0$, and $\simeq - 0.25$ at $x_1 \simeq 0.63$. \ Substituting for $k$ from
(\ref{equation determining k}), we find from (\ref{second bc in terms of x})
that for $\rho = 0.01840$, $x_1 \simeq 0.165$, for which we have $kp_1 \equiv
kp \left( x_1 \right) \simeq 0.043$, hence $A \simeq 0.96$, and $kq_1 \simeq -
0.020$, hence $b_1 \simeq 0.99 B \simeq 0.43 \kappa^{2 / 9}_{11}$. \ Thus
working to first order in $kp$ and $kq$ has been justified. \ For $\rho \simeq
0.11$, corresponding to an $\bar{H}^6$ boundary with no vacuum Yang-Mills
fields, if the Riemann$^2$ term in (\ref{Yang Mills term}) is the start of the
Lovelock-Gauss-Bonnet combination, we find $x_1 \simeq 0.498$, for which $kp_1
\simeq 0.24$, hence $A \simeq 0.81$, and $kq_1 \simeq - 0.16$, hence $b_1
\simeq 0.97 B \simeq 0.42 \kappa^{2 / 9}_{11}$.

From (\ref{equation determining k}) we find that $\frac{b_1}{B}$, as a
function of $x_1$, decreases smoothly from a peak value of 1 at $x_1 = 0$, to
a minimum value $\simeq 0.967$ at $x_1 \simeq 0.56$, which is around the
largest value of $x_1$ for which it is reasonable to work to first order in
$kp$ and $kq$.

The Giudice-Rattazzi-Wells (GRW) estimate of the expansion parameter for
graviton loop corrections in $11$ dimensions, in the sense that perturbation
theory must fail when the expansion parameter is $> 1$ {\cite{Giudice Rattazzi
Wells}}, is:
\begin{equation}
  \label{GRW estimate} \frac{\kappa_{11}^2}{945 \left( 2 \pi \right)^4} \left(
  \frac{E}{2 \pi} \right)^9 \simeq \left( 0.03 \kappa_{11}^{2 / 9} E
  \right)^9,
\end{equation}
where $E$ is the relevant energy of the process. \ For a wave of wavelength $2
\pi b_1$, $E$ is $\frac{1}{b_1}$, so the GRW estimate suggests that $b_1$ must
be larger than around $0.03 \kappa_{11}^{2 / 9}$. \ Thus for $b_1 \simeq 0.42
\kappa^{2 / 9}_{11}$, this requirement is satisfied by a substantial margin.

From just before (\ref{Breit Wigner cross section}), on page
\pageref{Breit Wigner cross section}, the first hypothesis
considered in section \ref{Introduction} leads to the
expectation that $0 < A \leq 0.72$ if $\frac{3 A}{B}=1.8 \textrm{  TeV}$, in
order to satisfy the experimental constraint \cite{Franceschini et al}
that $\kappa_{11}^{-2/9} \geq
0.36\textrm{  TeV}$, with $B \simeq 0.43 \kappa_{11}^{2/9}$ from section
\ref{The bulk}, while from just before the start of section \ref{Horava Witten
theory}, if $\frac{A}{B}=1.8 \textrm{  TeV}$, then the requirement
that $A\leq 1$, which follows from $k < 0$, leads to $\kappa_{11}^{-2/9}\geq
0.78\textrm{  TeV}$, hence
$M_{11}\geq 3.3\textrm{  TeV}$.

To study the Kaluza-Klein modes of the supergravity multiplet we have to
expand $\Gamma^{\left( \mathrm{{bos}} \right)}_{\mathrm{{HW}}}$, in
(\ref{Gamma HW}), including $\Gamma^{\left( 8, \mathrm{{bos}}
\right)}_{\mathrm{{SG}}}$, in (\ref{Gamma 8 SG}), to quadratic order in
small fluctuations about the solution found above. \ For a first estimate I
shall instead consider a massless scalar field $\Phi$ in the bulk, which is
intended to represent a small fluctuation of a component of any of the
supergravity fields, and retain only its classical action. \ Dropping
initially also $R \Phi$ and $H^2 \Phi$ terms, the equation for the small
fluctuation $\Phi$ is then:
\begin{equation}
  \label{small fluctuation equation} - \frac{1}{\sqrt{- G}} \partial_I \left(
  \sqrt{- G} G^{I J} \partial_J \Phi \right) = 0
\end{equation}
Trying an ansatz $\Phi \left( x^{\mu}, x^a, y \right) = \varphi \left( x^{\mu}
\right) \psi \left( x^a, y \right)$, we find from (\ref{small fluctuation
equation}) that:
\begin{dmath}
  \label{separation of variables} - \frac{a^2}{b^2 \psi \left( x^c, y \right)
  \sqrt{h}} \partial_a \left( \sqrt{h} h^{a b} \partial_b \psi \left( x^c, y
  \right) \right) - \frac{a^2}{a^4 b^6 \psi \left( x^c, y \right)} \partial_y
  \left( a^4 b^6 \partial_y \psi \left( x^c, y \right) \right) =
  \frac{1}{\varphi \left( x^{\mu} \right) \sqrt{- g}} \partial_{\mu} \left(
  \sqrt{- g} g^{\mu \nu} \partial_{\nu} \varphi \left( x^{\mu} \right) \right)
\end{dmath}
The left-hand side of (\ref{separation of variables}) is independent of
$x^{\mu}$ and the right-hand side is independent of $x^a$ and $y$, hence each
side must be a constant. \ The left-hand side is a positive operator on a
compact manifold so must be a non-negative constant $\frac{A^2}{B^2}m^2 \geq
0$. \ Away from the neighbourhood of the small boundary, the bulk is a
compact hyperbolic 7-manifold $\bar{H}^7$ with intrinsic volume $\bar{V}_7
\sim 10^{34}$ and curvature radius $B \simeq 0.43 \kappa_{11}^{2/9}$, and
$a = A$, so for most of the low-lying KK modes of the bulk, the KK
masses will to a first approximation not be affected by the presence of the
small boundary. \ So for most of the low-lying KK modes of the $d=11$
supergravity multiplet, $m^2$ will to a first approximation be an intrinsic
eigenvalue of the appropriate quantum-corrected differential operator, for
example the negative of the quantum-corrected Laplace-Beltrami operator, on
the corresponding closed $\bar{H}^7$ with curvature radius 1, with the
leading quantum corrections following from (\ref{Gamma 8 SG}).

For the negative of the Laplace-Beltrami operator $\Delta$, with quantum
corrections neglected, the left-hand side for the lightest massive modes is
estimated from (\ref{lower bound on lambda 1}), on page \pageref{lower bound
on lambda 1}, as $\frac{\left( n - 1 \right)^2}{2^2} \frac{A^2}{B^2} =
9 \frac{A^2}{B^2}$. \ So we find:
\begin{equation}
  \label{mass at boundary of lightest KK modes} - \frac{1}{\sqrt{- g}}
  \partial_{\mu} \left( \sqrt{- g} g^{\mu \nu} \partial_{\nu} \varphi \left(
  x^{\mu} \right) \right) + 9 \frac{A^2}{B^2} \varphi \left( x^{\mu} \right)
  = 0
\end{equation}
Thus for the first hypothesis in section \ref{Introduction}, the mass of the
lightest massive gravitational KK modes, as seen on the boundary, is $3
\frac{A}{B}$.

Most of the modes in the bulk couple only with gravitational strength to the
fields on the boundary, because the values on the boundary of their normalized
wavefunctions are suppressed by the reciprocal of the square root of the large
volume of the bulk. \ However if, as suggested in section \ref{Introduction},
there are around $10^{30}$ approximately degenerate KK modes of the bulk with
masses around 1.8 TeV as seen on the boundary, the modes in the bump could
equivalently be a relatively small number of linear combinations of these
modes, whose normalized wavefunctions are large near the boundary and small
away from the boundary, and these modes would have correspondingly large
couplings to the fields on the boundary.

To keep the number $\rho$ defined in (\ref{definition of rho})
positive, while introducing enough
vacuum Yang-Mills fields to break $E_8$ to $\mathrm{{SU}} \left( 3
\right) \times \mathrm{{SU}} \left( 2 \right) \times \mathrm{U} \left( 1
\right)_Y$ and produce the chiral fermions, it might be helpful if the
boundary had a large enough intrinsic volume for the $E_8$ breaking and chiral
fermions to arise from quantized Yang-Mills fluxes that are ``dilute'', in
the sense that only a small proportion of the possible quantized Yang-Mills
fluxes are present, and these have flux numbers $\pm 1$.  However if the
boundary is an $\bar{H}^6$, its intrinsic volume $\bar{V}_6$ is approximately
fixed by the value $\alpha_U$ of the QCD fine structure constant $\alpha_s =
\frac{g^2_s}{4 \pi}$ at unification, together with the fact that from
the discussion after (\ref{second bc in terms of x}), $b_1 = b\left(y_1
\right) \simeq B \simeq 0.43 \kappa_{11}^{2/9}$.

I shall assume that the Yang-Mills $E_8$ at $y_1$ is
broken to the Standard Model $\mathrm{{SU}} \left( 3 \right) \times
\mathrm{{SU}} \left( 2 \right) \times \mathrm{U} \left( 1 \right)_Y$ by
topologically stabilized vacuum gauge fields in the Cartan subalgebra of the
$E_8$. \ These were shown in subsection 5.3 of {\cite{CCHT}} to be
restricted by Dirac quantization conditions to lie on a lattice of isolated
points in the Cartan subalgebra. \ I shall assume that the unbroken
$\mathrm{{SU}} \left( 3 \right) \times \mathrm{{SU}} \left( 2
\right) \times \mathrm{U} \left( 1 \right)_Y$ is naturally embedded within an
$\mathrm{{SU}} \left( 9 \right)$ subgroup of $E_8$, and from subsection
5.2 of {\cite{CCHT}}, the $E_8$ generators $T^{\mathcal{A}}$ in (\ref{Yang
Mills term}) are normalized so
that the structure constants of the $\mathrm{{SU}} \left( 3 \right)$
subgroup have the same normalization as in the definition of $g_s$ in
subsection 9.1 of {\cite{PDG}}, corresponding to $\mathrm{{SU}} \left( 3
\right)$ generators $t^{\alpha}$ normalized to $\mathrm{{tr}} \left(
t^{\alpha} t^{\beta} \right) = \frac{1}{2} \delta^{\alpha \beta}$, so
performing the trace over the $E_8$ generators using $\mathrm{{tr}}
T^{\mathcal{A}} T^{\mathcal{B}} = 30 \delta^{\mathcal{A}\mathcal{B}}$, we find
that
\begin{equation}
  \label{alpha U} \alpha_U = \frac{\kappa^2_{11}}{b^6_1 \bar{V}_6} \left(
  \frac{4 \pi}{\kappa_{11}} \right)^{2 / 3} = \frac{5.405}{\bar{V}_6} \left(
  \frac{\kappa^{2 / 9}_{11}}{b_1} \right)^6 \simeq \frac{860}{\bar{V}_6},
\end{equation}
where $\bar{V}_6$ is the volume of the boundary in the metric $h_{a b}$,
which is the intrinsic volume of the boundary, if the boundary is an
$\bar{H}^6$.

In the Dienes-Dudas-Gherghetta version of accelerated unification $\alpha_U
\simeq \frac{1}{52}$ {\cite{DDG1, DDG2}}, and in the Arkani-Hamed-Cohen-Georgi
version, $\alpha_U \simeq \frac{1}{24}$ {\cite{ACG}}. \
If the boundary is an $\bar{H}^6$, then by (\ref{alpha U}) these imply
respectively $\bar{V}_6 \simeq 45000$ and $\bar{V}_6 \simeq 21000$, which by
the generalized Gauss-Bonnet theorem (\ref{generalized Gauss Bonnet theorem}),
on page \pageref{generalized Gauss Bonnet theorem}, correspond respectively to
Euler numbers $\chi \left( \bar{H}^6 \right) \simeq - 2700$ and $\chi \left(
\bar{H}^6 \right) \simeq - 1300$.

\begin{figure}[t]
\setlength{\unitlength}{1cm}
\begin{picture}(12.0,4.05)
\thicklines
\put(2.39,3.67){$u_1$}
\put(2.51,3.28){\circle*{0.265}}
\put(2.51,3.4){\line(1,0){1.40}}
\put(2.51,3.29){\line(1,0){1.40}}
\put(2.51,3.18){\line(1,0){1.40}}
\put(3.82,3.67){$u_2$}
\put(3.94,3.28){\circle*{0.265}}
\put(4.05,3.29){\line(1,0){1.40}}
\put(5.39,3.67){$u_3$}
\put(5.51,3.28){\circle*{0.265}}
\put(5.56,3.29){\line(1,0){1.48}}
\put(7.05,3.67){$u_4$}
\put(7.17,3.28){\circle*{0.265}}
\put(7.25,3.29){\line(1,0){1.43}}
\put(8.72,3.67){$u_8$}
\put(8.84,3.28){\circle*{0.265}}
\put(8.84,3.29){\line(1,0){1.48}}
\put(8.84,3.23){\line(0,-1){1.56}}
\put(10.28,3.67){$u_7$}
\put(10.4,3.28){\circle*{0.265}}
\put(10.25,3.33){\line(-1,-1){1.47}}
\put(10.35,3.23){\line(-1,-1){1.50}}
\put(10.45,3.16){\line(-1,-1){1.50}}
\put(8.20,1.65){$u_6$}
\put(8.86,1.75){\circle*{0.265}}
\put(10.28,0.55){$u_9$}
\put(10.4,1.05){\circle*{0.265}}
\put(7.08,0.05){$u_5$}
\put(7.2,0.54){\circle*{0.265}}
\put(8.98,1.64){\line(5,-2){1.56}}
\put(10.41,3.25){\line(0,-1){2.28}}
\put(7.15,3.17){\line(0,-1){2.50}}
\multiput(7.33,0.54)(0.18,0.03){17}{\circle*{0.06}}
\put(8.45,0.18){$\frac{1+\sqrt{5}}{2}$}
\end{picture}
\caption{Coxeter diagram for the $d = 6$ hyperbolic Coxeter polytope $T_3$}
\label{Coxeter diagram for T3}
\end{figure}

In 4 dimensions the Davis manifold {\cite{Davis manifold}} with $\chi = 26$,
which is an orientable spin manifold {\cite{Ratcliffe Tschantz Davis
manifold}}, can be constructed from a hyperbolic Coxeter simplex by a variant
of Selberg's Lemma {\cite{Selberg, Everitt Maclachlan}}. \ There are no
hyperbolic Coxeter simplexes in 5 or more dimensions {\cite{Lanner}}, and no
hyperbolic Coxeter polytopes with $n + 2$ $\left( n - 1 \right)$-dimensional
faces in 6 or more dimensions {\cite{Kaplinskaya, Esselmann}}. \ There are
exactly 3 hyperbolic Coxeter polytopes with 9 5-dimensional faces in 6
dimensions {\cite{Tumarkin}}, and the Coxeter diagram for one of these, which
I will call $T_3$, is shown in Figure \ref{Coxeter diagram for T3}. \ Twice
the Gram matrix for $T_3$ is:
\begin{equation}
  \label{twice T3 Gram matrix} \mathcal{G} \equiv \left(
  \begin{array}{ccccccccc}
    2 & - k & 0 & 0 & 0 & 0 & 0 & 0 & 0\\
    - k & 2 & - 1 & 0 & 0 & 0 & 0 & 0 & 0\\
    0 & - 1 & 2 & - 1 & 0 & 0 & 0 & 0 & 0\\
    0 & 0 & - 1 & 2 & - 1 & 0 & 0 & - 1 & 0\\
    0 & 0 & 0 & - 1 & 2 & 0 & 0 & 0 & - 2 k\\
    0 & 0 & 0 & 0 & 0 & 2 & - k & - 1 & - 1\\
    0 & 0 & 0 & 0 & 0 & - k & 2 & - 1 & - 1\\
    0 & 0 & 0 & - 1 & 0 & - 1 & - 1 & 2 & 0\\
    0 & 0 & 0 & 0 & - 2 k & - 1 & - 1 & 0 & 2
  \end{array} \right),
\end{equation}
where $k \equiv \frac{1 + \sqrt{5}}{2} = 2 \mathrm{\cos} \frac{\pi}{5} \simeq
1.6180$ is an algebraic integer that satisfies the equation:
\begin{equation}
\label{equation for k}
 k^2 - k - 1 = 0.
\end{equation}
$\mathcal{G}$ has two linearly independent eigenvectors with
eigenvalue 0, which may be chosen as:
\[ n_{\left( 1 \right)} \equiv \left( \begin{array}{ccccccccc}
     - 3 k - 2, & - 4 k - 2, & - 3 k - 1, & - 2 k, & - k, & k + 1, & k + 1, &
     1, & 0
   \end{array} \right) \]
\begin{equation}
  \label{eigvecs with eigval 0} n_{\left( 2 \right)} \equiv \left(
  \begin{array}{ccccccccc}
    - 5 k - 3, & - 6 k - 4, & - 4 k - 3, & - 2 k - 2, & - 1, & k + 1, & k + 1,
    & 0, & 1
  \end{array} \right)
\end{equation}

If we regard $\mathcal{G}_{i j}$ as a ``metric'' in 9 dimensions, then for any
contravariant 9-vector $v^i$ such that $v^i v_i \neq 0$, where $v_i \equiv v^j
\mathcal{G}_{j i}$, the matrix $r \left( v \right)^i \, \!_j \equiv \delta^i
\, \!_j - 2 \frac{v^i v_j}{v^k v_k}$ satisfies $r \left( v \right)^i \, \!_k r
\left( v \right)^k \, \!_j = \delta^i \, \!_j$ and $\mathcal{G}_{k l} r \left(
v \right)^k \, \!_i r \left( v \right)^l \, \!_j =\mathcal{G}_{i j}$, and is
thus a reflection matrix with respect to the ``metric'' $\mathcal{G}_{i j}$. \
In particular, if we define $u_{\left( i \right)}^j$, $1 \leq i \leq 9$, to be
the vector whose $i \,$th component is 1 and whose other components are 0,
then since $u_{\left( i \right)}^j u_{\left( i \right) j} = 2$ for all $1 \leq
i \leq 9$, and all the matrix elements of $\mathcal{G}$ are in the ring of
algebraic integers $\mathbf{Z} \left[ k \right]$, the matrices $r \left(
u_{\left( i \right)} \right)^j \, \!_k$, $1 \leq i \leq 9$, are 9 reflection
matrices whose matrix elements are in $\mathbf{Z} \left[ k \right]$. \ We
now define $\tilde{u}^j_{\left( i \right)} \equiv u^j_{\left( i \right)}$, $1
\leq i \leq 7$, $\tilde{u}^j_{\left( 8 \right)} \equiv u^j_{\left( 8 \right)}
- n^j_{\left( 1 \right)}$, and $\tilde{u}^j_{\left( 9 \right)} \equiv
u^j_{\left( 9 \right)} - n^j_{\left( 2 \right)}$. \ Then $\tilde{u}_{\left( i
\right)}^j \tilde{u}_{\left( i \right) j} = 2$ for all $1 \leq i \leq 9$, and
the last two components of $\tilde{u}_{\left( i \right)}^j$ are 0 for all $1
\leq i \leq 9$. \ Let indices $a, b, c, \ldots$ run from 1 to 7,
$\mathcal{H}_{a b}$ be the $7 \times 7$ symmetric matrix $\mathcal{G}_{a b}$,
and $\bar{u}^a_{\left( i \right)}$, $1 \leq i \leq 9$, be the 7-vectors
$\tilde{u}_{\left( i \right)}^a$. \ Then for all $1 \leq i \leq 9$, the
matrices $r \left( \bar{u}_{\left( i \right)} \right)^a \, \!_b$ are $7 \times
7$ reflection matrices with respect to the ``metric'' $\mathcal{H}_{a b}$, and
their matrix elements are in $\mathbf{Z} \left[ k \right]$. \ Let:
\begin{equation}
  \label{transformation matrix T} S \equiv \left( \begin{array}{ccccccc}
    1 & \frac{k}{2} & \frac{3 k + 1}{5} & \frac{2 k + 1}{2} & 5 k + 3 & 0 &
    0\\
    0 & 1 & \frac{2 k + 4}{5} & k + 1 & 6 k + 4 & 0 & 0\\
    0 & 0 & 1 & \frac{k + 2}{2} & 4 k + 3 & 0 & 0\\
    0 & 0 & 0 & 1 & 2 k + 2 & 0 & 0\\
    0 & 0 & 0 & 0 & 1 & 0 & 0\\
    0 & 0 & 0 & 0 & 0 & 1 & \frac{k}{2}\\
    0 & 0 & 0 & 0 & 0 & 0 & 1
  \end{array} \right)
\end{equation}
Then $S^T \mathcal{H}S = \mathrm{{diag}} \left( \begin{array}{ccccccc}
  2, & \frac{3 - k}{2}, & \frac{6 - 2 k}{5}, & \frac{2 - k}{2}, & - 2 k, & 2,
  & \frac{3 - k}{2}
\end{array} \right)$, so $\mathcal{H}$ has signature $\left( 6, 1 \right)$,
and $\mathcal{H}_{a b}$ is the metric on 7-dimensional Minkowski space in a
non-standard coordinate system. \ The dot products $\bar{u}^a_{\left( i
\right)} \bar{u}_{\left( j \right) a} =\mathcal{H}_{a b} \bar{u}^a_{\left( i
\right)} \bar{u}^b_{\left( j \right)}$ are equal to the corresponding elements
of $\mathcal{G}_{i j}$ for all $1 \leq i, j \leq 9$, so the matrices $r \left(
\bar{u}_{\left( i \right)} \right)^a \, \!_b$ provide a representation of the
Coxeter group $\Gamma \left( T_3 \right)$ corresponding to the Coxeter diagram
$T_3$. \ Let superscript $^g$ denote the replacement of $k$ by its Galois
conjugate $k^g = \frac{1 - \sqrt{5}}{2} \simeq - 0.6180$, which is the other
solution of (\ref{equation for k}). \ Then the diagonal matrix elements of
$S^{g T} \mathcal{H}^g S^g$ are $> 0$, so $\mathcal{H}^g$ has signature
$\left( 7, 0 \right)$, so $\Gamma \left( T_3 \right)$ is of the arithmetic
type {\cite{Borel Harish Chandra}} reviewed in {\cite{Witte Morris}} and
subsection 3.1 of {\cite{CCHT}}.

A finite index torsionless normal subgroup $\Lambda$ of $\Gamma \left( T_3
\right)$ can be obtained by the same variant of Selberg's lemma
{\cite{Selberg}} as used in {\cite{Everitt Maclachlan}} to obtain the
4-dimensional Davis manifold from a compact hyperbolic Coxeter
4-simplex. \ $\Lambda$
consists of the elements of $\Gamma \left( T_3 \right)$ whose matrices in the
representation just constructed are equal $\textrm{{mod}}\, \sqrt{5}$ to
the unit matrix. \ To calculate the index of $\Lambda$ in $\Gamma$, which
gives the ratio of the intrinsic volume of the corresponding compact
hyperbolic 6-manifold to the intrinsic volume of the compact hyperbolic
Coxeter polytope $\xi \left( T_3 \right)$, we note that $\frac{1 +
\sqrt{5}}{2} + \sqrt{5} \frac{\sqrt{5} - 1}{2} = 3$, hence $k = 3\,
\textrm{{mod}}\, \sqrt{5}$. \ Thus the factor group $\Gamma / \Lambda$,
which is the group of all the distinct elements of $\Gamma\, \textrm{{mod}}\,
\sqrt{5}$, is obtained by replacing $k$ by 3 in the $7 \times 7$ matrix
representations of the 9 generators of $\Gamma$ constructed above, then
reducing the resulting integer matrix elements $\textrm{mod 5}$ and
doing the matrix multiplications with $\textrm{mod 5}$ arithmetic.

Using GAP {\cite{GAP}} we find $\mathrm{{Size}} \left( \Gamma / \Lambda
\right) = \text{460687500000000} = 2^8 3^4 5^{12} 7 \, 13 \simeq 4.6 \times
10^{14}$. \ Using the same procedure for the 6-dimensional compact hyperbolic
Coxeter polytope with 9 5-dimensional faces whose Coxeter diagram is shown in
Figure 4.3 of {\cite{Tumarkin}}, whose Gram matrix times 2 has matrix elements
in $\mathbf{Z} \left[ k \right]$ after dividing 2 rows and 2 columns by
$\sqrt{2}$, and which is also of arithmetic type, we find that in that case,
$\mathrm{{Size}} \left( \Gamma / \Lambda \right) = \text{914004000000000}
= 2^{11} 3^4 5^9 7 \, 13 \, \, 31 \simeq 9.1 \times 10^{14}$. \ The third
6-dimensional compact hyperbolic Coxeter polytope with 9 5-dimensional faces
is obtained from the Coxeter polytope $\xi \left( T_3 \right)$ by cutting it
in half along its hyperplane of reflection symmetry. \ By {\cite{Belolipetsky
2}} and Table 2 of {\cite{Belolipetsky 1}} the magnitudes of the fractional
Euler numbers of these three 6-dimensional compact hyperbolic Coxeter
polytopes are not less than $\frac{67}{1152000} \simeq 5.8 \times 10^{- 5}$,
so the intrinsic volumes of the compact hyperbolic 6-manifolds constructed in
this way are too large by a factor of not less than around $10^7$. \ However
there are other methods of constructing torsionless subgroups of hyperbolic
Coxeter groups {\cite{Everitt Maclachlan, Everitt, Conder Maclachlan, Long}}
that might provide examples of smaller intrinsic volume.

While version 3 of this article was being prepared, the XENON100
collaboration published the results of the first 100.9 days of data taking
in the XENON100 direct search for dark matter, finding no evidence for dark
matter interacting with the ultra-low background innermost 48 kg of their
62 kg liquid xenon target, and excluding spin-independent elastic
WIMP-nucleon scattering cross-sections above $7.0\times 10^{-45}$ cm$^2$ for
a WIMP mass of 50 GeV at 90\% confidence level \cite{XENON100}.  The model
considered in this article could provide a variety of dark matter candidates
with masses up to around a TeV that interact only gravitationally with
ordinary matter, and are thus consistent with these limits, if there were
from 1 to around 10 other HW boundaries distributed around the compact
hyperbolic 7-manifold $\bar{H}^7$, possibly with different $E_8$ breakings
on each of them.

\begin{center}
{\bf Acknowledgements}
\end{center}

\noindent I would like to thank the organizers of the 2007 CERN BSM Institute,
in particular Nima Arkani-Hamed, Savas Dimopolous, and Christophe Grojean, for
arranging for me to give a talk and spend a very enjoyable and useful week at
CERN with financial support, Asimina Arvanitaki, Savas Dimopoulos, Philip
Schuster, Jesse Thaler, Natalia Toro, and Jay Wacker for very interesting
discussions, Greg Moore for a helpful email about flux quantization, Kasper
Peeters for correspondence about Cadabra both on and off the mailing list, and
Peter Woit \cite{Peter Woits blog} and ``Jester'' \cite{Jesters blog} for
providing efficient digests of what's happening
in high energy physics. \ The calculations made heavy use of Maxima
{\cite{Maxima}} and Cadabra \cite{Cadabra 1} and also used GAP \cite{GAP},
the diagram was drawn with help from TexPict \cite{TexPict}, the bibliography
was sequenced with help from Ordercite \cite{Ordercite},
and the article and the notebooks the work was done in were written with
GNU TeXmacs {\cite{TeXmacs}}
running in KDE 3.5 {\cite{KDE}} in Debian GNU/Linux {\cite{Debian}}.


\vspace{0.5cm}

\begin{thebibliography}{99}

\bibitem{AC1088}
The ATLAS Collaboration, ``Search for new physics in multi-body
final states at high invariant masses with ATLAS,''
ATLAS-CONF-2010-088 August 21, 2010.\\
http://cdsweb.cern.ch/record/1299103/files/ATLAS-CONF-2010-088.pdf

\bibitem{DDG1}
K.R. Dienes, E. Dudas, and T. Gherghetta, ``Extra Spacetime
Dimensions and Unification,'' Phys. Lett. \textbf{B436} (1998)
55-65, arXiv:hep-ph/9803466.

\bibitem{DDG2}
K.R. Dienes, E. Dudas, and T. Gherghetta, ``Grand Unification at
Intermediate Mass Scales through Extra Dimensions,'' Nucl. Phys.
\textbf{B537} (1999) 47-108, arXiv:hep-ph/9806292.

\bibitem{ACG}
N.~Arkani-Hamed, A.~G.~Cohen and H.~Georgi, ``Accelerated unification,''
arXiv:hep-th/0108089.
        
\bibitem{KMST}
N.~Kaloper, J.~March-Russell, G.~D.~Starkman, M.~Trodden,
``Compact hyperbolic extra dimensions: Branes, Kaluza-Klein modes and
cosmology,'' Phys.\ Rev.\ Lett.\ {\bf 85} (2000) 928-931,
arXiv:hep-ph/0002001.

\bibitem{CJS}
E.~Cremmer, B.~Julia and J.~Scherk,
``Supergravity theory in 11 dimensions,''
  Phys.\ Lett.\  {\bf B76} (1978) 409-412.  Scanned version from KEK: \\
  \verb#http://ccdb4fs.kek.jp/cgi-bin/img_index?7805106#

\bibitem{KP}
A.~Kehagias and H.~Partouche,
``On the exact quartic effective action for the type iib superstring,''
Phys.\ Lett.\  B {\bf 422} (1998) 109,
arXiv:hep-th/9710023.

\bibitem{Richards 2}
D.~M.~Richards,
``The One-Loop $H^2R^3$ and $H^2(DH)^2R$ Terms in the Effective Action,''
JHEP {\bf 0810} (2008) 043, arXiv:0807.3453 [hep-th].

\bibitem{Mostow 1}
G. D. Mostow, ``Quasi-conformal mappings in $n$-space and the rigidity of the
hyperbolic space forms,'' Publ. Math. IHES \textbf{34} (1968) 53-104.

\bibitem{Mostow 2}
G. D. Mostow, ``Strong rigidity of locally symmetric spaces,'' Ann. of Math.
Studies, {\bf{78}} (1973) 1-195.

\bibitem{Prasad}
G. Prasad, ``Strong rigidity of rank 1 lattices,'' Invent. Math. \textbf{21}
(1973) 255 - 286.

\bibitem{Thurston}
William Thurston, The geometry and topology of 3-manifolds, Princeton
University lecture notes (1978-1981). \\
http://www.msri.org/publications/books/gt3m/

\bibitem{Gromov Hyperbolic}
M. Gromov, ``Hyperbolic manifolds according to Thurston and J\o rgensen,''
S\'eminaire Bourbaki, {\bf{32e ann\'ee, 546}} (1979/80) 40-53.\\
\verb#http://www.ihes.fr/~gromov/PDF/1[29].pdf#

\bibitem{Nicolai Townsend van Nieuwenhuizen}
H.~Nicolai, P.~K.~Townsend and P.~van Nieuwenhuizen,
``Comments On Eleven-Dimensional Supergravity,''
Lett.\ Nuovo Cim.\  {\bf 30} (1981) 315.  Scanned version from KEK: \\
\verb#http://ccdb4fs.kek.jp/cgi-bin/img_index?8006023#

\bibitem{Sagnotti Tomaras}
A.~Sagnotti and T.~N.~Tomaras,
``Properties Of Eleven-Dimensional Supergravity,''
CALT-68-885.  Scanned version from KEK: \\
\verb#http://ccdb4fs.kek.jp/cgi-bin/img_index?8203146#

\bibitem{Bautier Deser Henneaux Seminara}
K.~Bautier, S.~Deser, M.~Henneaux and D.~Seminara,
``No cosmological D = 11 supergravity,''
Phys.\ Lett.\ B {\bf 406} (1997) 49-53, arXiv:hep-th/9704131.

\bibitem{Nahm}
W.~Nahm,
``Supersymmetries and their representations,''
Nucl.\ Phys.\  {\bf B135} (1978) 149.  Scanned version from KEK: \\
\verb#http://ccdb4fs.kek.jp/cgi-bin/img_index?197709213#

\bibitem{Michelson Lawson}
H.B. Lawson and M.L. Michelson, \emph{Spin Geometry}, Princeton Univ. Press,
Princeton, 1989.

\bibitem{Hirzebruch et al}
A. Borel and F. Hirzebruch, ``Characteristic classes and homogeneous
spaces I,'' American Journal of Mathematics {\bf 80} (1958) 97-136.

\bibitem{Milnor Stasheff}
J.W. Milnor and J.D. Stasheff, \emph{Characteristic Classes}, Princeton
Univ. Press and Univ. of Tokyo Press, Princeton, 1974.

\bibitem{Gromov Volume Bounded Cohomology}
M. Gromov, ``Volume and bounded cohomology,'' Publ. Math. IH\'ES {\bf 56}
(1982), 5-100,\\
\verb#http://www.ihes.fr/~gromov/PDF/4[35].pdf#

\bibitem{Gromov Volume and Bounded Cohomology}
M. Gromov, theorem 2 in W. Ballmann, M. Gromov, and V. Schroeder,
\emph{Manifolds of Nonpositive Curvature}, Birkhauser, 1985.

\bibitem{Wang}
H. C. Wang, ``Topics on totally discontinuous groups,'' in {\emph{Symmetric
Spaces}}, edited by W. Boothby and G. Weiss, M. Dekker (1972) 460-487.

\bibitem{Burger Gelander Lubotzky Mozes}
M. Burger, T. Gelander, A. Lubotzky, and S. Mozes, ``Counting hyperbolic
manifolds,'' Geometric and Functional Analysis {\bf 12} (2002) 1161-1173.\\
\verb#http://www.ma.huji.ac.il/~alexlub/PAPERS/Counting hyperbolic#\\
\verb#manifolds/Counting hyperbolic manifolds.pdf#

\bibitem{Davis manifold}
M. W. Davis, ``A hyperbolic 4-manifold,'' Proc. Amer. Math. Soc. \textbf{93}
(1985) 325-328.

\bibitem{Ratcliffe Tschantz Davis manifold}
J. G. Ratcliffe and S. T. Tschantz, ``On the Davis hyperbolic 4-manifold,''
Topology Appl. {\bf 111} (2001), 327-342.

\bibitem{Cao Meyerhoff}
C. Cao and R. Meyerhoff, ``The Orientable Cusped Hyperbolic 3-Manifolds of
Minimum Volume,'' Invent. Math. {\bf 146} (2001), 451-478.

\bibitem{Wess Zumino}
J.~Wess and B.~Zumino,
``Supergauge Invariant Extension of Quantum Electrodynamics,''
Nucl.\ Phys.\ {\bf B78} (1974) 1-13.
     
\bibitem{Gliozzi Scherk Olive}
F.~Gliozzi, J.~Scherk and D.~I.~Olive,
``Supersymmetry, Supergravity Theories And The Dual Spinor Model,''
  Nucl.\ Phys.\ B {\bf 122} (1977) 253 - 290.  Scanned version from KEK: \\
  \verb#http://ccdb4fs.kek.jp/cgi-bin/img_index?197701150#

\bibitem{Lu}
J.~X.~Lu,
``Remarks on M theory coupling constants and M-brane tension quantizations,''
arXiv:hep-th/9711014.

\bibitem{Horava Witten 1}
P. Ho\v{r}ava and E. Witten, ``Heterotic And Type I String Dynamics
From Eleven Dimensions,'' Nucl. Phys. \textbf{B460} (1996) 506-524,
arXiv:hep-th/9510209.

\bibitem{Horava Witten 2}
P. Ho\v{r}ava and E. Witten, ``Eleven-Dimensional Supergravity
on a Manifold with Boundary,'' Nucl. Phys. \textbf{B475} (1996)
94-114, arXiv:hep-th/9603142.

\bibitem{Moss 1}
I.~G.~Moss,
``Boundary terms for eleven-dimensional supergravity and M-theory,''
Phys.\ Lett.\ B {\bf 577} (2003) 71-75, arXiv:hep-th/0308159.

\bibitem{Moss 2}
I.~G.~Moss,
``Boundary terms for supergravity and heterotic M-theory,''  Nucl.\ Phys.\ B
{\bf 729} (2005) 179-202, arXiv:hep-th/0403106.

\bibitem{Moss 3}
I.~G.~Moss,
``A new look at anomaly cancellation in heterotic M-theory,''
Phys.\ Lett.\ B {\bf 637} (2006) 93-96, arXiv:hep-th/0508227.

\bibitem{Moss 4}
I.~G.~Moss,
``Higher order terms in an improved heterotic M theory,''
  JHEP {\bf 0811} (2008) 067, arXiv:0810.1662 [hep-th].

\bibitem{Borel Harish Chandra}
A. Borel and Harish-Chandra, ``Arithmetic Subgroups of Algebraic Groups,''
Annals of Mathematics {\bf{75}} (1962) 485-535.

\bibitem{Witte Morris}
D. Witte Morris: Introduction to Arithmetic Groups, 2003 draft version: \\
\verb#http://people.uleth.ca/~dave.morris/lectures/ArithGrps/# \\
\verb#Morris-ArithGrps-Feb03.pdf#

\bibitem{CCHT}
C. Austin, ``TeV-scale gravity in Ho\v{r}ava-Witten theory on a compact complex
hyperbolic threefold,'' arXiv:0704.1476 [hep-th].

\bibitem{Gromov Piatetski Shapiro}
M. Gromov and I. Piatetski-Shapiro, ``Non-arithmetic groups in Lobachevsky
spaces,'' Inst. Hautes \'Etudes Sci. Publ. {\bf{66}} (1988) 93-103.\\
\verb#http://www.ihes.fr/~gromov/PDF/7[59].pdf#

\bibitem{SnapPea}
http://www.geometrygames.org/SnapPea/

\bibitem{SnapPy}
M. Culler, N. M. Dunfield, and J. R. Weeks. SnapPy, a computer program for
studying the geometry and topology of 3-manifolds,
http://snappy.computop.org

\bibitem{Snap}
\verb#http://www.ms.unimelb.edu.au/~snap/#

\bibitem{Costa}
S.S. e Costa, ``A description of several coordinate systems for hyperbolic
spaces,'' arXiv:math-ph/0112039.

\bibitem{Hyakutake Ogushi 2}
Y.~Hyakutake and S.~Ogushi,
``Higher derivative corrections to eleven dimensional supergravity via local
supersymmetry,''
JHEP {\bf 0602} (2006) 068, arXiv:hep-th/0601092.

\bibitem{PDG}
K. Nakamura et al., (the Particle Data Group), ``The Review of Particle
Physics,'' J. Phys. G {\bf 37} (2010) 075021.  http://pdg.lbl.gov/

\bibitem{ADD1}
N. Arkani--Hamed, S. Dimopoulos and G. Dvali, ``The Hierarchy
Problem and New Dimensions at a Millimeter,'' Phys. Lett.
\textbf{B429} (1998) 263-272, arXiv:hep-ph/9803315.

\bibitem{AADD}
I.~Antoniadis, N.~Arkani-Hamed, S.~Dimopoulos and G.~R.~Dvali,
``New dimensions at a millimeter to a Fermi and superstrings at a TeV,''
  Phys.\ Lett.\ B {\bf 436} (1998) 257 - 263, arXiv:hep-ph/9804398.

\bibitem{ADD2}
N. Arkani--Hamed, S. Dimopoulos and G. Dvali, ``Phenomenology,
Astrophysics and Cosmology of Theories with Sub-Millimeter
Dimensions and TeV Scale Quantum Gravity,'' Phys. Rev. \textbf{D59}
(1999) 086004, arXiv:hep-ph/9807344.

\bibitem{RS1}
L.~Randall and R.~Sundrum,
``A large mass hierarchy from a small extra dimension,''
Phys.\ Rev.\ Lett.\  {\bf 83} (1999) 3370-3373, arXiv:hep-ph/9905221.

\bibitem{Franceschini et al}
R. Franceschini, G.F. Giudice, P.P. Giardino, P. Lodone, and A. Strumia,
``LHC bounds on large extra dimensions,'' arXiv:1101.4919v2 [hep-ph].

\bibitem{Diagrammar}
G. 't Hooft and M, Veltman, ``DIAGRAMMAR,'' CERN report 73-9 (1973),
reprinted in G. ’t Hooft, \emph{Under the Spell of Gauge Principle}, World
Scientific, Singapore (1994).
http://cdsweb.cern.ch/record/186259/files/p1.pdf

\bibitem{Lee Les Houches}
B.W. Lee, ``Gauge theories,'' in \emph{Les Houches 1975: Methods in Field
Theory}, R.~Balian and
J. Zinn-Justin (Eds.), Elsevier, Amsterdam, 1976, p. 79.

\bibitem{Tseytlin Field Redefinitions}
A.~A.~Tseytlin,
``Ambiguity in the Effective Action in String Theories,''
Phys.\ Lett.\ B {\bf 176} (1986) 92-98.  Scanned version from KEK:\\
\verb#http://ccdb4fs.kek.jp/cgi-bin/img_index?8702274#

\bibitem{Duff Liu Minasian}
M.~J.~Duff, J.~T.~Liu and R.~Minasian,
``Eleven-dimensional origin of string / string duality: A one-loop test,''
Nucl.\ Phys.\ B {\bf 452} (1995) 261-282, arXiv:hep-th/9506126.

\bibitem{Witten 5 branes}
E.~Witten, ``Five-brane effective action in M-theory,''
J.\ Geom.\ Phys.\  {\bf 22} (1997) 103-133, arXiv:hep-th/9610234.

\bibitem{Freed Harvey Minasian Moore}
D.~Freed, J.~A.~Harvey, R.~Minasian and G.~W.~Moore, ``Gravitational anomaly
cancellation for M-theory fivebranes,'' Adv. Theor. Math. Phys.  {\bf 2}
(1998) 601-618, arXiv:hep-th/9803205.

\bibitem{Bilal Metzger}
A.~Bilal and S.~Metzger,
``Anomaly cancellation in M-theory: A critical review,''
Nucl.\ Phys.\ {\bf B675} (2003) 416-446, arXiv:hep-th/0307152.

\bibitem{Harvey}
J.~A.~Harvey, ``TASI 2003 lectures on anomalies,'' arXiv:hep-th/0509097.

 \bibitem{de Alwis}
S.~P.~de Alwis,
``Anomaly cancellation in M-theory,''
Phys.\ Lett.\ B {\bf 392} (1997) 332-334, arXiv:hep-th/9609211.

\bibitem{Conrad}
J.~O.~Conrad,
``Brane tensions and coupling constants from within M-theory,''
Phys.\ Lett.\ B {\bf 421} (1998) 119-124, arXiv:hep-th/9708031.

\bibitem{Faux Lust Ovrut}
M.~Faux, D.~L\"ust, B.~A.~Ovrut,
``Intersecting orbifold planes and local anomaly cancellation in M theory,''
  Nucl.\ Phys.\  {\bf B554} (1999)  437-483,\\
arXiv:hep-th/9903028.

\bibitem{Bilal Derendinger Sauser}
A. Bilal, J.-P. Derendinger, and R. Sauser, ``M-Theory on
$S^1/\mathbf{Z}_2$ : new facts from a careful analysis,'' Nucl. Phys. B
{\bf 576} (2000) 347-374, arXiv:hep-th/9912150.

\bibitem{Harmark}
T.~Harmark,
``Coupling constants and brane tensions from anomaly cancellation in M
theory,'' Phys.\ Lett.\  {\bf B431} (1998)  295-302, arXiv:hep-th/9802190.

\bibitem{Meissner Olechowski}
K.~A.~Meissner, M.~Olechowski,
``Anomaly cancellation in M theory on orbifolds,''
Nucl.\ Phys.\  {\bf B590} (2000)  161-172, arXiv:hep-th/0003233.

\bibitem{Hyakutake Ogushi 1}
Y.~Hyakutake and S.~Ogushi,
``$ R^4 $ corrections to eleven dimensional supergravity via supersymmetry,''
Phys.\ Rev.\  {\bf D74} (2006) 025022, arXiv:hep-th/0508204.

\bibitem{Hyakutake}
Y.~Hyakutake,``Toward the determination of $R^3 F^2$ terms in $M$-theory,''
Prog. Theor. Phys.  {\bf 118} (2007) 109, arXiv:hep-th/0703154.

\bibitem{Hyakutake 2}
Y.~Hyakutake,
``Higher derivative corrections in M-theory via local supersymmetry,''
arXiv:0710.2673 [hep-th].

\bibitem{Metsaev}
R.~R.~Metsaev,
``Eleven dimensional supergravity in light cone gauge,''
Phys.\ Rev.\  {\bf D71} (2005) 085017, arXiv:hep-th/0410239.

\bibitem{Stephenson}
P.~M.~Stevenson,
``Optimized Perturbation Theory,''
  Phys.\ Rev.\  {\bf D23} (1981)  2916-2944.

\bibitem{Rohm Witten}
R.~Rohm, E.~Witten,
``The Antisymmetric Tensor Field in Superstring Theory,''
  Annals Phys.\  {\bf 170} (1986)  454-489.

\bibitem{Witten flux quantization}
E.~Witten,
``On flux quantization in M-theory and the effective action,''
J.\ Geom.\ Phys.\  {\bf 22} (1997) 1-13, arXiv:hep-th/9609122.

\bibitem{HST}
W.~L.~Freedman {\it et al.},
``Final Results from the Hubble Space Telescope Key Project to Measure the
Hubble Constant,'' Astrophys.\ J.\  {\bf 553} (2001) 47-72,
arXiv:astro-ph/0012376.

\bibitem{0302209 Spergel et al}
D. N. Spergel et al., ``First Year Wilkinson Microwave Anisotropy Probe (WMAP)
Observations: Determination of Cosmological Parameters'' Astrophys.
J. Suppl. {\bf{148}} (2003) 175, arXiv:astro-ph/0302209.

\bibitem{de Sitter radius}
J.~L.~Tonry {\it et al.}  [Supernova Search Team Collaboration],
``Cosmological Results from High-z Supernovae,''
  Astrophys.\ J.\  {\bf 594} (2003) 1-24, arXiv:astro-ph/0305008.

\bibitem{Yau Isoperimetric}
S.-T. Yau, ``Isoperimetric constants and the first eigenvalue of a compact
Riemannian manifold,'' Ann. Scient. \'Ecole Norm. Sup. {\bf 8} (1975)
487-507.

\bibitem{Agmon}
S. Agmon, ``On the spectral theory of the Laplacian on noncompact hyperbolic
manifolds,'' Journ\'ees \'Equations aux d\'eriv\'ees partielles (1987)
1-16.\\
\verb#http://www.numdam.org/item?id=JEDP_1987____A17_0#

\bibitem{Brooks Makover 1}
R. Brooks and E. Makover, ``The first eigenvalue of a Riemann surface,''
Electronic Research Announcements of the American Mathematical Society
{\bf 5} (1999) 76-81.\\
http://www.ams.org/journals/era/1999-05-11/S1079-6762-99-00064-5/

\bibitem{Brooks Makover 2}
R. Brooks and E. Makover, ``Riemann surfaces with large first eigenvalue,''
Journal d'Analyse Math\'ematique {\bf 83} (2001) 243-258.

\bibitem{Brooks Makover 3}
R. Brooks and E. Makover, ``Belyi surfaces,'' IMCP {\bf 15} (2001) 37-46.

\bibitem{Brooks Makover 4}
R. Brooks and E. Makover, ``Random Construction of Riemann Surfaces,''
J. Differential Geom. {\bf 68} (2004) 121-157, arXiv:math/0106251.

\bibitem{Cheeger}
Jeff Cheeger, ``A lower bound for the smallest eigenvalue of the Laplacian,''
in \emph{Problems in analysis (Papers dedicated to Salomon Bochner, 1969)},
pp. 195-199, Princeton Univ. Press, Princeton, 1970.

\bibitem{Cheeger Wikipedia}
\verb#http://en.wikipedia.org/wiki/Cheeger_constant#

\bibitem{Donnelly}
H. Donnelly, ``The differential form spectrum of hyperbolic space,''
Manuscripta Math. {\bf 33} (1981) 365-385.

\bibitem{Sarnak}
P. Sarnak, ``Arithmetic and geometry of some hyperbolic three manifolds,''
Acta Mathematica {\bf 151} (1983) 253-295.

\bibitem{Luo}
W. Luo, Z. Rudnick, and P. Sarnak, ``On the generalized Ramanujan Conjectures
for $\mathrm{GL}(n)$,'' Proc. Symp. Pure Math. {\bf 66-2} (1999) 301-311.

\bibitem{Rudnick Sarnak}
S.-y. Koyama, ``The First Eigenvalue Problem and Tensor Products of Zeta
Functions,'' Proceedings of the Japan Academy, Ser. A, Mathematical Sciences
{\bf 80} (2004-05) 35-39.

\bibitem{Orlando Park}
D.~Orlando and S.~C.~Park,
``Compact hyperbolic extra dimensions: a \mbox{M-theory} \mbox{solution} and
its implications for the LHC,'' JHEP {\bf 1008} (2010) 006,\\
arXiv:1006.1901 [hep-th].

\bibitem{Mazzeo Phillips}
R. Mazzeo and R.S. Phillips, ``Hodge theory on hyperbolic manifolds,''
Duke Math. J. {\bf 60} (1990), 509-559.

\bibitem{Donnelly Xavier}
H. Donnelly and F. Xavier, ``On the Differential Form Spectrum of Negatively
Curved Riemannian Manifolds,'' American Journal of Mathematics {\bf 106}
(1984) 169-185.\\
\verb#http://www.nd.edu/~fxavier/Publications/Donnely_Xavier_84.pdf#

\bibitem{Colbois Courtois 1}
B. Colbois and G. Courtois, ``Les valeurs propres inf\'erieures \`a
$\frac{1}{4}$ des surfaces de Riemann de petit rayon d'injectivit\'e,''
Comment. Math. Helv. {\bf 64} (1989) 349-362.

\bibitem{Colbois Courtois 2}
B. Colbois and G. Courtois, ``Convergence de vari\'et\'es et convergence du
spectre du Laplacien,'' Ann. Sci. \'Ecole Norm. Sup. {\bf 24} (1991)
507-518.

\bibitem{Bunke}
U. Bunke, ``The spectrum of the Dirac operator on the hyperbolic space,''
Math. Nachr. {\bf 153} (1991) 179-190.

\bibitem{Bar}
C. B\"ar, ``The Dirac Operator on Hyperbolic Manifolds of Finite Volume,''
J. Diff. Geom. {\bf 54} (2000) 439-488, arXiv:math/0010233 [math.DG].

\bibitem{Xue}
X. Xue, ``On the Betti numbers of a hyperbolic manifold,'' Geometric And
Functional Analysis {\bf 2} (1992) 126-136.

\bibitem{Cremmer Scherk}
E.~Cremmer and J.~Scherk,
``Spontaneous dynamical breaking of gauge symmetry in dual models,''
  Nucl.\ Phys.\  {\bf B72} (1974)  117-124.

\bibitem{Kalb Ramond}
M.~Kalb and P.~Ramond,
``Classical direct interstring action,''
  Phys.\ Rev.\  D {\bf 9} (1974) 2273-2284.

\bibitem{Inoue}
K.T. Inoue, ``Numerical Study of Length Spectra and Low-lying Eigenvalue
Spectra of Compact Hyperbolic 3-manifolds,'' Class. Quant. Grav. {\bf 18}
(2001) 629-652, arXiv:math-ph/0011012.

\bibitem{Leininger et al}
C.J. Leininger, D.B. McReynolds, W.D. Neumann, and A.W. Reid, ``Length and
eigenvalue equivalence,'' International Mathematics Research Notices 2007
(2007), rnm135-24, arXiv:math/0606343 [math.GT].

\bibitem{Perkins}
D.H. Perkins, \emph{Introduction to High Energy Physics}, third edition,
Addison-Wesley Publishing Company, Inc., 1987.

\bibitem{Giudice Rattazzi Wells}
G.~F.~Giudice, R.~Rattazzi and J.~D.~Wells,
``Quantum gravity and extra dimensions at high-energy colliders,''
Nucl.\ Phys.\  {\bf B544} (1999) 3-38, arXiv:hep-ph/9811291.

\bibitem{MSTW}
A.D. Martin, W.J. Stirling, R.S. Thorne, G. Watt, ``Parton distributions for
the LHC,'' Eur. Phys. J. {\bf C63} (2009) 189-285, arXiv:0901.0002
[hep-ph].\\
\verb#http://projects.hepforge.org/mstwpdf/plots/#\\
\verb#mstw2008lo68cl_allpdfs.eps#

\bibitem{Lueck}
W. L\"uck, ``Approximating $L^2$-invariants by their finite dimensional
analogues,'' Geom. and Func. Anal., {\bf 4} (1994) 455-481.\\
http://wwwmath.uni-muenster.de/users/lueck/publ/lueck/r.pdf

\bibitem{Clair Whyte}
B. Clair and K. Whyte, ``Growth of Betti Numbers,'' Topology
{\bf 42} (2003) 1125-1142, arXiv:math/0111120 [math.GT].

\bibitem{Wikipedia injectivity radius}
\verb#http://en.wikipedia.org/wiki/#\\
\verb#Glossary_of_Riemannian_and_metric_geometry#

\bibitem{Allendoerfer Weil}
C. B. Allendoerfer and A. Weil,
``The Gauss-Bonnet theorem for riemannian polyhedra,'' Trans. Amer. Math. Soc.
{\bf{53}} (1943) 101-129.

\bibitem{Chern}
S.-S. Chern, ``On the curvatura integra in a Riemannian manifold,'' Annals of
Mathematics {\bf 46} (1945) 674-684.

\bibitem{Selberg}
A. Selberg, ``On discontinuous groups in higher-dimensional symmetric spaces'',
in \emph{Contributions to function theory}, Tata Institute, Bombay (1960)
147-164.

\bibitem{Dalitz}
R.H. Dalitz, ``CXII. On the analysis of $\tau$-meson data and the nature of
the $\tau$-meson,'' Phil. Mag. {\bf 44} (1953) 1068-1080.

\bibitem{Fabri}
E. Fabri, ``A study of $\tau$-meson decay,'' Nuovo Cim. {\bf 11} (1954)
479-491.

\bibitem{Green Schwarz Witten}
M. B. Green, J, Schwarz and E. Witten, \emph{Superstring theory},
Vol. 1: \emph{Introduction}, Vol. 2: \emph{Loop amplitudes, anomalies and
phenomenology}, Cambridge University Press, 1987.

\bibitem{Israel}
W.~Israel,
``Singular hypersurfaces and thin shells in general relativity,''
Nuovo Cim.\ {\bf B44} (1966) 1.  Erratum: Nuovo Cim. \textbf{B48}, (1967) 463.

\bibitem{Chamblin Reall}
H.~A.~Chamblin and H.~S.~Reall,
``Dynamic dilatonic domain walls,''
Nucl.\ Phys.\  {\bf B562} (1999) 133-157, arXiv:hep-th/9903225.

\bibitem{Dyer Hinterbichler}
E.~Dyer and K.~Hinterbichler,
``Boundary Terms, Variational Principles and Higher Derivative Modified
Gravity,'' Phys.\ Rev.\  {\bf D79} (2009)  024028,\\
arXiv:0809.4033 [gr-qc].

\bibitem{Schwinger}
J. Schwinger, I. A. S. (Princeton) lectures, unpublished.

\bibitem{DeWitt}
B. S. DeWitt, in : \emph{Relativity, Groups and Topology}, eds. B. S. DeWitt
and C. DeWitt, Gordon and Breach, New York, 1964.

\bibitem{Nambu Jona Lasinio 1}
Y.~Nambu, G.~Jona-Lasinio,
``Dynamical Model of Elementary Particles Based on an Analogy with
Superconductivity. 1.,''
  Phys.\ Rev.\  {\bf 122} (1961)  345-358.

\bibitem{Nambu Jona Lasinio 2}
Y.~Nambu, G.~Jona-Lasinio,
``Dynamical Model Of Elementary Particles Based On An Analogy With
Superconductivity. Ii,''
  Phys.\ Rev.\  {\bf 124} (1961)  246-254.

\bibitem{Tseytlin}
A.~A.~Tseytlin,
``$R^4$ terms in 11 dimensions and conformal anomaly of $(2,0)$ theory,''
  Nucl.\ Phys.\  {\bf B584} (2000)  233-250, arXiv:hep-th/0005072.

\bibitem{Green Schwarz}
M.~B.~Green and J.~H.~Schwarz,
``Anomaly Cancellation In Supersymmetric $ D=10 $ Gauge Theory And Superstring
Theory,'' Phys.\ Lett.\ B {\bf 149} (1984) 117-122.
Scanned version from KEK: \\
\verb#http://ccdb4fs.kek.jp/cgi-bin/img_index?8412338#

\bibitem{Deser Seminara 1}
S.~Deser and D.~Seminara,
``Counterterms/M-theory corrections to D = 11 supergravity,''
Phys.\ Rev.\ Lett.\  {\bf 82} (1999) 2435-2438, arXiv:hep-th/9812136.

\bibitem{Deser Seminara 2}
S.~Deser and D.~Seminara,
``Tree amplitudes and two-loop counterterms in D = 11 supergravity,''
Phys.\ Rev.\  {\bf D62} (2000) 084010, arXiv:hep-th/0002241.

\bibitem{Peeters Plefka Stern}
K.~Peeters, J.~Plefka, and S.~Stern,
``Higher-derivative gauge field terms in the M-theory action,''
  JHEP {\bf 0508} (2005)  095, arXiv:hep-th/0507178.

\bibitem{Green Schwarz IIA}
M.~B.~Green and J.~H.~Schwarz,
``Covariant Description Of Superstrings,''
Phys.\ Lett.\   {\bf B136} (1984) 367-370.  Also in Bohm, A. et al.,
\emph{Dynamical groups and spectrum generating algebras}, vol. 2, 885-888, and
in Schwarz, J.H. (ed.) \emph{Superstrings}, Vol. 1, 372-375.

\bibitem{Witten Various Dimensions}
E.~Witten,
``String theory dynamics in various dimensions,''
Nucl.\ Phys.\   {\bf B443} (1995) 85-126, arXiv:hep-th/9503124.

\bibitem{Gross Sloan}
D.~J.~Gross and J.~H.~Sloan,
``The Quartic Effective Action for the Heterotic String,''
  Nucl.\ Phys.\  {\bf B291} (1987)  41-89.   Scanned version from KEK:\\
\verb#http://ccdb4fs.kek.jp/cgi-bin/img_index?200033932#

\bibitem{Policastro Tsimpis}
G.~Policastro and D.~Tsimpis,
``$R^4$, purified,''
  Class.\ Quant.\ Grav.\  {\bf 23} (2006)  4753-4780, arXiv: hep-th/0603165.

\bibitem{Cadabra 1}
K. Peeters,
Cadabra: A field-theory motivated approach to computer algebra.\\
http://cadabra.phi-sci.com/

\bibitem{Cadabra 2}
K. Peeters, ``Symbolic field theory with Cadabra,''
Computeralgebra Rundbrief {\bf 41} (2007) 16.

\bibitem{Cadabra 3}
K. Peeters, ``Introducing Cadabra: a symbolic computer algebra system for
field theory problems,'' arXiv:hep-th/0701238.

\bibitem{Cadabra 4}
L. Brewin, ``A brief introduction to Cadabra: a tool for tensor computations
in General Relativity,''  Comput. Phys. Commun. {\bf 181} (2010) 489-498,\\
arXiv:0903.2085 [gr-qc].

\bibitem{Cadabra 5}
K. Peeters, ``Cadabra: reference guide and tutorial,''
preprint AEI-2006-038.

\bibitem{Cadabra 6}
K. Peeters, ``A field-theory motivated approach to symbolic computer
algebra,'' Comp. Phys. Commun. {\bf 176} (2007) 550-558,
arXiv:cs/0608005 [cs.SC].

\bibitem{Lukas Ovrut Waldram}
A.~Lukas, B.~A.~Ovrut and D.~Waldram,
``On the four-dimensional effective action of strongly coupled heterotic
string theory,'' Nucl.\ Phys.\ B {\bf 532} (1998) 43-82, arXiv:hep-th/9710208.

\bibitem{Palais}
R.S. Palais, ``The principle of symmetric criticality,'' Comm. Math.
Physics {\bf 69} (1979) 19-30.
http://projecteuclid.org/euclid.cmp/1103905401

\bibitem{Deser Franklin}
S.~Deser, J.~Franklin, and B.~Tekin,
``Shortcuts to spherically symmetric solutions: A Cautionary note,''
  Class.\ Quant.\ Grav.\  {\bf 21} (2004)  5295-5296,\\
arXiv:gr-qc/0404120.

\bibitem{Torre}
C.~G.~Torre,
``Symmetric Criticality in Classical Field Theory,''
arXiv:1011.3429 [math-ph].

\bibitem{Guven}
R.~Gueven,
``Black $p$-brane solutions of $D = 11$ supergravity theory,''
Phys.\ Lett.\  {\bf B276} (1992) 49-55.

\bibitem{Duff Stelle}
M.~J.~Duff and K.~S.~Stelle,
``Multi-membrane solutions of $D = 11$ supergravity,''
Phys.\ Lett.\  {\bf B253} (1991) 113-118.  Scanned version from KEK: \\
\verb#http://ccdb4fs.kek.jp/cgi-bin/img_index?9010197#

\bibitem{Moore}
G.~W.~Moore,
``Anomalies, Gauss laws, and Page charges in M-theory,''
  Comptes Rendus Physique {\bf 6} (2005) 251-259, arXiv:hep-th/0409158.

\bibitem{h7bulk}
http://chrisaustin.info/h7bulk.html

\bibitem{Maxima}
Maxima, a Computer Algebra System.
http://maxima.sourceforge.net/

\bibitem{Bousso Polchinski}
R.~Bousso, J.~Polchinski,
``Quantization of four form fluxes and dynamical neutralization of the
cosmological constant,'' JHEP {\bf 0006} (2000)  006,\\
arXiv:hep-th/0004134.

\bibitem{Anderson}
M.T. Anderson, ``Dehn filling and Einstein metrics in higher dimensions,''
J.Diff.Geom. {\bf 73} (2006) 219-261, arXiv:math/0303260 [math.DG].

\bibitem{Wilczek embedding spin connection}
F. Wilczek, ``Geometry and Interaction of Instantons,'' in \emph{Quark
Confinement and Field Theory: Proceedings  of a Conference at the University
of Rochester, Rochester, NY, June 14-18, 1976}, Stump and Weingarten, eds.,
Wiley-Interscience, NY (1977) 211-219.

\bibitem{Charap Duff}
J.~M.~Charap and M.~J.~Duff,
``Space-Time Topology and a New Class of Yang-Mills Instanton,''
  Phys.\ Lett.\  {\bf B71} (1977)  219-221.

\bibitem{Witten Shelter Island}
E. Witten, ``Fermion Quantum Numbers in Kaluza-Klein Theory,'' in
\emph{Shelter Island II: Proceedings of the 1983 Shelter Island Conference
on Quantum Field Theory and the Fundamental Problems of Physics}, R.
Jackiw, N.N. Khuri, S. Weinberg, and E. Witten, eds., MIT Press, Cambridge,
Massachusetts, USA, 1985.

\bibitem{Candelas Horowitz Strominger Witten}
P. Candelas, G. Horowitz, A. Strominger and E. Witten, ``Vacuum
configurations for superstrings,'' Nucl. Phys. \textbf{B258} (1985)
46.

\bibitem{Everitt Maclachlan}
B. Everitt and C. Maclachlan, ``Constructing Hyperbolic Manifolds,''
in \emph{Computational \& Geometric Aspects of Modern Algebra},
M. Atkinson and N. Gilbert, eds., Proc ICMS Workshop, Edinburgh 1998, LMS
Lecture Note Series {\bf 275} (2000) 78-86, arXiv:math/9907139 [math.GT].

\bibitem{Lanner}
F. Lann\'er, ``On complexes with transitive groups of automorphisms,'' Comm.
Sem. Math. Univ. Lund {\bf 11} (1950) 1–71.

\bibitem{Kaplinskaya}
I. M. Kaplinskaja, ``Discrete groups generated by reﬂections in the faces of
simplicial prisms in Lobachevskian spaces,'' Math. Notes {\bf 15} (1974)
88–91.

\bibitem{Esselmann}
F. Esselmann, ``\"Uber kompakte hyperbolische Coxeter-Polytope mit wenigen
Facetten,'' Universit\"at Bielefeld, SFB 343, Preprint No. 94-087.

\bibitem{Tumarkin}
P. Tumarkin, ``Compact hyperbolic Coxeter $n$-polytopes with $n + 3$
facets,'' Electron. J. Combin. {\bf 14} (2007) R69,
arXiv:math/0406226 [math.MG].

\bibitem{GAP}
The GAP Group, GAP -- Groups, Algorithms, and Programming, Version 4.4.12;
2008.  http://www.gap-system.org.

\bibitem{Belolipetsky 2}
M. Belolipetsky, ``Addendum to: On volumes of arithmetic quotients of\\
$\mathrm{SO}\left(1,n\right)$,'' Ann. Scuola Norm. Sup. Pisa Cl. Sci. (5)
{\bf 6} (2007) 263-268,\\
arXiv:math/0610177 [math.NT].

\bibitem{Belolipetsky 1}
M. Belolipetsky, ``On volumes of arithmetic quotients of
$\mathrm{SO}\left(1,n\right)$,'' Ann. Scuola Norm. Sup. Pisa Cl. Sci.
(5) {\bf 3} (2004) 749-770, arXiv:math/0306423 [math.NT].

\bibitem{Everitt}
B. Everitt, ``Coxeter Groups and Hyperbolic Manifolds,''
Mathematische Annalen {\bf 330} (2004) 127-150,
arXiv:math/0205157 [math.GT].

\bibitem{Conder Maclachlan}
M. Conder and C. Maclachlan, ``Compact hyperbolic 4-manifolds of small
volume,'' Proc. Amer. Math. Soc. {\bf 133} (2005) 2469-2476.\\
http://www.ams.org/journals/proc/2005-133-08/S0002-9939-05-07634-3/

\bibitem{Long}
C. Long, ``Small volume closed hyperbolic 4-manifolds,'' Bull. London Math.
Soc. {\bf 40} (2008) 913-916.

\bibitem{XENON100}
XENON100 Collaboration, ``Dark Matter Results from 100 Live Days of XENON100
Data,'' arXiv:1104.2549 [astro-ph.CO].

\bibitem{Peter Woits blog}
\verb#http://www.math.columbia.edu/~woit/wordpress/#

\bibitem{Jesters blog}
http://resonaances.blogspot.com/

\bibitem{TexPict}
R. Rib\'o, Tex Pictures, a program to create easy drawings to include in
\LaTeX documents.
http://www.tex.ac.uk/ctan/graphics/texpict/

\bibitem{Ordercite}
G. Salam, Ordercite, a program to establish whether your bibliography is in
the same order as the citations to it.\\
\verb#http://www.lpthe.jussieu.fr/~salam/ordercite/#

\bibitem{TeXmacs}
J. van der Hoeven, GNU TeXmacs, a free ``what you see is what you want''
editing platform with special features for scientists.
http://www.texmacs.org/

\bibitem{KDE}
http://www.kde.org/

\bibitem{Debian}
http://www.debian.org/

\end{thebibliography}
\end{document}